\documentclass[twocolumn,times,tighten,twocolappendix]{aastex63}

\usepackage{amsmath}

\received{2021 May 30}
\revised{2021 September 14}
\accepted{2021 September 20}

\submitjournal{ApJ}

\begin{document}

\title{An IFU View of the Active Galactic Nuclei in MaNGA Galaxy Pairs}

\correspondingauthor{Y. Sophia Dai}
\email{ydai@nao.cas.cn}

\author[0000-0003-3087-318X]{Gaoxiang Jin}
\affiliation{Chinese Academy of Sciences South America Center for Astronomy (CASSACA), National Astronomical Observatories(NAOC),
20A Datun Road, Beijing 100012, China}
\affiliation{University of Chinese Academy of Sciences (UCAS), 
Beijing, 100049, China}

\author[0000-0002-7928-416X]{Y. Sophia Dai}
\affiliation{Chinese Academy of Sciences South America Center for Astronomy (CASSACA), National Astronomical Observatories(NAOC),
20A Datun Road, Beijing 100012, China}

\author[0000-0002-1370-6964]{Hsi-An Pan}
\affiliation{Max-Planck-Institut f\"ur Astronomie, K\"onigstuhl 17, D-69117 Heidelberg, Germany}

\author[0000-0001-7218-7407]{Lihwai Lin}
\affiliation{Institute of Astronomy and Astrophysics, 
Academia Sinica, Taipei 10617, Taiwan}

\author[0000-0002-8711-8970]{Cheng Li}
\affiliation{Department of Astronomy, Tsinghua University, Beijing 100084, China}

\author[0000-0001-5615-4904]{Bau-Ching Hsieh}
\affiliation{Institute of Astronomy and Astrophysics, 
Academia Sinica, Taipei 10617, Taiwan}

\author[0000-0002-3073-5871]{Shiyin Shen}
\affiliation{Key Laboratory for Research in Galaxies and Cosmology, Shanghai Astronomical Observatory, Chinese Academy of Sciences, 80 Nandan Road, Shanghai 200030, China}

\author[0000-0001-6763-5869]{Fang-Ting Yuan}
\affiliation{Key Laboratory for Research in Galaxies and Cosmology, Shanghai Astronomical Observatory, Chinese Academy of Sciences, 80 Nandan Road, Shanghai 200030, China}

\author[0000-0002-9767-9237]{Shuai Feng}
\affiliation{College of Physics, Hebei Normal University, 20 South Erhuan Road, Shijiazhuang 050024, China}
\affiliation{Hebei Key Laboratory of Photophysics Research and Application, Shijiazhuang 050024, China}
\affiliation{Key Laboratory for Research in Galaxies and Cosmology, Shanghai Astronomical Observatory, Chinese Academy of Sciences, 80 Nandan Road, Shanghai 200030, China}

\author[0000-0003-0202-0534]{Cheng Cheng}
\affiliation{Chinese Academy of Sciences South America Center for Astronomy (CASSACA), National Astronomical Observatories(NAOC),
20A Datun Road, Beijing 100012, China}

\author[0000-0003-1094-5190]{Hai Xu}
\affiliation{Chinese Academy of Sciences South America Center for Astronomy (CASSACA), National Astronomical Observatories(NAOC),
20A Datun Road, Beijing 100012, China}

\author[0000-0001-6511-8745]{Jia-Sheng Huang}
\affiliation{Chinese Academy of Sciences South America Center for Astronomy (CASSACA), National Astronomical Observatories(NAOC),
20A Datun Road, Beijing 100012, China}

\author[0000-0002-9808-3646]{Kai Zhang}
\affiliation{Lawrence Berkeley National Laboratory, 1 Cyclotron Road, Berkeley, CA 94720, USA}

\begin{abstract}
The role of active galactic nuclei (AGNs) during galaxy interactions and how they influence the star formation in the system are still under debate. 
We use a sample of 1156 galaxies in galaxy pairs or mergers (hereafter `pairs') from 
the MaNGA survey. 
This pair sample is selected by the velocity offset, projected separation, and morphology,
and is further classified into four cases along the merger sequence based on morphological signatures. 
We then identify a total of 61 (5.5\%) AGNs in pairs based on the emission-line diagnostics. 
No evolution of the AGN fraction is found, either along the merger sequence or compared to isolated galaxies (5.0\%). 
We observe a higher fraction of passive galaxies in galaxy pairs, 
especially in the pre-merging cases, and associate the higher fraction to their environmental dependence.
The isolated AGN and AGN in pairs show similar distributions in their global stellar mass, star formation rate (SFR), and central [\ion{O}{3}] surface brightness.
AGNs in pairs show radial profiles of increasing specific SFR and declining Dn4000 from center to outskirts, and no significant difference from the isolated AGNs. 
This is clearly different from star-forming galaxies (SFGs) in our pair sample,
which show enhanced central star formation, as reported before. 
AGNs in pairs have lower Balmer decrements at outer regions, possibly indicating less dust attenuation. 
Our findings suggest that AGNs likely follow an inside-out quenching
and the merger impact on the star formation in AGNs is less prominent than in SFGs.
\end{abstract}

\keywords{galaxies: evolution -- galaxies: interactions -- galaxies: active -- galaxies: star formation -- galaxies: Seyfert}

\section{Introduction} \label{sec:intro}
Galaxy-galaxy interaction plays an important role in the evolution of galaxies. 
Theoretically, the merging of galaxies will result in
the in-fall of gas towards the center and trigger central star formation \citep{1991ApJ...370L..65B,2018MNRAS.479.3952B}. 
Several numerical simulations \citep[e.g.][]{2000MNRAS.311..576K,2005Natur.433..604D,2006ApJS..163....1H,2006ApJ...652..864H,2016A&A...592A..62G,2017MNRAS.469.4437C} also predicted the emergence of active galactic nuclei (AGN) in 
galaxy mergers. 
According to simulations, the merging of two gas-rich \citep[$M_{gas}\,=\,20\%\,M_{*}$, e.g.][]{2006ApJS..163....1H} equal-mass galaxies will drive the gas into the center owing to the loss of angular momentum. 
The supply of infalling gas to the center will fuel both nuclear starbursts and the growth of supermassive black holes (SMBHs), 
which would experience several peaks from the first encounter to the final coalescence \citep[e.g.][]{2006ApJ...652..864H}.

To study the merger effects from an observational view, ideally we should build an ongoing merger sample along the merger sequence.
There are several approaches to build a merger sample,
but all have pros and cons. 
For example, by selecting galaxy pairs through projected separation and velocity offset \citep[e.g.][]{2002ApJ...565..208P,2004ApJ...617L...9L}, 
one can build statistically significant galaxy pair samples from large spectroscopic surveys \citep[e.g.][]{2008AJ....135.1877E,2015MNRAS.451.3249A,2016RAA....16...43S,2019ApJ...880..114F}.
But this method requires spectroscopic redshifts and these samples often suffer from incompleteness issues \citep[e.g.][]{2008ApJ...685..235P}.
Visual classification \citep[e.g.][]{2013MNRAS.435.2835W,2015ApJS..221...11K}
is a powerful tool to select the late-stage and post merging systems
which could be missed in spectroscopic pairs.
Machine learning is a recent, effective method \citep[e.g.][]{2018MNRAS.476.3661D,2018MNRAS.479..415A,2019A&A...626A..49P,2019MNRAS.483.2968W}.
However, due to the limitation of training sets \citep[e.g.][]{2019MNRAS.490.5390B},
machine learning method is yet to achieve a high accuracy compared to visual classifications. 
Most current works, including this one, 
still adopt the physical selection of pairs followed by visual classification. 
Combining these two methods includes both galaxy pairs and late-stage mergers. 
Hereafter for convenience we refer to both galaxy pairs and merger systems as galaxy `pairs'.

In galaxy pairs, the enhancement of star formation has been widely observed,
often based on the comparison of the star formation rate (SFR)
with isolated control galaxies. 
These enhancements have been found in various SFR indicators, 
including stronger emission-lines \citep[e.g.][]{1987AJ.....93.1011K,2000ApJ...530..660B,2003MNRAS.346.1189L,2006AJ....132..197W,2008MNRAS.385.1903L,2010AJ....139.1857W},
bluer colors \citep[e.g.][]{1978ApJ...219...46L,2005AJ....130.2043P,2007ApJ...660L..51L,2010AJ....140.1975S,2011MNRAS.412..591P},
and stronger infrared emission \citep[e.g.][]{1991ApJ...374..407X,1996ARA&A..34..749S,2006AJ....132.2243G,2011A&A...535A..60H}. 
The level of enhancements varies with pairs' mass ratios \citep{2008AJ....135.1877E} 
or with different morphologies
\citep[e.g.][]{2010ApJ...713..330X,2012A&A...548A.117Y}.

According to simulations and theoretical predictions, 
mergers are expected to facilitate the accretion onto the central SMBHs, 
and trigger AGNs \citep[e.g.][]{2006ApJS..163....1H}. 
Observationally, enhanced AGN luminosity is found in galaxy pairs. 
Compared to isolated galaxies, 
[\ion{O}{3}] luminosity, proxy for AGN luminosity,
is found to increase by 0.7 to 0.9 dex in pairs \citep[e.g.][]{2012ApJ...745...94L,2013MNRAS.435.3627E,2018A&A...618A.149A}. 
Similarly, AGNs in galaxy pairs are found to have a higher X-ray detection rate (58\%) 
than AGN in isolated galaxies (17\%) \citep{2020ApJ...900...79H}.
In addition, mergers appear to play a dominant role 
in the triggering and fueling of high-luminosity AGNs.
The most luminous AGNs are often found to be associated with signatures of merging,
such as tidal tails, asymmetric morphology, bridges, and shells.
Using deep and high-resolution $Hubble\ Space\ Telescope$ imaging, 
interaction features have been found in various quasar samples:
in more than 80\% of AGNs selected from the FIRST-2MASS red quasar survey \citep{2008ApJ...674...80U,2015ApJ...806..218G}, 
in 4 out of the 5 nearby early-type quasars from \citet{2008ApJ...677..846B}, 
in 57\% of the Palomar-Green quasars \citep{2009ApJ...701..587V},
and in 62\% of the hosts of highly-obscured AGNs \citep{2016ApJ...822L..32F}.
\citet{2018ApJ...853...63D} found that in CANDLES/COSMOS field, compared with X-ray AGNs,
infrared selected AGNs are more likely to have disturbed morphologies. 

Another evidence for merger triggered AGN activities is the 
increased AGN fractions in galaxy pairs. 
However, this conclusion is still elusive as 
different results have been found.
For instance, in optically-selected AGNs in galaxy pairs,
the AGN fraction enhancement has been reported in several studies to be 1.4$\times$ to 2.4$\times$ in some studies \citep[e.g.][]{1985AJ.....90..708K,2007AJ....134..527W,2011MNRAS.418.2043E,2013MNRAS.435.3627E},
but not in other samples \citep[e.g.][]{2001AJ....122.2243S,2006MNRAS.371..786C,2007MNRAS.375.1017A,2008AJ....135.1877E,2010MNRAS.401.1552D}.
For X-ray selected AGNs, 
AGN fraction enhancement has been reported found in 
\citet{2011ApJ...743....2S,2014AJ....148..137L,2020MNRAS.499.2380S},
but not in a much larger sample of \citet{2020ApJ...904..107S}.  
As for AGNs that are infrared color selected,
most works have found a higher AGN fraction in pairs than in isolated control sample \citep[e.g.][]{2014MNRAS.441.1297S,2017MNRAS.464.3882W,2018PASJ...70S..37G,2019MNRAS.487.2491E,2020A&A...637A..94G}. 
In addition, \citet{2016A&A...592A..30A} found more radio AGNs in pair or cluster environments.
Recently, \citet{2021ApJ...909..124S} built a multi-wavelength sample including optical, X-ray, infrared, 
and radio selected AGNs, but found no AGN fraction excess in galaxy pairs.
Various factors, such as different pair selections, control sample selections, the sizes of the sample,
the redshift bins, and the different merger stages   
could all contribute to the diverse observation results.

Most previous works focused on galaxies' global properties,
due to the lack of spatially resolved spectra.
Integral field unit (IFU) observations offer a new opportunity
to study thousands of nearby galaxies in sub-galactic scales \citep[e.g.][]{2020ARA&A..58...99S}.
For instance, the MaNGA \citep[Mapping Nearby Galaxies at Apache Point Observatory, ][]{2015ApJ...798....7B} survey is 
one of the largest IFU survey, 
which has observed $\sim$10\,000 galaxies at redshift of $\sim$0.02-0.1.
IFU surveys have two advantages to study galaxy pairs:
1. their high spatial sampling spectra allow the confirmation of the
accurate velocity offset between galaxies,
and we can identify the pairs
with very small projected separations;
2. IFU observations also offer two-dimensional dynamical information,
so that we can analyze the resolved properties of the pair systems.
Several works have taken advantage of IFU 
to study the spatial extent of star formation in star-forming galaxy (SFG) pairs or mergers
\citep[e.g.][]{2009ApJ...698.1437K,2013MNRAS.432..285S,2014A&A...567A.132W,2015A&A...579A..45B,2018A&A...613A..13Y,2019MNRAS.482L..55T,2019ApJ...881..119P,2021ApJ...909..120S},
and have found enhanced SFR at different radii of interacting SFGs. 
Specifically,
\citet{2019MNRAS.482L..55T} found a centrally-peaked SFR enhancement 
and general metallicity suppression in MaNGA star-forming post-mergers;
while \citet{2019ApJ...881..119P} found that 
the SFR enhancement in SFGs pairs emerges after the first encounter. 
In addition, morphologically, \citet{2021MNRAS.501...14L} found a higher pair fraction in 
both SFGs and quiescent galaxies that show misaligned gas-stellar rotation. \citet{2020ApJ...892L..20F} showed that galaxies in pairs have higher kinematic asymmetry.

IFU studies on the AGN properties in galaxy pairs,
on the other hand, are still lacking. 
Using earlier MaNGA data, \citet{2018ApJ...856...93F} found 14 AGN binaries
and discovered an increase of binary AGN systems in pairs with smaller separations,
but no analysis on the resolved properties was performed. 
\citet{2019MNRAS.482..194B} showed that MaNGA AGNs have centrally suppressed star formation.
Whether star formation (SF) is enhanced or suppressed in pairs with AGN remains an open question.

With the MaNGA survey,
we now construct the largest IFU sample of galaxy pairs with AGNs. 
In this paper, we aim to study the sub-galactic properties including star formation, age, and extinction
in galaxy pairs with one or both AGNs, 
to understand the merger effects on the star formation condition of these AGN host galaxies. 
In addition, we will also classify our sample into different merger cases
in order to study the change of AGN fraction and galaxies' resolved SF properties along the merger sequence. 
This paper is structured as follows. Sec.~\ref{sec:data} is the data overview, 
pair sample selection, merger sequence definition,
and control sample selection.
In Sec.~\ref{sec:agnclass}, we select the AGNs in our sample
and study their fractions along the merger sequence. 
In Sec.~\ref{sec:analyse}, we present the global and resolved properties of our sample,
and compare them with isolated control samples.
We compare our work with previous works and discuss the selection biases and caveats in Sec.~\ref{sec:discuss}.
We summarize our results in Section \ref{sec:conclusion}. 
Throughout this paper, we use the AB magnitude system \citep{1983ApJ...266..713O},
the Salpeter initial mass function \citep[IMF,][]{1955ApJ...121..161S},
and adopt a $\rm \Lambda CDM$ cosmology 
with $\rm \Omega = 0.3$, $\rm \Lambda = 0.7$ 
and $\rm H_{0} = 70\ km \ s^{-1} Mpc^{-1}$.

\section{Sample Overview} \label{sec:data}
\subsection{The MaNGA Data} \label{subsec:manga}

MaNGA is one of the major surveys of SDSS--IV \citep[The fourth-generation Sloan Digital Sky Survey;][]{2017AJ....154...28B},
which aims to obtain resolved spectroscopy for $\sim$10,000 nearby galaxies \citep{2015AJ....150...19L},
using 17 science IFUs \citep{2015AJ....149...77D} over the 2.5 m Sloan Telescope's $3\degr$ diameter field of view (FOV).
These IFUs vary in diameter from 12$\arcsec $ to 32$\arcsec $ (19 to 127 fibers).
Each 2$\arcsec $ fiber has a spatial resolution of $\sim$1kpc at the peak redshift of $z\sim 0.03$.
MaNGA's observed wavelength range (3600 -- 10300 \AA)
can cover most strong nebular lines out to $z\sim 0.4$.
This includes the important lines used in the BPT excitation diagnostic diagram \citep{1981PASP...93....5B}, 
which is widely used to identify galaxy types between AGNs and SFGs.
MaNGA has spectral resolution that varies from $R\sim 1400$ at 4000 \AA \ to $R\sim 2600$ at 9000 \AA \ \citep{2016AJ....151....8Y}.
Target galaxies are covered out to at least 1.5 $R_{e}$ \citep{2016AJ....152..197Y}.
MaNGA's parent sample is made of 641,409 galaxies with spectroscopic data from
NASA-Sloan-Atlas\footnote{NSA; M. Blanton; \url{http://www.nsatlas.org/}\label{footnote:nsa}},
based on the SDSS DR7 main galaxy sample \citep{2009ApJS..182..543A}.
Detailed target selection for MaNGA can be found in \citet{2017AJ....154...86W}.

Our sample is drawn from the public data release MaNGA Product Launch-6 (MPL-6 and SDSS DR15),
which contains 4691 IFU observations within the survey's first 4 years of operation.
The emission-lines and spectral indices are from Data Analysis Pipeline \citep[{\tt DAP},][]{2019AJ....158..160B,2019AJ....158..231W},
the official high level data product of MaNGA.
DAP uses the stellar templates from MILES library \citep[Medium-resolution Isaac Newton Telescope library of empirical spectra,][]
{2006MNRAS.371..703S,2011A&A...532A..95F} and adopts the {\tt pPXF} \citep[penalized pixel-fitting,][]{2004PASP..116..138C,2017MNRAS.466..798C}
as the spectral-fitting routine. 
The integrated and resolved dust-corrected stellar masses are taken from {\tt Pipe3D}\footnote{\url{https://data.sdss.org/datamodel/files/MANGA_PIPE3D/MANGADRP_VER/PIPE3D_VER}}
\citep{2016RMxAA..52...21S,2016RMxAA..52..171S,2018RMxAA..54..217S},
another model-derived MaNGA data product.

\subsection{Identification of Galaxy Pair Systems}
\label{subsec:pairselect}
MaNGA galaxies and most of their neighbors have spectroscopic redshifts from SDSS single fiber spectra. 
We adopt a two-step pair selection, first based on 
projected distances and velocity offsets,
and then the late-stage mergers are visually selected.
These two steps allow us to select galaxy pairs 
from the incoming merging phase till the final coalescence.
Similar to other pair selections \citep[e.g.][]{2002ApJ...565..208P,2004ApJ...617L...9L,2019ApJ...881..119P},
galaxies in our pair sample are required to have 
a close spectroscopic companion at a projected separation $\rm \Delta d \,<\,50\, kpc\, h^{-1}$ (i.e. 71.4\,kpc)
and a line-of-sight velocity difference $\rm \Delta v \, < \, 500\, km\, s^{-1}$. 
This method misses mergers at their late merging stage 
due to the lack of the redshifts of the companions, 
or mergers in the coalescence stage, where only one source is identified. 
Therefore, we also visually check all MPL-6 galaxies 
and identify the missing late-stage pairs or mergers 
based on their morphology from the SDSS $gri$ images.
Out of the 4622 (of 4691) unique MaNGA MPL-6 targets,
we eventually identify 994 unique galaxy pair systems,
with a total of 1156 galaxies covered in MaNGA.
This is the same parent sample as in \citet{2019ApJ...881..119P}. 
Among these 994 galaxy pairs/mergers, 
46 pairs have individual IFU coverage for both member galaxies;
116 pairs have both members covered in the same IFU cube;
125 are mergers in late-stage coalescence covered with one single IFU cube;
and the remaining 707 pair systems have only one member galaxy with a MaNGA IFU coverage.

\subsection{Merger Sequence Definition}\label{subsec:mergersequence}
Simulations have predicted that the merging of two galaxies 
would experience several passages before the final coalescence \citep[e.g.][]{1972ApJ...178..623T,1988ApJ...331..699B,1992ARA&A..30..705B}.
Thus, the projected separation alone is not sufficient
to define the merger sequence.
Therefore, we combine the kinematic information with the morphological features to classify the merger stages
and divide our sample into four cases, 
to represent the possible merger sequences,
same as \citet{2019ApJ...881..119P}. 
The classification follows the following criteria: 
\begin{itemize}
  \item{Case 1 \textendash \ Well-separated pairs which do not show any morphology distortion (i.e. incoming pairs, before the first pericenter passage).}
  \item{Case 2 \textendash \ Close pairs showing strong signs of interaction, such as tital tails and bridges (i.e. likely at the first pericenter passage).}
  \item{Case 3 \textendash \ Well-separated pairs, showing weak morphology distortion (i.e. approaching the apocenter or just passing the apocenter).}
  \item{Case 4 \textendash \ Two components strongly overlapping with each other and showing strong morphological distortion (i.e. final coalescence phase), or single galaxies with obvious tidal features such as tails and shells (post-mergers).}
\end{itemize}

\begin{figure}
  \epsscale{1.15}
  \plotone{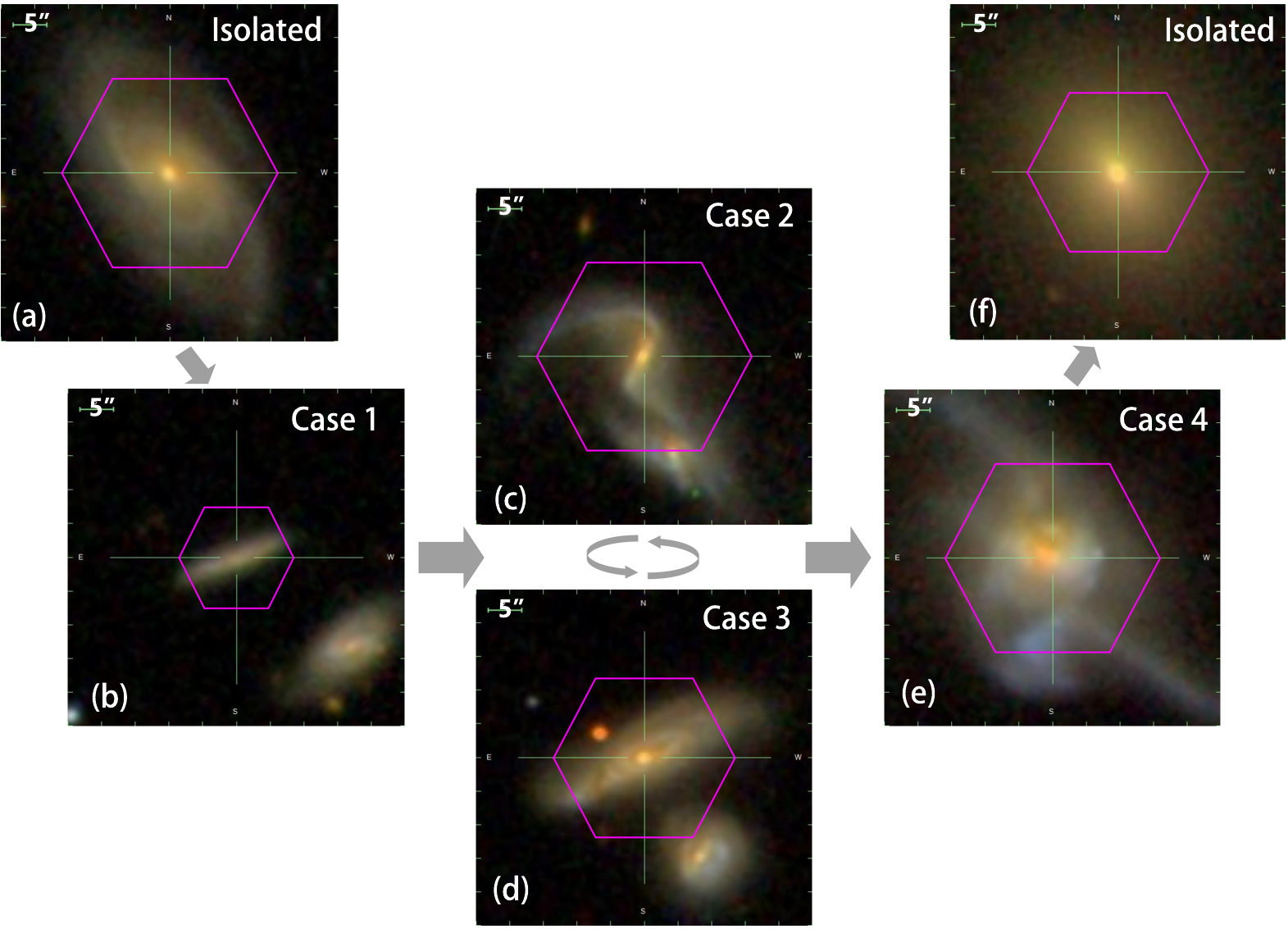}
  \caption{Illustration of the possible merger sequence.
  Examples of the SDSS $gri$-composite color images for the four merging cases (b, c, d, e) and isolated galaxies (a, f).
  The MaNGA Plate-IFU numbers are (a):9500-12702; (b):8485-3704; (c):8241-12705; (d):8082-9102; (e):9507-12704; (f):8984-9101, respectively.
  The magenta hexagons mark the corresponding IFU's FOV.
  The arrows indicate the possible merger sequence among our four merger cases and between isolated galaxies.
  \label{fig:4CASES}}
  \end{figure}

Each pair candidate is visually classified by four expert classifiers
and divided into the above four cases.
The classification result is decided by the majority if possible,
otherwise it is discussed on an individual basis by all inspectors
(this only applies to $\sim$5\% of objects in our sample).
We note that although Case 1 and Case 4 represent the incipient and final stages 
of the merging galaxies,
the relative chronological order of Case 2 and Case 3 is not clear.
Therefore, in order to avoid confusions, we refer to them as `Case' instead of `Stages'. 
We reach a final parent sample of 441 unique pair systems in Case 1, 
119 in Case 2, 
265 in Case 3, 
and 169 in Case 4.
Figure \ref{fig:4CASES} shows illustrations of examples 
of SDSS $gri$-composite images, for each case and for isolated galaxies.
The galaxy morphology in each case is consistent with
the morphological signatures of the Toomre Sequence (\citealt{1977egsp.conf..401T},
also see \citealt{2002ApJS..143..315V} and \citealt{2015A&A...582A..21B}),
and in simulated mergers 
(e.g., Figure 8 in \citealt{2012ApJ...746..108T} and 
Figure 2 in \citealt{2015MNRAS.448.1107M}). 
Factors other than morphology may affect the SF properties in pairs,
such as the encounter geometry \citep[e.g.][]{2007A&A...468...61D},
mass ratio \citep[e.g.][]{2008MNRAS.384..386C},
gas-richness \citep[e.g.][]{2015MNRAS.449.3719S,2018MNRAS.476.2591V},
and the relative morphological types of the member galaxies \citep[e.g.][]{2016ApJS..222...16C}.
We do not control these factors in our merger sequence classification due to
our limited sample size.

\subsection{Control Samples}\label{subsec:controlsample}
To better estimate the effect of merging for various physical parameters,
control samples of isolated galaxies are needed.
In order to define the various control samples, we
firstly select galaxies without a physical companion
($\rm \Delta d \,>\,150\, kpc\, h^{-1}$  or $\rm \Delta v \, > \, 500\, km\, s^{-1}$) in MaNGA.
This way we construct a parent sample of 2317 isolated galaxies.
The SFR is known to increase with the stellar mass, as shown in the star formation main sequence \citep[e.g.][]{2007ApJ...660L..43N}.
To make a fair comparison of the SF conditions in the various samples with limited mass effect,
we further define the control samples on a  mass-controlled basis. 
Based on the galaxy types and similar stellar mass requirement,
we build a series of control samples from the parent isolated sample, namely, 
the isolated AGN sample,
the isolated SFG sample,
and the isolated passive galaxy sample (for various galaxies' definition, see Sec~\ref{subsec:bpt}). 
Table~\ref{tab:control} summarizes the various subsamples used in the following analysis,
as well as the Kolmogorov-Smirnov (K-S) test probability p values of the relative stellar mass distributions to the pair subsamples. 
Given the intrinsic different mass distributions between the AGN and SFGs, 
to make a fair comparison of their radial profiles in Sec.~\ref{subsec:suppression}, 
we further require that both the AGN and SFG subsamples
to have a stellar mass between $\rm 10^{10.0} - 10^{11.0}\, M_{\sun}$.
Similarly, for the comparison between AGN and the passive galaxies (retired galaxies and lineless galaxies), 
we also require the similar mass distribution 
and limit their stellar mass to be between $\rm 10^{9.9} - 10^{11.6}\, M_{\sun}$.

\begin{deluxetable*}{cccccc}
  \tablecaption{Information of the various subsamples and their relevant control samples}
  \tablewidth{0pt}
  \tablehead{
  \colhead{Related figure(s)} & \colhead{Subsample} & \colhead{Number of galaxies} & \colhead{Mass range} & \colhead{Median Mass}  & \colhead{K-S test p} \\
  \colhead{ } & \colhead{ } & \colhead{(total)} & \colhead{log($M_{*}/M_{\sun}$)} & \colhead{log($M_{*}/M_{\sun}$)} & \colhead{ } 
  }
  \startdata
  {Paired AGN vs. Isolated AGN} & {\bf AGNs in pairs} & {\bf 61} & {\bf 9.94--11.55} & {\bf 10.93} & $/$   \\
  {(Figure~\ref{fig:sfms}, \ref{fig:o3sb}, \ref{fig:profile}, \ref{fig:profile_case})}& Isolated AGNs & 116  & 9.93--11.70 & 10.80 & 0.05  \\
\hline
  {} & {\bf AGNs in pairs} & {\bf 34} & {\bf 10.11--10.97} & {\bf 10.73} & $/$   \\
  {AGN vs. SFG} & Isolated AGNs & 36  & 10.11--10.97 & 10.73 & 0.99  \\
  {(Figure~\ref{fig:agnsf})} & SFGs in pairs & 50 & 10.11--10.97 & 10.68 & 0.38  \\
  {} & Isolated SFGs & 101 & 10.12--10.97 & 10.66 & 0.40  \\
  \hline
  {} & {\bf AGNs in pairs} & {\bf 61} & {\bf 9.94--11.55} & {\bf 10.93} & $/$   \\
  {AGN vs. passive galaxy} & Isolated AGNs & 71 & 10.11--11.40 & 10.84 & 0.33  \\
  {(Figure~\ref{fig:agnrg})} & Passive in pairs & 311 & 9.95--11.55 & 11.00 & 0.50  \\
  {} & Isolated Passive & 352 & 9.94--11.55 & 10.94 & 1.00  \\
  \enddata
  \tablecomments{Information of the control samples used in different analysis. 
  From left to right: the names of the subsamples, total numbers of galaxies, mass ranges, median stellar masses, and K-S test p values as compared to the paired AGN subsample (bold font).}
  \label{tab:control}
  \end{deluxetable*}

\section{AGN Classification} \label{sec:agnclass}
\subsection{Emission-line Classification} \label{subsec:bpt}
Nebular emission-lines from the narrow line region (NLR) of an AGN 
show different flux ratios from those from \ion{H}{2} regions, 
and are widely used to classify AGNs from SFGs.
With MaNGA's high quality spectra,
we adopt various emission-line diagnostics to classify the AGNs.
In this work, we use both the original [\ion{N}{2}]-BPT \citep{1981PASP...93....5B} 
and the modified [\ion{S}{2}]-BPT diagrams \citep{1987ApJS...63..295V}, 
which utilize combinations of 
the [\ion{O}{3}]$_{\lambda 5007}$/${\rm H\beta}$ $vs$ [\ion{N}{2}]$_{\lambda 6584}$/${\rm H\alpha}$, 
[\ion{O}{3}]$_{\lambda 5007}$/${\rm H\beta}$ $vs$ [\ion{S}{2}]$_{\lambda\lambda 6716,6731}$/${\rm H\alpha}$ 
line ratios. 
We also adopt the 
${\rm H\alpha}$ equivalent width (EW) $vs$ [\ion{N}{2}]$_{\lambda 6584}$/${\rm H\alpha}$ diagram \citep[WHAN,][]{2010MNRAS.403.1036C} 
to single out `retired galaxies' (RGs).

Given the fact that SMBHs locate in the center of galaxies,
we use the mean value of the central 3$\times$3 spaxels ($\rm 1.5\arcsec \times 1.5\arcsec$) of the galaxy,
for emission-line based  classifications.
Therefore, the galaxy types reported here 
represent the galaxies' nuclear properties. 
`AGN-like' spaxels in the outskirts of a galaxy will not be considered. 
For instance, if a galaxy does not have emission line features in the central region, 
but have strong star-forming regions in the disk or outskirt, 
it will still be classified as a lineless galaxy.

\begin{figure*}
  \plotone{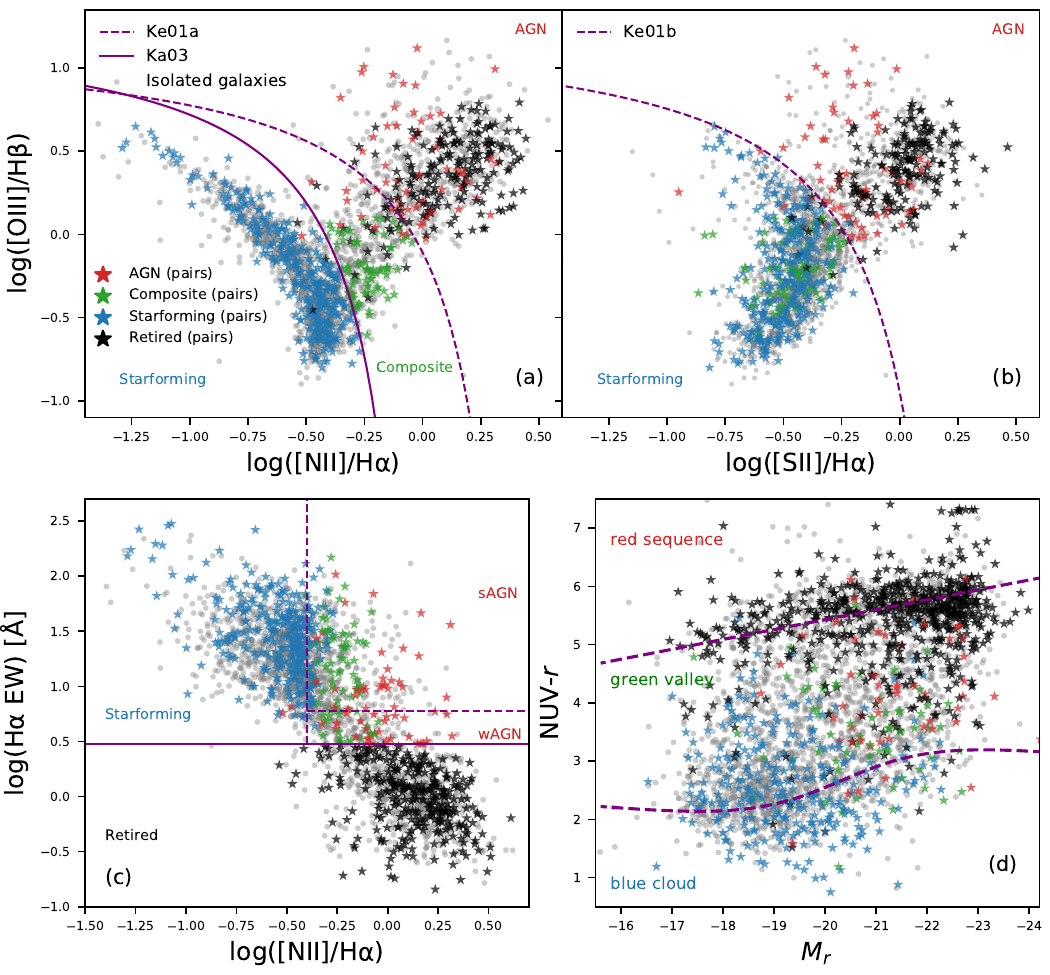}
  \caption{The (a): original and (b): modified BPT diagrams. 
  (c): the WHAN diagram and (d):
  NUV-$r$ $vs$ $M_{r}$ color-magnitude diagram using k-corrected magnitudes from the NSA catalog. 
  The gray dots are all the BPT-classified isolated MaNGA MPL-6 galaxies. 
  Galaxies in pairs are plotted as colored stars,
  with red for AGNs, green for composite galaxies,
  blue for SFGs,
  and black for retired galaxies. 
  For (a) and (b), the dividing curves are from
  \citet{2001ApJ...556..121K} and \citet{2003MNRAS.346.1055K}
  and relevant galaxy types are marked in the corresponding panels.
  In (c), the dividing lines are from \citet{2011MNRAS.413.1687C}, which classifies the galxies into SFGs,
  strong AGNs, weak AGNs, and retired galaxies.
  In (d), SFGs (blue), RGs (black), and AGNs (red) tend to lie in the
  `blue cloud', `red sequence', and `green valley' regions, respectively.
  The dividing lines are from \citet{2007ApJS..173..293W}, 
  corrected for the underestimated NUV flux by 0.3 mag found in nearby galaxies$^{\ref{footnote:nsa}}$.
  \label{fig:bptall}}
  \end{figure*}

In the [\ion{N}{2}]-BPT diagram
(Figure \ref{fig:bptall}, a),
the dashed curve (Equation(\ref{niike01}), Ke01a)
marks the starburst line from \citet{2001ApJ...556..121K}
and the solid curve marks the empirical separation 
between AGNs and SFGs from Equation(\ref{niika03}) from \citet{2003MNRAS.346.1055K}.
Star-forming galaxies locate below the Ke01a curve and
AGN-like galaxies locate above the Ka03 curve.
Galaxies between these two curves are considered to have radiation contribution
from both the star formation and the SMBH accretion \citep{2009MNRAS.397..135K},
and are considered as `composite' galaxies.
We find that most of the composite galaxies
are indeed in the star-forming region of the [\ion{S}{2}]-BPT diagram. 
To avoid the contamination from SFGs, 
we restrict our AGN selections to only the AGN regions (see Figure \ref{fig:bptall}),
and do not include the composite galaxies in the [\ion{N}{2}]-BPT diagram. 
The separation curves for the [\ion{N}{2}]- and [\ion{S}{2}]-BPT diagrams are summarized below:
\begin{equation}
  {\rm log([O\,III]/H\beta)=\frac{0.61}{log([N\,II]/H\alpha)-0.47}+1.19 \ ;\ Ke01a}
  \label{niike01}
\end{equation}
\begin{equation}
  {\rm log([O\,III]/H\beta)=\frac{0.61}{log([N\,II]/H\alpha)-0.05}+1.30 \ ;\ Ka03}
  \label{niika03}
\end{equation}
\begin{equation}
  {\rm log([O\,III]/H\beta)=\frac{0.72}{log([S\,II]/H\alpha)-0.32}+1.30 \ ;\ Ke01b}
  \label{siike01}
\end{equation}

Since the hot evolved stellar populations such as post-AGB stars
can also produce similar line ratios in the AGN region of the BPT diagrams \citep{1994A&A...292...13B,2012ApJ...747...61Y}, 
we further remove the `inactive' galaxies from our AGN sample
based on the WHAN diagram. 
WHAN diagram is based on the fact that 
the equivalent width of $\rm H\alpha$ ($\rm H\alpha\ EW $) is a robust proxy for measuring the photo-ionization by stellar populations older than 100 Myr
\citep{2011MNRAS.413.1687C}. 
We adopt the suggested empirical division between RGs 
and AGNs at 3\,\AA, and only keep galaxies with $\rm H\alpha$
EW $\geqslant$ 3\,\AA \ in our final AGN sample. 

To summarize, our AGN selections in the nuclear region 
follow these criteria:

1. We require all emission-lines used in the BPT diagram (H$\alpha$, H$\beta$, [\ion{O}{3}], [\ion{N}{2}] or [\ion{S}{2}])
to have a signal-noise-ratio (S/N) greater than 5. 
If a galaxy's central region has a well fitted continuum but 
the S/N of $\rm H\alpha$ is lower than 5, or 
includes weak or no $\rm H\alpha$ emission, 
it will be classified as a lineless galaxy.

2. We use the WHAN diagram to select retired galaxies (RGs),
defined as galaxies with nuclear $\rm H\alpha$ EW $<$ 3\AA,
regardless of their positions in the BPT diagrams.

3. For galaxies with $\rm H\alpha$ EW $\rm \geqslant$ 3\AA,
we classify the galaxy as an AGN if it falls in either the [\ion{N}{2}]-AGN or the [\ion{S}{2}]-AGN regions. 
In our final sample of 61 AGNs in pairs, 
a total of 43 galaxies are classified as AGNs 
by both BPT criteria,
while 6 are AGNs only selected in the [\ion{N}{2}]-BPT diagram, 
and 12 are selected in the [\ion{S}{2}]-BPT diagram only.

4. We then use the [\ion{N}{2}]-BPT diagram to 
classify composite galaxies and SFGs.

\startlongtable
\begin{deluxetable*}{ccccccccc}
\tablecaption{Parameters of the 61 AGNs in galaxy pairs\label{tab:info}}
\tablewidth{0pt}
\tabletypesize{\scriptsize}
\tablehead{
\colhead{Plate-IFU} & \colhead{RA} & \colhead{DEC} & \colhead{$z$} &
\colhead{Merger Case} & \colhead{log($M_{*}$)} & \colhead{log(SFR)} &
\colhead{Morphology} & \colhead{$\rm \Sigma_{[OIII]}$}\\
\colhead{} & \colhead{$\degr$} & \colhead{$\degr$} & \colhead{} &
\colhead{} & \colhead{log($M_{\sun}$)} & \colhead{log($M_{\sun}\, yr^{-1}$)} &
\colhead{} & \colhead{log($erg\, s^{-1}\, kpc^{-2}$)}
}
\decimalcolnumbers
\startdata
7975-12702 & 323.5212 & 10.4219 & 0.0774 & 1           & 10.73     & 0.147  & E          & 38.32    \\
8132-6101  & 111.7337 & 41.0267 & 0.1294 & 3           & 11.55     & 0.980  & SBb        & 39.92    \\
8247-6101  & 136.0896 & 41.4817 & 0.0245 & 2           & 10.77     & -0.551 & E          & 37.83    \\
8256-12704 & 166.1294 & 42.6246 & 0.1261 & 1           & 11.42     & 0.532  & E          & 40.06    \\
8249-3704  & 137.8748 & 45.4683 & 0.0268 & 3           & 10.46     & -0.384 & SBa        & 39.05    \\
8329-3701  & 213.4322 & 43.6625 & 0.0893 & 1           & 11.08     & -0.028 & E          & 38.76    \\
8459-3702  & 146.7091 & 43.4238 & 0.0722 & 3           & 11.25     & 0.736  & Sa         & 39.07    \\
8452-12705 & 157.9377 & 46.6717 & 0.0249 & 1           & 10.36     & 0.323  & SABc       & 37.35    \\
8465-12704 & 198.1419 & 48.3666 & 0.0558 & 1           & 10.95     & 0.019  & Sa         & 39.30    \\
8447-9102  & 207.4544 & 40.5374 & 0.0961 & 2           & 11.11     & 0.338  & Sb         & 39.04    \\
8486-12705 & 238.1414 & 46.3399 & 0.0606 & 1           & 11.21     & 0.213  & Sab        & 38.87    \\
8464-6101  & 186.1810 & 44.4108 & 0.1256 & 4           & 11.54     & 1.446  & S0a        & 40.64    \\
8330-12702 & 203.8530 & 38.0952 & 0.0649 & 3           & 10.81     & 0.670  & S0a        & 38.12    \\
8603-6101  & 247.1593 & 39.5513 & 0.0304 & 4           & 11.23     & -0.645 & E          & 38.23    \\
8612-12705 & 255.1016 & 38.3517 & 0.0358 & 2           & 10.94     & 0.236  & SBa        & 38.26    \\
8156-12701 & 54.3896  & 0.1442  & 0.0481 & 4           & 10.66     & -0.947 & Sc         & 36.93    \\
8077-6103  & 39.4466  & 0.4051  & 0.0473 & 1           & 10.73     & 0.093  & Sa         & 38.77    \\
8146-12705 & 118.0532 & 28.7726 & 0.0637 & 3           & 11.05     & -0.115 & SBa        & 38.37    \\
8714-6102  & 119.1980 & 45.8879 & 0.0561 & 3           & 11.29     & 0.795  & SABb       & 38.93    \\
8711-12701 & 116.9431 & 51.6460 & 0.1009 & 1           & 11.34     & 0.413  & S0a        & 39.63    \\
8720-1901  & 121.1479 & 50.7086 & 0.0227 & 3           & 10.11     & -1.070 & S0         & 38.35    \\
8952-3703  & 205.4409 & 27.1063 & 0.0288 & 1           & 10.53     & -0.423 & SABbc      & 37.96    \\
8978-12705 & 249.5586 & 41.9388 & 0.0286 & 2           & 10.92     & 0.509  & Sc         & 38.26    \\
8595-12704 & 221.2231 & 51.3411 & 0.0890 & 2           & 11.42     & -0.179 & E          & 38.88    \\
8943-9101  & 156.4031 & 37.2223 & 0.0608 & 4           & 11.00     & 0.224  & Sa         & 38.68    \\
8939-12701 & 124.7068 & 22.9545 & 0.0919 & 1           & 11.35     & -0.043 & Sab        & 39.10    \\
8946-3703  & 170.5882 & 46.4305 & 0.0323 & 1           & 10.82     & -0.955 & S0         & 38.19    \\
9029-12704 & 247.2170 & 42.8120 & 0.0316 & 3           & 10.83     & -0.020 & SBb        & 38.47    \\
9039-6102  & 230.1022 & 32.8596 & 0.0620 & 3           & 11.26     & 0.689  & Sa         & 38.60    \\
9036-6102  & 239.1021 & 42.3955 & 0.0408 & 4           & 10.93     & 0.627  & Sb         & 38.67    \\
9047-6104  & 248.1409 & 26.3807 & 0.0586 & 3           & 11.39     & 1.240  & Sbc        & 39.45    \\
8154-9102  & 45.9602  & -0.2045 & 0.0276 & 1           & 10.69     & 0.655  & SBc        & 38.74    \\
9182-6102  & 119.4863 & 39.9934 & 0.0658 & 3           & 11.08     & 0.243  & S0a        & 39.96    \\
9193-12701 & 45.9546  & -1.1038 & 0.0136 & 3           & 10.87     & -0.380 & S0a        & 39.06    \\
8993-9102  & 165.9101 & 45.1800 & 0.0205 & 3           & 10.53     & -0.452 & SABbc      & 38.70    \\
9491-6102  & 119.9304 & 18.4677 & 0.0378 & 1           & 10.22     & -1.357 & Sb         & 36.79    \\
9486-9101  & 120.7992 & 39.8858 & 0.0410 & 1           & 11.19     & -0.516 & S0a        & 38.26    \\
8311-6104  & 205.2827 & 23.2821 & 0.0264 & 3           & 10.88     & 0.876  & SABb       & 39.45    \\
8309-6101  & 210.1903 & 51.7287 & 0.0697 & 1           & 11.16     & -0.166 & Sa         & 38.53    \\
9507-12704 & 129.6000 & 25.7545 & 0.0182 & 4           & 10.67     & 0.352  & S+S        & 38.74    \\
9507-12705 & 129.5207 & 25.3295 & 0.0282 & 3           & 10.63     & -0.198 & SABb       & 39.06    \\
9024-12705 & 223.8675 & 32.8400 & 0.0602 & 2           & 11.25     & 0.984  & SBbc       & 38.99    \\
9000-1901  & 171.4007 & 54.3826 & 0.0207 & 2           & 10.44     & 0.524  & S0         & 39.06    \\
9502-9101  & 128.3419 & 25.1049 & 0.0866 & 1           & 11.53     & 0.212  & S0         & 38.95    \\
9502-12703 & 129.5456 & 24.8953 & 0.0287 & 2           & 11.15     & 0.381  & SBb        & 40.00    \\
8985-12703 & 204.5544 & 32.8228 & 0.0245 & 1           & 10.51     & -0.008 & SBc        & 37.85    \\
9095-6102  & 243.4418 & 22.9190 & 0.0319 & 1           & 10.97     & 0.164  & Sab        & 38.18    \\
9088-9102  & 242.4723 & 26.6259 & 0.0779 & 4           & 11.49     & 0.576  & S          & 38.58    \\
9864-9101  & 213.9158 & 50.7138 & 0.0498 & 4           & 10.68     & -0.163 & Irr        & 38.22    \\
9870-6103  & 233.2283 & 44.5387 & 0.0371 & 1           & 10.80     & -0.046 & SABa       & 38.55    \\
9043-3704  & 230.9032 & 28.6431 & 0.0841 & 1           & 11.35     & -0.290 & S0         & 38.83    \\
9888-3701  & 236.0080 & 27.6993 & 0.0322 & 3           & 10.76     & 0.171  & SABa       & 38.03    \\
9888-12701 & 235.4758 & 28.1340 & 0.0332 & 3           & 11.18     & -0.257 & SBb        & 38.59    \\
8156-12701 & 54.3903  & 0.1448  & 0.0481 & 4           & 10.52     & -1.003 & Sc         & 38.31    \\
8322-12702 & 200.0916 & 30.4451 & 0.0476 & 1           & 10.96     & -0.181 & E          & 38.88    \\
8549-12705 & 241.9053 & 45.0655 & 0.0442 & 4           & 10.62     & -0.033 & Sbc        & 38.72    \\
8711-12701 & 116.9417 & 51.6489 & 0.1009 & 1           & 10.78     & 0.227  & S0a        & 39.92    \\
8943-9101  & 156.4018 & 37.2214 & 0.0608 & 4           & 10.22     & -1.382 & Sa         & 39.13    \\
9039-9101  & 229.0024 & 34.3553 & 0.1253 & 2           & 10.82     & 1.116  & SABb       & 40.50    \\
9049-12701 & 246.6169 & 24.0270 & 0.0648 & 2           & 11.21     & 0.302  & Sab        & 38.00    \\
8601-12701 & 247.7213 & 41.2863 & 0.0939 & 2           & 9.94      & -1.740 & S0         & 39.41   
\enddata
\tablecomments{The information of all 61 MaNGA MPL-6 AGNs in galaxy pairs.
(1): MaNGA Plate-IFU number;
(2)\&(3): RA and DEC of target galaxy;
(4) redshift from MaNGA spectra;
(5): merger case classified in Sec.~\ref{subsec:mergersequence};
(6) stellar mass in unit of solar mass;
(7) star-formation rate;
(8) Visual morphological classification from MaNGA Visual Morphology Catalogue\footnote{\url{https://data.sdss.org/datamodel/files/MANGA_MORPHOLOGY/manga_visual_morpho/}}.
(9) [OIII] surface brightness of the central $\rm 1.5\arcsec \times 1.5\arcsec$ region.}
\end{deluxetable*}

The classification results for all MaNGA galaxies are listed in Table \ref{tab:bptresult}.
The results of the three diagnostic diagrams 
(BPT, modified-BPT, and WHAN, Figure \ref{fig:bptall} a, b, and c)
are generally consistent (88\%) with each other.
For comparison, we also plot the positions of our galaxies 
in the color-magnitude diagram (Figure \ref{fig:bptall} d).
In Figure \ref{fig:bptall} (d), 
we also draw the division lines from \citet{2007ApJS..173..293W} 
to guide the eyes of the three regions defined as: 
`red sequence', `green valley (GV)', and `blue cloud'.
We find that in Figure \ref{fig:bptall} (d),
SFGs (blue) and RGs (black) lie mostly in 
the `blue cloud' and `red sequence', respectively;
while AGNs (red) and composite galaxies (green) tend to 
lie in the `green valley',
indicating that the AGNs in our sample are in
possible transition from the blue cloud to the red sequence.
We note that our AGN sample,
like other BPT selected AGN samples,
is biased against AGNs with broad emission lines,
or in very dusty systems with significant extinction of the emissions from the NLR,
as well as radio AGNs without emission-lines \citep[see][for a review]{2017A&ARv..25....2P}.

\subsection{AGN Fractions Along the Merger Sequence}\label{subsec:agnfrac}
In this section we compare the AGN fractions
along the merger sequence (for definition, see Sec.~\ref{subsec:mergersequence})
from Case 1 to Case 4, as well as in the isolated galaxies.
MPL-6 includes 4620 unique IFU cubes,
out of which there are 116 IFU cubes that cover
two galaxies in a pair (see Sec~\ref{subsec:pairselect}). 
Therefore, in the full MPL-6 parent sample, we classify 4736 galaxies.
A total of 187 galaxies' {\tt DAP} products
are marked as not suitable for scientific use
due to the contamination of foreground stars,
uncertainties in redshift,
or other critical failures.
After removing these galaxies,
we classify the remaining 4549 galaxies (including 1115 galaxies in pairs) following the same criteria
listed in Sec.~\ref{sec:agnclass},
and find 239(5.3\%) AGNs, 385(8.5\%) AGN-starburst composites,
1654(36.4\%) SFGs, 1267(27.8\%) RGs, and 1004(22.1\%) lineless galaxies.
In our pair sample (1115 galaxies), 
the corresponding numbers and fractions are:
61(5.5\%), 74(6.6\%), 310(27.8\%), 313(28.1\%), 357(32.0\%).
We list the physical parameters of the 61 AGNs in pairs in Table~\ref{tab:info}.

This $\sim$5.3\% AGN fraction found in the full MPL-6 is consistent with previous MaNGA works
using the emission-line diagnostics
\citep{2017MNRAS.472.4382R,2018MNRAS.474.1499W,2018RMxAA..54..217S}, 
where an AGN fraction of $\sim$3-11\% was found in $\sim$2700 galaxies
from earlier MaNGA data release of MPL-5.
For MaNGA MPL-8, 
\citet{2020ApJ...901..159C} built a sample of 406 AGNs (283 are from MPL-6), 
compiled through a combination of
Wide-field Infrared Survey Explorer (WISE) mid-infrared color cuts,
Swift/BAT hard X-ray detection,
NVSS/FIRST 1.4 GHz radio sources,
and SDSS broad emission-lines.
Given the significantly different selection criteria,
there are only 21\% AGNs from \citet{2017MNRAS.472.4382R},
13\% AGNs from \citet{2018MNRAS.474.1499W},
23\% AGNs from \citet{2018RMxAA..54..217S},
and 22\% from our AGN sample 
that overlap with the \citet{2020ApJ...901..159C} sample.
Among the 283 MPL-6 AGNs in \citet{2020ApJ...901..159C}, 
222 are not in our sample,
which consists of 206 (93\%) radio (NVSS/FIRST) AGNs
with weak or no emission-lines, 
and 16 WISE or X-ray AGNs.  
The lack of radio AGNs in our sample indicates that radio AGNs are not necessarily line-emitters,
as demonstrated in local galaxies \citep[e.g.][]{2005MNRAS.362....9B}.

\begin{figure}
  \epsscale{1.15}
  \plotone{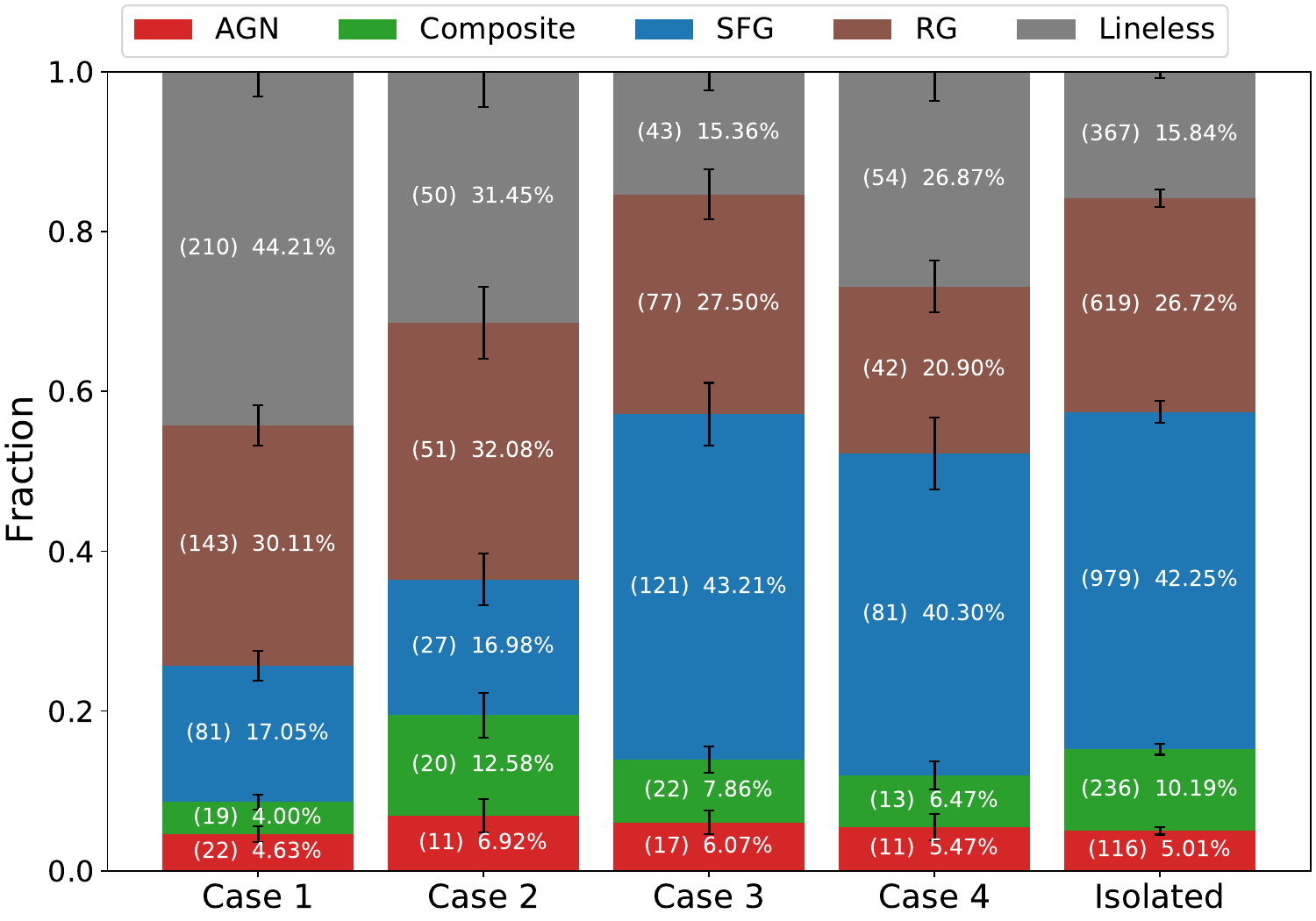}
  \caption{The fractions of the five galaxy types for pairs (in color) 
  in the 4 merger cases (left 4 panels, for definition, see Sec~\ref{subsec:mergersequence})
  and in isolated galaxies (right panel). 
  In each panel, the colored blocks from the bottom to top 
  represent AGNs (red), composite galaxies (green), 
  SFGs (blue), 
  retired galaxies (RG, brown),
  and lineless (gray) galaxies, respectively. 
  In brackets are the numbers of galaxies of that specific galaxy type, 
  followed by the percentages in the corresponding cases. 
  The error bars at the top of each color block mark the corresponding binomial errors. 
  Compared to isolated galaxies, no significant difference ($>$\,3$\sigma$) in the fraction of 
  galaxy types is found for any merger cases, 
  except for SFG in Case 1 \& 2, and Lineless galaxies in Case 1. 
  Overall, higher fractions of passive galaxies (Lineless+RGs) by $\sim$15\%--25\%
  are found in Case 1 \& 2.
  \label{fig:agnfraction}}
  \end{figure}

The fractions of each category in different merger cases are listed in 
Table \ref{tab:bptresult} and shown in Figure \ref{fig:agnfraction}.
Overall, no significant excess or trend in the AGN fractions is 
found between the four merger cases.
Case 1 has the lowest AGN fraction, 4.6$\rm \pm$1.0\%, 
while Case 2 has the highest, 6.9$\rm \pm$2.1\%, 
though in all four cases the AGN fractions are consistent within 3$\sigma$. 
The AGN fraction in isolated galaxies also has a 
comparable value of 5.0$\rm \pm$0.6\%.
Even after including the composite galaxies, 
the AGN fractions remain more or less constant for the various cases 
and with the isolated galaxies. 
Case 2 galaxies, which have the strongest distortion, 
show higher (AGN+Composite) fraction (19.5$\pm$3.5\%), 
as compared to the isolated galaxies (15.2$\pm$0.8\%). 
In Case 1 \& 2, we find fewer SFGs ($\sim$17\%) than in
Case 3 \& 4, as well as in isolated galaxies, which have an SFG fraction of $\sim$40\%.
We suspect that this is an environmental effect and will discuss this in more detail in Sec.~\ref{subsec:env}.
In Table \ref{tab:bptresult}, we also list the total fractions in MPL-6 galaxies for comparison. 

On the other hand, we find a clear difference of the fractions of passive (RGs+Lineless) galaxies.
Significantly higher fractions ($\sim$15\%--25\%) of passive galaxies are found in Case 1 (74$\pm$4\%)
and Case 2 (64$\pm$6\%), as compared to isolated galaxies (42$\pm$1\%), 
and $\sim$45\%--50\% in Case 3 and 4.
This reflects the selection bias towards more early type galaxies (ETGs)
in Case 1 and Case 2, because ETG pairs hardly show morphological distortions and will be classified as
either Case 1 (if separated) or Case 2 (if with overlap) based on the criteria in Sec.~\ref{subsec:mergersequence}.
In addition, we find a higher fraction of passive galaxies
in pairs (60$\pm$2\%) than in isolated galaxies (42$\pm$1\%),
possibly related to their environments
(see Sec~\ref{subsec:env} for more discussion).

\begin{deluxetable*}{ccccccc}
  \tablecaption{Galaxy types and fractions based on emission-line classifications for 
  different merger cases \label{tab:bptresult}}
  \tablewidth{0pt}
  \tablehead{
  \colhead{Case} & \colhead{AGN} & \colhead{Composite} & \colhead{SFG} & \colhead{RG} & \colhead{Lineless} & \colhead{Total}\\
  \colhead{(1)} & \colhead{(2)} & \colhead{(3)} & \colhead{(4)} & \colhead{(5)} & \colhead{(6)} & \colhead{(7)}
  }
  \startdata
  Case1 & 4.6$\pm$1.0\% (22) &  4.0$\pm$0.9\% (19) & 17.1$\pm$1.9\% (81)  & 30.1$\pm$2.5\% (143) & 44.2$\pm$3.1\% (210) & 475 \\
  Case2 & 6.9$\pm$2.1\% (11) & 12.6$\pm$2.8\% (20) & 17.0$\pm$3.3\% (27)  & 32.1$\pm$4.5\% (51)  & 31.5$\pm$4.4\% (50) & 159 \\
  Case3 & 6.1$\pm$1.5\% (17) &  7.9$\pm$1.7\% (22) & 43.2$\pm$3.9\% (121) & 27.5$\pm$3.1\% (77)  & 15.4$\pm$2.3\% (43) & 280 \\
  Case4 & 5.5$\pm$1.7\% (11) &  6.5$\pm$1.8\% (13) & 40.3$\pm$4.5\% (81)  & 20.9$\pm$3.2\% (42) & 26.9$\pm$3.7\% (54) & 201 \\
  Isolated & 5.0$\pm$0.5\% (116) & 10.2$\pm$0.7\% (236) & 42.3$\pm$1.4\% (979) & 26.7$\pm$1.1\% (619) & 15.8$\pm$0.8\% (367) & 2317 \\
  MPL-6 & 5.3$\pm$0.3\% (239) & 8.4$\pm$0.4\% (382) & 36.2$\pm$0.9\% (1649) & 27.8$\pm$0.8\% (1267) & 22.1$\pm$0.7\% (1004) & 4549 \\
  \enddata
  \tablecomments{Column 1: Case name, `Isolated' represents these galaxies without a physical companion 
  nor can be identified as merger (see Sec.~\ref{subsec:mergersequence}). 
  Column 2-6: The fraction and its binomal error for each galaxy type in percentage, 
  with the actual number of galaxies (AGN, composite, SFG, retired and lineless galaxies) listed in bracket.
  Column 7: The total number of galaxies for each case.}
  \end{deluxetable*}

\section{Galaxy properties}\label{sec:analyse}
In this section we present the different galaxy properties of our AGN pair sample,
and compare with the control sample of isolated AGNs.
We begin with their global properties, including stellar mass, star formation rate (SFR), and central
[\ion{O}{3}] surface brightness.
Then we compare the radial profiles of resolved specific SFR (sSFR), Dn4000, and the Balmer decrement.
The $\rm H\alpha$ and [\ion{O}{3}] fluxes used in this section 
are all dust corrected based on the $\rm H\alpha/H\beta$ flux ratios
with a reddening curve (R$_V$=3.1, gas environment) from \citet{2000ApJ...533..682C},
assuming the case B recombination \citep[$\rm H\alpha/H\beta = 2.86$,][]{2006agna.book.....O}.
The dust-corrected luminosities of $\rm H\alpha$ and [\ion{O}{3}] are calculated using Equation (\ref{eq:corrha}):
\begin{equation}
  {\rm L_{line} = 4\pi d^2 S_{line} 10^{0.79 k_{\lambda}\log(\frac{H\alpha}{2.86\times\,H\beta})}},
  \label{eq:corrha}
\end{equation}
where d is the luminosity distance from the NSA catalog;
S$\rm _{line}$ is the observed flux of $\rm H\alpha$ or [\ion{O}{3}]; 
k$\rm _{\lambda}$ is the correction factor from \citet{2000ApJ...533..682C}, and has a value of 2.4 for $\rm H\alpha$ and 3.5 for [\ion{O}{3}].

\subsection{The Global Properties}\label{subsec:global}
\subsubsection{Stellar Mass and Global SFR}\label{subsec:sfrvsm}
\begin{figure}
  \epsscale{1.21}
  \plotone{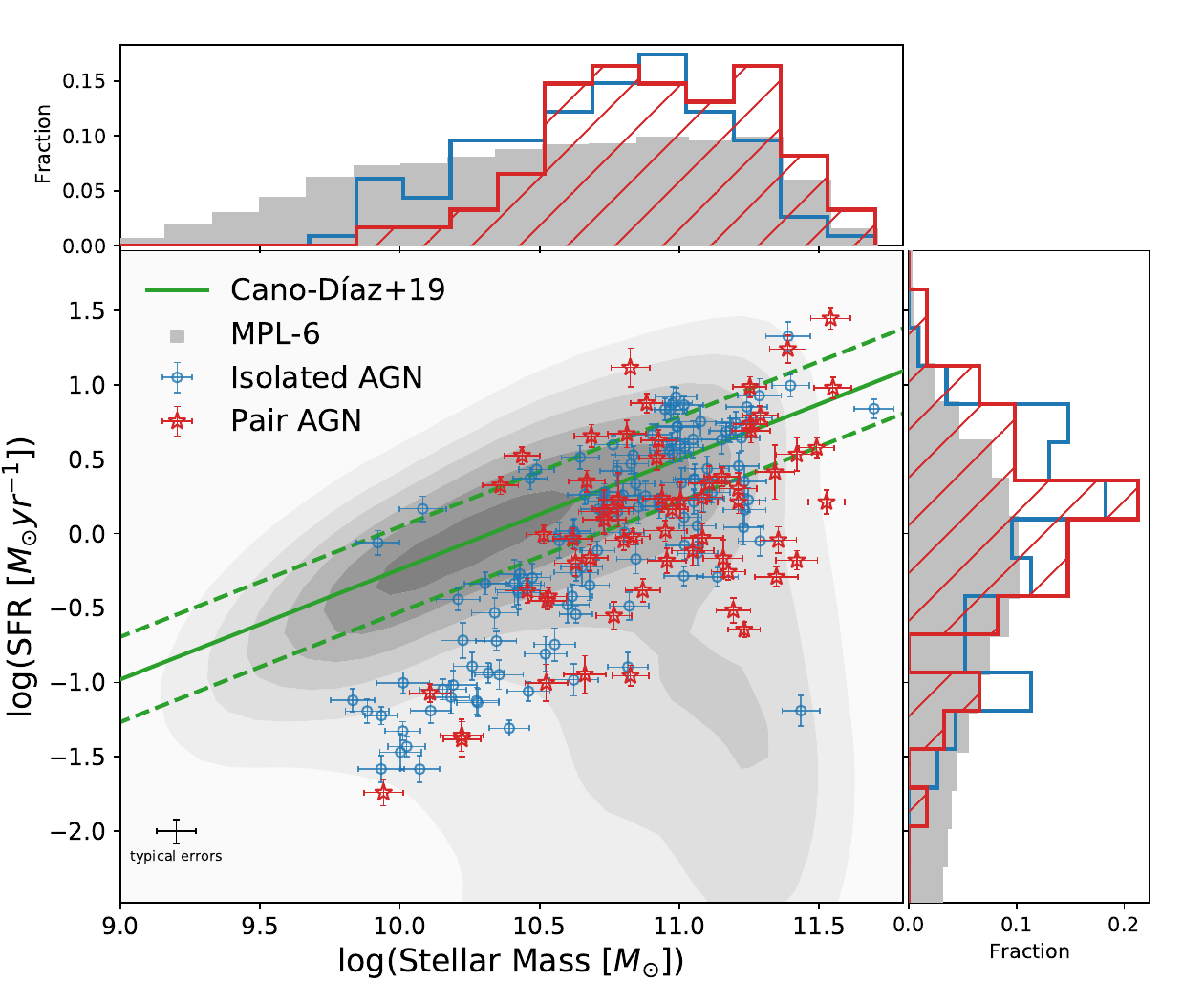}
  \caption{Distribution of $M_{*}$ versus SFR for AGNs in our sample (colored stars and circles) and the MaNGA MPL-6 galaxies (gray contour). 
  The contour is generated with a Gaussian-KDE fitting. 
  AGN in isolated galaxies and pairs are plotted as blue circles and red stars, respectively. 
  The solid and dashed green lines are the SFMS and 1 $\sigma$ offset for MaNGA MPL-5 galaxies from \citet{2019MNRAS.488.3929C}. 
  Also plot in the upper and right panels are the normalized distributions of $M_{*}$ and SFR:
  gray for all MaNGA MPL-6 galaxies, 
  blue for isolated AGNs, 
  and red for AGNs in pairs. 
  The typical errors are 0.08 dex for SFR and 0.07 dex for $M_{*}$. 
  In our sample, AGNs lie mainly on or below the SFMS, which is consistent with \citet{2019MNRAS.488.3929C}, and AGN in pairs tend to have slightly higher SFR and $M_{*}$.} 
  \label{fig:sfms}
  \end{figure}

The global stellar masses ($M_{*}$) and dust-corrected SFRs are taken from the {\tt Pipe3D} catalog (version 3.0.1).
We compare our pair sample with the MaNGA star formation main sequence (SFMS), based on the Pipe3D results,
as defined in  \citet{2019MNRAS.488.3929C}. 
In Figure \ref{fig:sfms}, 
we plot all MPL-6 galaxies (contour)
and mark the isolated AGNs as blue circles and AGNs in pairs as red stars.
The MPL-6 galaxies show two distinguished populations: one that mainly lies on the SFMS,
and another of quenched galaxies that extend to the high mass, low SFR region in Figure\,\ref{fig:sfms}. 
All our emission-line-selected MaNGA AGNs have $M_{*}$ larger than $10^{9.6}\,M_\sun$.
They lie on or below the SFMS, likely in transition between the SFMS and the quenched galaxies, 
while the less massive AGN hosts appear more quenched. 
This is consistent with the color-magnitude diagram in Figure \ref{fig:bptall} (d),
where most AGN host galaxies are in the `green valley' region.
The typical errors are 0.08 dex for SFR and 0.07 dex for  $M_{*}$. 

AGNs in pairs and isolated galaxies are well blended in Figure \ref{fig:sfms}, 
with no significant difference in SFR (+0.04 dex)
or stellar mass (+0.13 dex), 
though their median SFR ($\rm 10^{0.15}\,M_{\sun}\,yr^{-1}$) and $M_{*}$ ($\rm 10^{10.93}\,M_{\sun}$)
are higher than the full MPL-6 sample ($\rm 10^{-0.59}\,M_{\sun}\,yr^{-1}$ and $\rm 10^{10.62}\,M_{\sun}$). 
The lack of low mass AGN hosts on or above the SFMS may be a combined effect 
due to the nature of the AGN population as well as the selection effect from the BPT diagnostics. 
First, the number density of AGN is lower in low-mass host galaxies, as reported in \citet{2009MNRAS.397..135K}, 
so the limited sample size of the MaNGA survey may miss AGNs in low-mass galaxies.   
Second, the so-called `star formation dilution' effect in the BPT diagram
is stronger in low-mass, high-SFR host galaxies, as discussed in \citet{2015ApJ...811...26T}. 
We will discuss the AGN selection bias in more detail in Sec.~\ref{subsec:bpt}.

\subsubsection{Stellar Mass and [\ion{O}{3}] Surface Brightness}\label{sfrvsm}
\begin{figure}
  \epsscale{1.2}
  \plotone{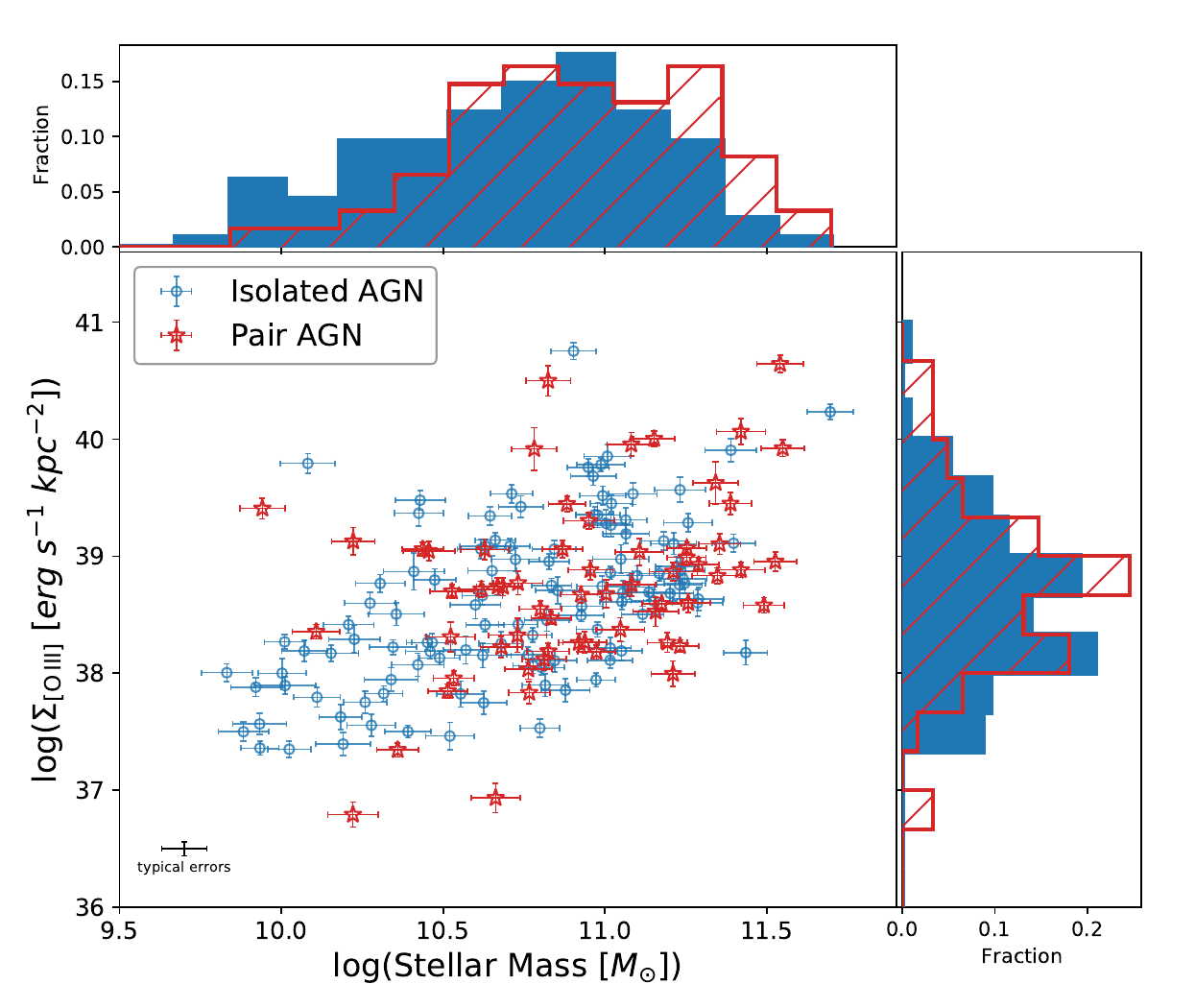}
  \caption{Central [\ion{O}{3}] luminosity surface density $\rm \Sigma$[\ion{O}{3}] ($\rm 1.5\arcsec \times 1.5\arcsec$)
  versus global stellar mass for AGNs 
  in MaNGA MPL-6, with red stars for AGNs in pairs, and blue circles for isolated AGNs. 
  Histograms are their normalized distributions and we find marginally higher (+0.13\,dex)
   $\rm \Sigma$[\ion{O}{3}] for paired AGNs.}
  \label{fig:o3sb}
  \end{figure}

A common proxy for the bolometric luminosity of AGN is the [\ion{O}{3}] luminosity \citep[e.g.][]{2014ARA&A..52..589H}.
With the IFU data, 
here we only focus on the central [\ion{O}{3}],
which is dominated by nuclear activity and likely less contaminated from extended SF activities.
Same as the BPT classification, we use the central $\rm 1.5\arcsec \times 1.5\arcsec$ spaxels 
to calculate the surface brightness of [\ion{O}{3}] ($\rm \Sigma$[\ion{O}{3}] = [\ion{O}{3}]/area).
We compare the central $\rm \Sigma$[\ion{O}{3}]
with the global stellar mass distribution in Figure \ref{fig:o3sb}.
AGN in pairs have a marginally higher $\rm \Sigma$[\ion{O}{3}]
than isolated AGNs
(+0.13 dex, median error in $\rm \Sigma$[\ion{O}{3}] is 0.06 dex). 
This is different from the result of \citet{2012ApJ...745...94L} using the SDSS single fiber data.
They found an global [\ion{O}{3}] luminosity enhancement of 0.5 to 0.7 dex in AGN pairs.
One possible cause of the difference is due to the lack of dust extinction correction
in the [\ion{O}{3}] luminosity measurements in \citet{2012ApJ...745...94L}. 
As discussed later in Sec.~\ref{subsec:profile},
in our sample, 
we found lower Balmer decrement thus lower extinction correction in AGNs in pairs than that in isolated AGNs.
After the extinction correction, the luminosity difference would be smaller
between pairs and isolated galaxies. 
On the other hand, most of our AGNs have a central $\rm \Sigma$[\ion{O}{3}] less than $\rm 10^{40}\,erg \, s^{-1} \, kpc^{-2}$,
indicating that they are mostly AGNs with moderate luminosities \citep{2003MNRAS.346.1055K}.

\subsection{Resolved properties}\label{subsec:resolved}
\subsubsection{Measurements}\label{subsec:measure}
To examine the difference between AGN in pairs and isolated galaxies,
in this section we investigate the resolved properties (measured as surface densities $\rm \Sigma_{x}$, in unit of $\rm x\,kpc^{-2}$) of 
the specific SFR ($\rm \Sigma_{SFR}$/$\Sigma_{M_{*}}$), Balmer decrement, and Dn4000, chosen to represent the resolved star formation, dust extinction, and age of the stellar population,
respectively.

The mass surface density ($\Sigma_{M_{*}}$) is from the {\tt Pipe3D} data cube,
calculated after fitting the spectra with a model of stellar populations using the GSF156 single-stellar population (SSP) library.
We obtain the $\rm \Sigma_{SFR}$ from the attenuation corrected H$\alpha$ luminosity
using the star formation law \citep[][]{2012ARA&A..50..531K}: 
\begin{equation}
  {\rm \log(\frac{SFR}{M_{\sun}\,yr^{-1}}) = \log (\frac{L_{H\alpha}}{erg\,s^{-1}}) - 41.01}.
  \label{eq:sflaw}
\end{equation}

This relation is based on the assumption that
the H$\alpha$ emission is produced by young stellar populations
(e.g. OB stars in \ion{H}{2} regions). 
Therefore, the H$\alpha$ contamination from AGN's narrow line region
will lead to overestimated SFR.
We disentangle the AGN's contribution through
different line ratios as compared to pure star-forming \ion{H}{2} regions.

The intrinsic [\ion{N}{2}]$\rm /H\alpha$ emitted by SF and AGN
can be predicted using different photoionization models
\citep[e.g.][]{2002ApJS..142...35K,2004ApJS..153....9G,2013ApJS..208...10D}.
\citet{2020MNRAS.499.5749J} presented a new 3D diagnostic diagram
which can be applied to estimate the contributions from AGN and SF based on a given model.
They used the best-fitting SF and AGN model for their MaNGA spaxel sample and derived a relation between AGN's contribution 
and the indicator $P_{1}$, which can be approximated as Equation~(\ref{eq:fagn}):
\begin{equation}\label{eq:fagn}
  f_{AGN}=\left\{
  \begin{aligned}
  &0, &\, P_{1} \leqslant -0.53 \\
  &0.14P_{1} ^{2} + 0.96 P_{1} +0.47, &\, -0.53<P_{1}<0.51 \\
  &1, &\, P_{1} \geqslant 0.51
  \end{aligned}
  \right.
  ,
  \end{equation}
where $f_{AGN}$ is AGN's contribution to the $\rm H\alpha$ flux ($\rm H\alpha _{AGN}/H\alpha _{total}$)
and $P_{1}$ equals to 0.63\,log([\ion{N}{2}]$\rm /H\alpha$)+0.51\, log([\ion{S}{2}]$\rm /H\alpha$)+0.59\,log([\ion{O}{3}]$\rm /H\beta)$.
We apply this relation to calculate the $f_{AGN}$ of all spaxels with enough S/N ($>$3),
and then decompose the $\rm H\alpha_{SF}$ to obtain the SFR through Equation~(\ref{eq:sflaw}).
We test this relation for all MaNGA spaxels with robust S/N regardless of their host galaxies' categories in Appendix~\ref{ap:decompose}.
Overall the $f_{AGN}$ per spaxel equals to 0 in the star-forming region and increases to 1 towards the edge of the AGN sequence.
Thus our SFR corrections will only affect the AGN or Composite spaxels, but does not affect SF spaxels, as expected.

The spaxel-by-spaxel Dn4000 and line flux values are from the MaNGA {\tt DAP} data product.
For each galaxy, the effective radius ($R_{e}$) is from the NSA catalog and was calculated from the $r$-band photometry. 
We then calculate the radial profiles by averaging the corresponding values in 6 equal radius bins 
from the center (0$R_{e}$) to MaNGA's reliable coverage (1.5$R_{e}$) with a bin size of 0.25$R_{e}$.

\subsubsection{Radial Profiles}\label{subsec:profile}
\begin{figure*}
  \plotone{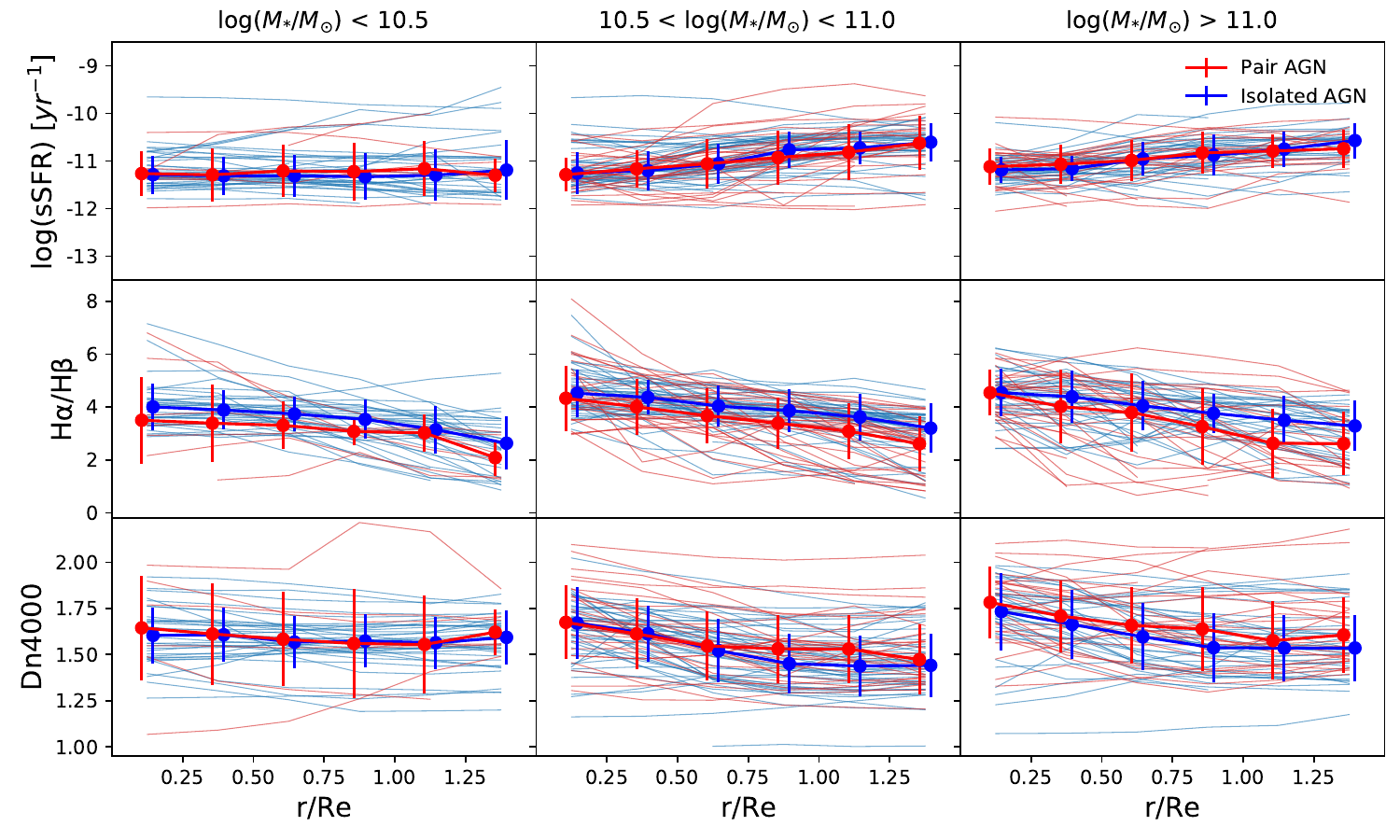}
  \caption{The radial profiles of sSFR (top), Balmer decrement (middle) and Dn4000 (bottom)
   for AGNs in pairs (red) and isolated AGNs (blue).
  The AGN sample is divided into three mass bins as listed at the top of each column.
  The thin lines are the radial profiles for each galaxy in the mass bin. 
  The thick lines with solid dots represent the median value in each radius bin from 0.00 to 1.50 $Re$, with a binsize of 0.25 $Re$, 
  and manually offset in the x direction to guide the eye. 
  The error bars indicate the 1$\sigma$ scatter of individual galaxies around the median value.
  No significant difference is found between AGNs in pairs and isolated AGNs.
  Both show an increasing sSFR, as well as decreasing Balmer decrement and Dn4000 radial profiles. 
  This is consistent with the inside-out quenching scenario. 
  The only exception is in the low mass bins (log($M_{*}/$$\rm M_{\sun}$) $<$ 10.5), 
  where AGNs show flat sSFR and Dn4000 radial profiles.  
    } \label{fig:profile}
\end{figure*}

Previous work by \citet{2018MNRAS.477.3014B} showed that 
the sSFR radial profiles are mass dependent,
with low mass MaNGA main sequence galaxies having higher
and more flat sSFR than high mass ones.
We first divide all AGN hosts into 3 mass bins of
log($M_{*}$/$\rm M_{\sun}$) $<$ 10.5,
10.5 $<$ log($M_{*}$/$\rm M_{\sun}$) $<$ 11.0, and
log($M_{*}$/$\rm M_{\sun}$) $>$ 11.0.
The numbers of AGNs in each mass bin are 32, 64, and 53, respectively.
Every galaxies' radial profiles of the sSFR, Balmer decrement, and Dn4000 are shown in Figure \ref{fig:profile},
with the AGNs in pairs in red and the isolated AGNs in blue. 
The solid dots and thick lines 
are generated using the median values in the corresponding radius bins, 
and the error bars indicate the 1$\sigma$ scatter of individual galaxies around the median values.
The larger uncertainties for AGNs in pairs are related to the relatively smaller sample size.
From the radial profiles, we observe that:

(a) The sSFR radial profiles show no significant difference between AGNs in pairs or isolated AGNs in all mass bins. 
In the lowest mass bin
(log($M_{*}/$$\rm M_{\sun}$)\,$<$\,10.5),
AGNs in pairs have a flat sSFR radial profile. 
At log($M_{*}/$$\rm M_{\sun}$)\,$>$\,10.5, both AGNs in pairs and isolated galaxies 
have an increasing sSFR towards larger radius, indicating an inside-out quenching. 

(b) The Balmer decrements decrease from the inside to the outside, indicating more dust attenuation in the nuclear region.
As in the case of sSFR, the Balmer decrement is 
also flatter in the lowest mass bin for AGN in pairs, but not as flat as the sSFR and Dn4000 in the same mass bin.
AGNs in pairs tend to have lower Balmer decrements than isolated AGNs, 
though not significant ($<$1$\sigma$).

(c) Overall, the Dn4000 radial profiles decrease towards larger radii, indicating younger stellar populations in the outer regions of the galaxy. 
Again, this trend is less obvious in low mass AGNs, which is consistent with the flatter trend of sSFR radial profiles (top panel). 

(d) For all AGN host galaxies more massive than $\rm 10^{10.5}\,M_{\sun}$,
the radial profiles of the sSFR, Balmer decrement, and Dn4000 do not change as the stellar mass increases. 
At the lowest mass bin ($M_{*}$\,$<$\,$\rm 10^{10.5}\,M_{\sun}$), however, 
the radial profiles of the AGN, regardless in pairs or isolated galaxies,
show flatter sSFR with lower absolute values,
and flatter Dn4000 profiles,
which is again consistent with the inside-out picture. 
These results are different from similar analysis for SFGs.
Elevated sSFRs are found in the SFG pairs regardless of their mass values,
especially in the nuclear region \citep[e.g.][]{2019ApJ...881..119P,2021ApJ...909..120S}. 
Higher sSFR are found in low-mass SFGs \cite[e.g.][]{2018MNRAS.477.3014B},
which is opposite to what we find in our AGN host galaxies.

\begin{figure*}
  \plotone{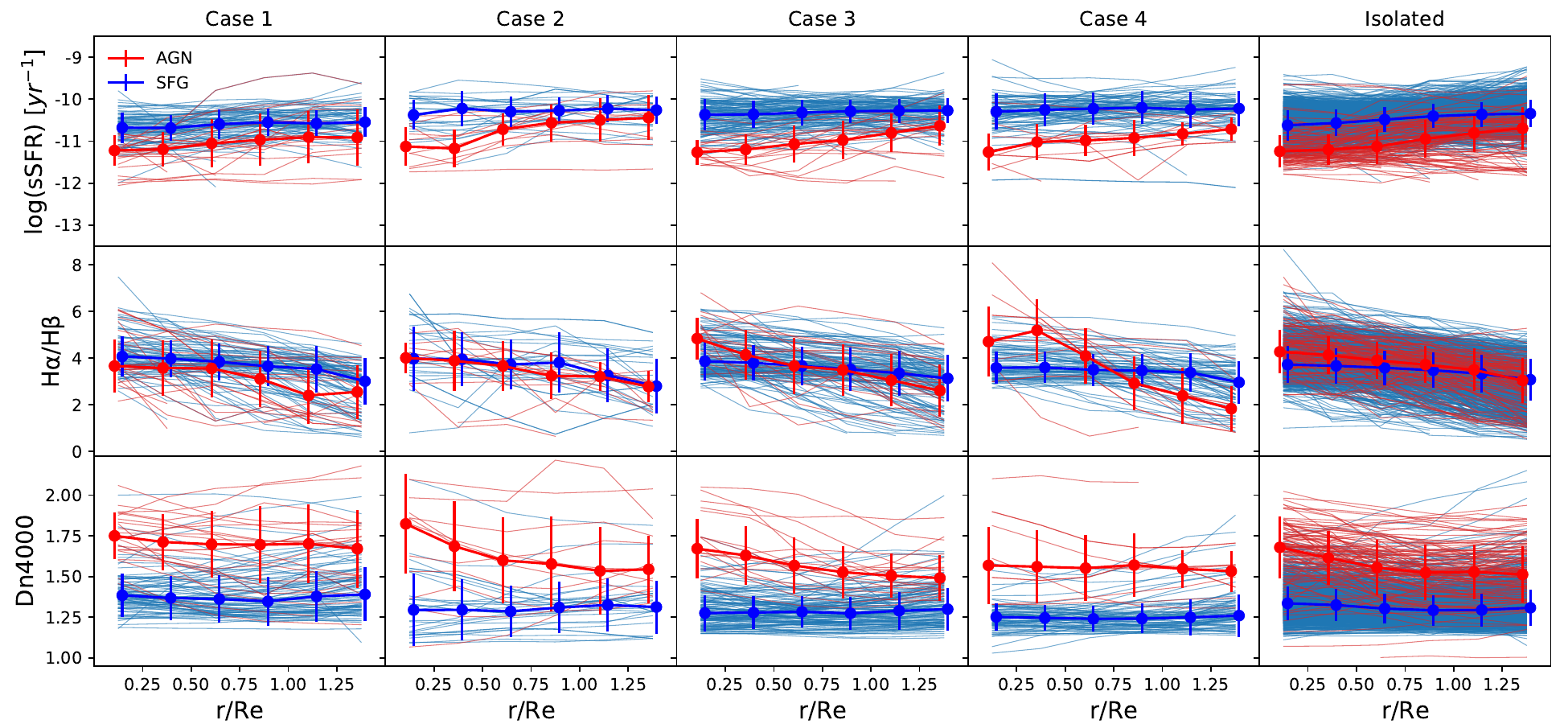}
  \caption{The radial profiles of sSFR, Balmer decrement, and Dn4000, 
  for AGNs (red) and SFGs (blue) divided into different merger cases, and in isolated galaxies.
  We find no significant evolution along the merger sequence or with the isolated galaxies.
  Both paired AGNs and SFGs show similar trends with their isolated counterparts. 
  AGNs have an overall increasing sSFR, decreasing Balmer decrement and decreasing Dn4000 as the radius increases, 
  while the SFGs have overall flat sSFR and Dn4000 radial profiles. 
  This suggests that AGNs are more quenched and have older central stellar populations than SFGs, 
  regardless if they are in pairs or in isolated galaxies. The total galaxy numbers for each subsample are listed in Table~\ref{tab:bptresult}. 
  } \label{fig:profile_case}
\end{figure*}

Figure \ref{fig:profile_case} shows the similar radial profiles as Figure \ref{fig:profile}, 
except that galaxies are separated by their merger cases, 
with red and blue lines represent the AGNs and SFGs, respectively.
For AGNs, 
the overall sSFR radial profile is increasing,
consistent with the declining Dn4000 radial profiles, 
and their trends indicate a centrally depressed SFR along older stellar populations,
suggesting more quenched nucleus regions.
The Balmer decrement radial profiles also decrease from the inside to the outside,
indicating more dust attenuation in the central region.
From Case 1 to Case 4, the radial profiles of these parameters do not show any significant evolution.
Central Balmer decrements are slightly higher in Case 3 and 4, consistent with the scenario of dustier later merger stages.
On the other hand, SFGs show clearly higher and flatter sSFR radial profiles, 
similar Balmer decrement radial profiles, 
and lower and flatter Dn4000 radial profiles than AGNs,
both pairs and isolated galaxies.
These findings are consistent with the star forming nature of the SFGs, 
which have ongoing star formation both in the nucleus and the outskirts. 
We will discuss the difference between the AGN and SFG radial 
profiles in more detail in Sec.~\ref{subsec:suppression}.

\subsection{Comparison to Star-forming and Passive Galaxies}\label{subsec:suppression}
\begin{figure*}
  \plotone{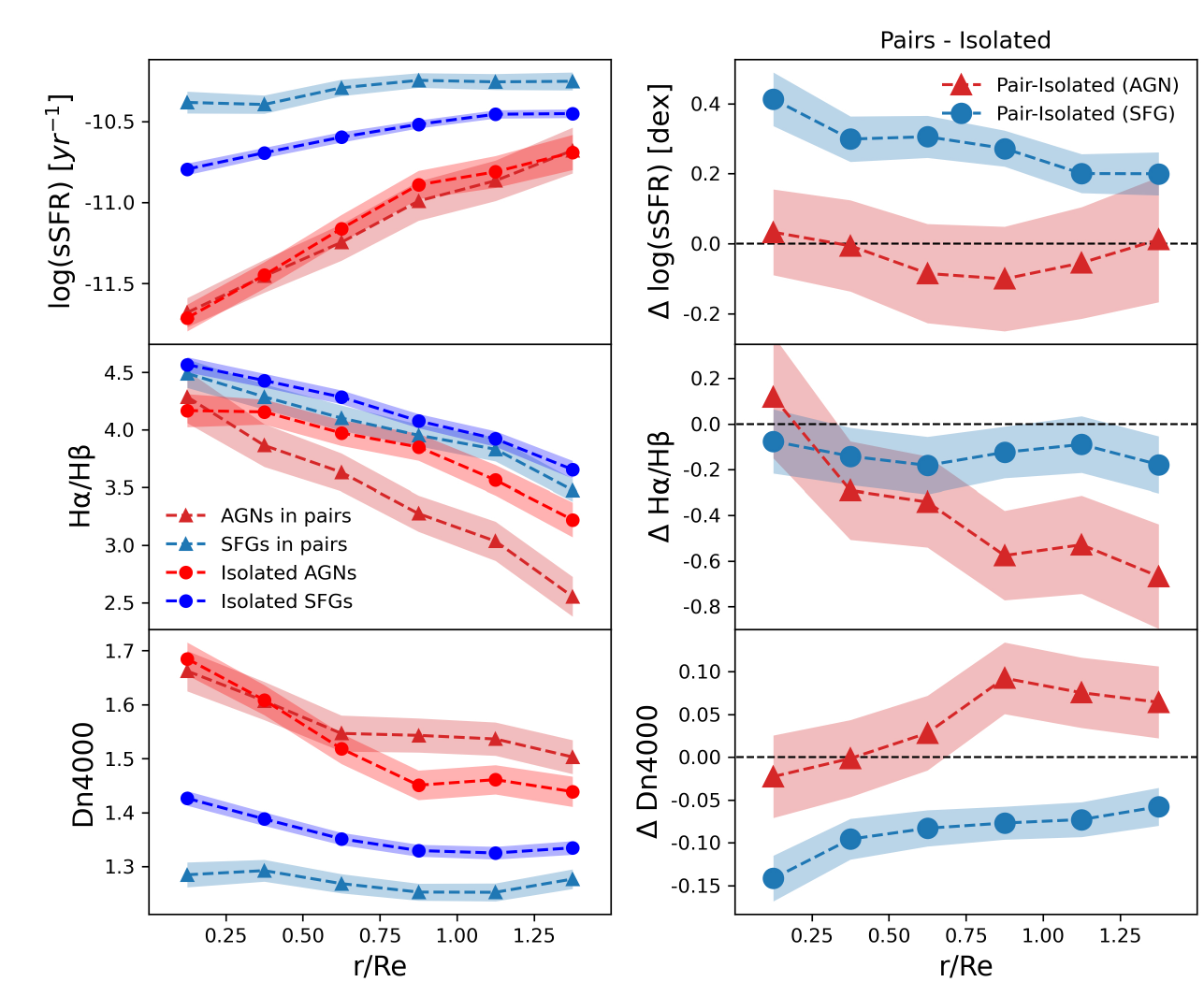}
  \caption{Comparison of the radial profiles between mass-controlled AGNs (red) and SFGs (blue). 
  From top to bottom: the sSFR, Balmer decrement, and Dn4000 as a function of radius for AGNs and SFGs in pairs (triangles) 
  and isolated ones (dots). 
  Left panel shows the absolute values and the right panel shows the difference between pairs and isolated galaxies. 
  The shadows represent the standard deviation of the mean at each radius bin. 
  AGNs have lower sSFR and higher Dn4000 at all radii, regardless of whether they are in pairs or isolated galaxies. 
  Unlike SFGs in pairs, which show enhanced sSFR and suppressed Dn4000 towards the galaxy center,
  AGNs do not show any significant difference in the radial profile between pairs and isolated ones.
  }\label{fig:agnsf}
  \end{figure*}

In this section we compare the differences among AGNs, SFGs, and passive galaxies.
Similar to Sec.~\ref{subsec:profile},
we calculate the radial profiles of each galaxy 
and use the median value to generate the stacked profiles in Figure \ref{fig:agnsf}.
Here the shadowed errors are obtained by calculating the standard deviation divided by $\sqrt{N}$, 
where $N$ is the number of values at each radius bin.

Using the mass-controlled subsamples defined in Sec.~\ref{subsec:controlsample},
we firstly compare the resolved properties between AGNs and SFGs.
The radial profiles of SFGs and AGNs in both pairs and isolated galaxies 
are shown in the left panel of Figure \ref{fig:agnsf}.
In both galaxy pairs and isolated galaxies,
SFGs have higher sSFR and lower Dn4000 values than AGN hosts at all radii,
and have flatter radial profiles.
This indicates more star formation and younger stellar populations in SFGs, 
as expected. 
AGNs (red triangle and circles in Figure \ref{fig:agnsf}, left) have lower sSFR and higher Dn4000 values 
in the central regions, 
consistent with the `inside-out' quenching scenario. 
For the Balmer decrement radial profiles,
we find that the absolute value is almost the same in the center for AGNs and SFGs, 
but decreases faster in AGNs towards the outskirts, especially for AGNs in pairs.
This is consistent with the more dusty outskirts in SFGs.

We further compare the differential radial profiles for pairs and isolated galaxies
in the right panel of Figure \ref{fig:agnsf}. 
Compared to isolated SFGs,
SFGs in pairs show clearly enhanced sSFR and suppressed Dn4000, which is more obvious in the central regions, 
consistent with previous findings \citep[e.g.][]{2019ApJ...881..119P,2021ApJ...909..120S}. 
Compared to paired AGNs,  
isolated AGNs have a marginally increasing sSFR towards the larger radii. 

The Balmer decrement is almost the same in paired and isolated SFGs, 
but decreases in AGNs from the center to the outskirts,
with ${\rm \Delta\, (H\alpha/H\beta)}$ dropping from 0 to -0.6.
We note that other than different dust attenuation, the intrinsic $\rm H\alpha/H\beta$ value can also vary
in different regions of a galaxy due to different interstellar medium (ISM) environment.
For example, higher temperature or higher electron density can also result in lower Balmer decrements \citep{2006agna.book.....O}. 
Thus we refrain ourselves from over-interpreting the trend in the Balmer decrement. 

\begin{figure*}
  \plotone{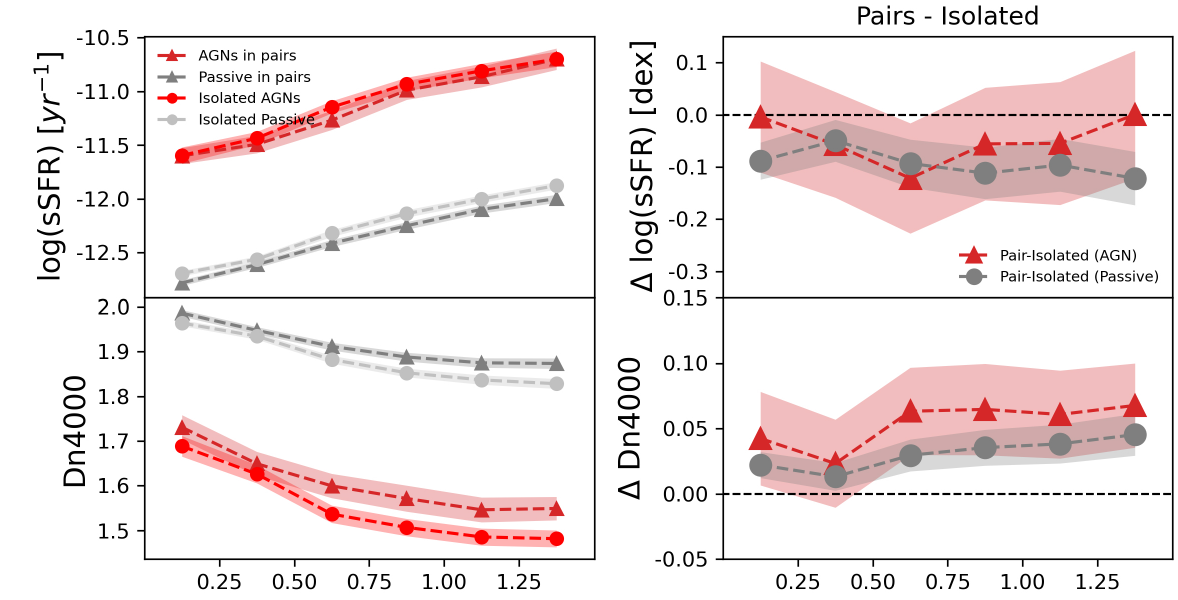}
  \caption{Comparison of the radial profiles of the sSFR (top) and Dn4000 (bottom)
  between the mass-controlled AGNs (red) and passive galaxies (grey).  
  Galaxy in pairs are marked with triangles,
  and isolated ones in dots.
  The sSFR of passive galaxies is calculated by the Dn4000-sSFR relation described in Appendix~\ref{ap:dn4kssfr}.
  AGNs have higher sSFR and lower Dn4000 at all radii than isolated galaxies, 
  regardless of whether they are in pairs or isolated. 
  No difference is found between the isolated galaxies and galaxies in pairs.
  The total galaxy numbers for each subsample is listed in Table~\ref{tab:control}. 
  }\label{fig:agnrg}
  \end{figure*}

We make similar comparison with the passive galaxies in Figure~\ref{fig:agnrg},
with a mass-controlled passive subsample as defined in
Sec.~\ref{subsec:controlsample}.
Given the low S/N of $\rm H\alpha$ and other emission lines,
it is difficult to derive the SFR of most spaxels in the passive galaxies. 
Dn4000 (or D4000) has been used to study the SFR in passive galaxies,
by using single fiber spectra \citep[e.g.][]{2004MNRAS.351.1151B,2007ApJS..173..267S,2016MNRAS.457.2703R},
or IFU spectra \citep[e.g.][]{2018MNRAS.476..580S,2019ApJ...877..132W,2020MNRAS.492...96B}.
Inspired by this, we derive our own resolved 
sSFR vs Dn4000 relation,
using the MaNGA spaxels with robust $\rm H\alpha$-based sSFR in all MPL-6 galaxies (Appendix~\ref{ap:dn4kssfr}).
Since Balmer decrements are not reliable in spaxels with low S/N of $\rm H\alpha$ and $\rm H\beta$ or no emission lines,
in Figure~\ref{fig:agnrg} we only compare the sSFR and Dn4000 radial profiles of the passive galaxies with AGNs.

The radial profiles of AGNs and passive galaxies show similar declining sSFR trends as radius decreases (Figure \ref{fig:agnrg}, left),
consistent with the inside-out quenching scenario. 
Regardless of the similar trends, passive galaxies still have lower sSFR by $\sim$1.2\,dex, 
and higher Dn4000 by $\sim$0.3.
This indicates that AGN host galaxies,
despite having lower sSFR than the SFGs (Figure~\ref{fig:agnsf}),
are still not as quenched as passive galaxies.
AGNs are more likely in transition between SFG and passive galaxies.
Comparing isolated and paired passive galaxies,
we find no difference with AGNs in their $\Delta$\,log(sSFR) and $\Delta$Dn4000 (Figure \ref{fig:agnrg}, right). 
The differential radial profiles between AGN and passive galaxies are generally flat within 3$\sigma$,
suggesting no interaction-triggered star formation activities in both populations.

In summary, unlike SFGs, AGNs and passive galaxies in pairs do not show SFR enhancement as compared to isolated galaxies.
One explanation of the less impact on star formation in AGN hosts may be the lack of sufficient gas,
similar to the passive galaxies.

\section{Discussions} \label{sec:discuss}

\subsection{Comparison to previous studies} 
\label{review}
Several previous works using single-fiber spectra from large surveys 
have found no AGN fraction evolution based on BPT-selected AGNs
among different galaxy merger cases.
These studies found
the same AGN fraction in galaxy pairs and isolated control sample
\citep[e.g.][]{2001AJ....122.2243S,2006MNRAS.371..786C,2007MNRAS.375.1017A,2008AJ....135.1877E,2010MNRAS.401.1552D},
no enhanced [\ion{O}{3}] luminosity in AGNs within galaxy pairs \citep[e.g.][]{2008MNRAS.385.1915L},
and no increase in neighbor numbers for higher [\ion{O}{3}] luminosity AGNs \citep[e.g.][]{2015MNRAS.448L..72S}.
Our sample is the first to study AGN fractions
along the merger sequence based on IFU data,
and we find no change of the IFU-classified AGN fraction
for different merger cases and isolated galaxies.

On the other hand, using galaxy pair samples from the IFU surveys,
\citet{2015A&A...579A..45B}, \citet{2019MNRAS.482L..55T},
\citet{2019ApJ...881..119P}, and \citet{2021ApJ...909..120S}
have studied the spatially resolved sSFR
of SFGs in pair or merger systems.
Despite the different sample selections,
a unanimous conclusion is that in SFGs,
galaxy interactions trigger stronger SF enhancements
in the center than in the disk.
The radial profile of the SFGs in our sample is also plot in Figure \ref{fig:agnsf}, 
and are consistent with previous studies with higher SF enhancement in the center. 
The AGNs in pairs, however, 
do not show any SF enhancement as compared to isolated AGNs.
\citet{2019ApJ...881..119P} perform the analysis along the merger sequence and
found the enhancement evolves in different merger cases.
The enhancement of central sSFR emerges after the `pre-merger' phase (Case 1).
We use the same parent pair sample and focus on the AGN pairs. 
We find that unlike SFGs, 
the evolution of AGNs' properties along the merger sequence
is not statistically significant (Figure \ref{fig:profile_case}).

The global and resolved properties of our AGNs are in general agreement 
with an inside-out quenching scenario,
as proposed in several earlier MaNGA works.
Compared to that in outer regions, the decrease of SFR in the central regions contributes more to galaxy quenching \citep[e.g.][]{2018MNRAS.474.2039E,2018ApJ...854..159P,2019ApJ...870...19G}. 
Based on our AGNs' GV-like colors (Figure \ref{fig:bptall}, d),
location below the SFMS (Figure \ref{fig:sfms}),
and sSFR and Dn4000 radial profiles being in between the SFGs and the passive galaxies (Figure \ref{fig:agnsf}, ~\ref{fig:agnrg}),
our BPT-selected AGNs are likely to be experiencing the transition
from SFGs to quiescent galaxies. 
The locations on the main sequence and the color-magnitude diagram of our AGN sample are
similar to previous BPT-selected AGNs in MaNGA\citep{2018ApJ...856...93F, 2018RMxAA..54..217S}. 
This possible transition can be explained by gas consumption,
by either previous star formation,
or AGN triggered outflows.
If the gas has been consumed already in the AGN systems,
then the lack of sSFR enhancement in AGN samples,
as observed in our sample,
can be naturally explained.
This is also supported by the 
lower global gas fractions than SFGs found in MaNGA AGNs
\citep[e.g. xCOLD GASS survey][]{2017ApJS..233...22S},
and lower $\rm H_{2}$ mass in AGNs than normal SFGs
at comparable star formation efficiencies \citep{2017ApJ...851...18L}. 

\subsection{Selection bias from the environment} \label{subsec:env}
In Sec~\ref{subsec:agnfrac}, we find a higher fraction of passive galaxies in pairs.
It is known that massive, bright, and passive early-type galaxies
tend to locate in a dense, clustering environment \citep[e.g.][]{2005ApJ...630....1Z,2006MNRAS.368...21L,2009MNRAS.399..966S,2011MNRAS.412..825D}.
In this section, we discuss the environmental influence ON our MaNGA pair sample in this section.

\begin{figure}
  \epsscale{1.18}
  \plotone{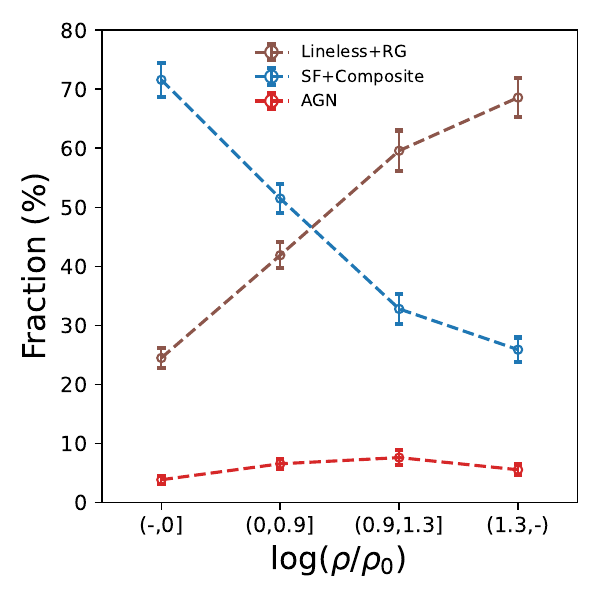}
  \caption{Fractions of isolated galaxies dividing into four local density bins.
  Error bars represent the binomial errors of each fraction.
  $\rho_{0}$ is the average cosmic mean density,
  equals to $\rm 7.16 \times 10^{10}\,M_{\sun}h^{-1}$(Mpc$/h)^{-3}$.
  The passive fraction becomes higher and (SFG+Composite) fraction becomes lower in denser environment, but the AGN fraction does not change significantly.
  The total numbers for each galaxy type can be found in Table~\ref{tab:bptresult}.}
  \label{fig:denfrac}
  \end{figure}

Different environmental indicators such as neighbouring galaxies and halo occupation distribution
represent the galaxy environment at different scales \citep[see][for a review]{2012MNRAS.419.2670M}.
Previous work by \citet{2004MNRAS.353..713K} has shown that star formation
mainly depends on galaxies' local environment.
So here we adopt the local mass density ($\rho$) from 
the MaNGA-GEMA\footnote{\url{https://data.sdss.org/datamodel/files/MANGA_GEMA/GEMA_VER}} catalog to investigate the influence of the galaxy environment.
The local mass density uses the halo-domain method developed by \citet{2009MNRAS.394..398W}
for the SDSS DR7 galaxy group catalog \citep[][]{2007ApJ...671..153Y}, 
which reconstructs the cosmic density field 
by calculating the Gaussian-kernel smoothed density at each galaxy's position in a scale of 1\,Mpc/$h$.
We compare the galaxy fraction for different galaxy types (Lineless+RG, SFG+composite, and AGN)
with their local mass densities (Figure \ref{fig:denfrac}).
The local mass densities are divided into four bins:
log($\rho/\rho_{0}$) = (-$\infty$,0], (0,0.9], (0.9,1.3] and (1.3,+$\infty$),
where $\rho_{0}$ is the average cosmic mean density,
equals to $\rm 7.16 \times 10^{10}\,M_{\sun}h^{-1}$(Mpc$/h)^{-3}$.
The fraction of passive galaxies increases with the local mass density bin 
from 25$\pm$2\% (log($\rho/\rho_{0}$)\,$<$\,0) to 69$\pm$3\% (log($\rho/\rho_{0}$)\,$>$\,1.3).
This is consistent with the observations that 
passive galaxies tend to locate in a denser environment \citep[e.g.][]{2002MNRAS.332..827N}.
In contrast, 
the fraction of SFG$+$composite galaxies
decreases as the local density increases \citep[e.g.][]{2004MNRAS.353..713K,2017ApJ...838...87C}. 
The AGN fraction remains more or less the same
from the lowest density (3.9$\pm$0.7\%) 
to the highest density (5.6$\pm$0.9\%), 
with a peak (7.6$\pm$1.2\%) at the median density of log($\rho/\rho_{0}$) = 0.9-1.3.

\begin{figure*}
  \plotone{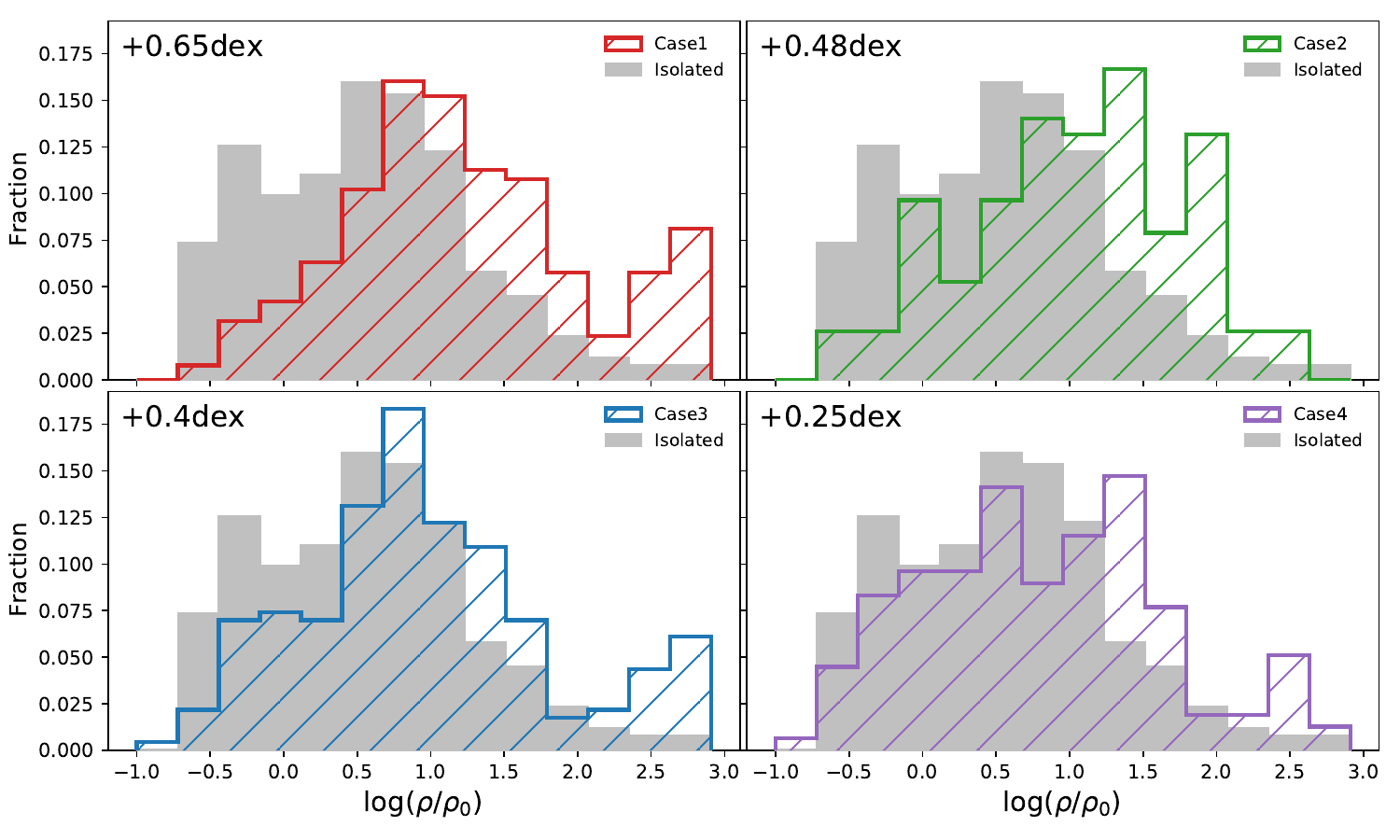}
  \caption{Histograms of the local mass density for galaxies in four merger case and isolated ones.
  The distribution offsets are listed in the top left of each panel.
  Case 1 and Case 2 show clearly higher local density distributions than isolated galaxies.
    The actual numbers for each pair case can be found in Table~\ref{tab:bptresult}. }
  \label{fig:denhist}
  \end{figure*}

We have shown that the more passive MaNGA galaxies live in denser environment,
we then compare the local mass density distributions between isolated galaxies and paired galaxies. 
In Figure \ref{fig:denhist} we plot the log($\rho/\rho_{0}$) distribution
for our four cases, along with the differences of the mean density in pairs and isolated galaxies.
The distribution of log($\rho/\rho_{0}$) is clearly higher 
for galaxies in Case 1 and Case 2 pairs, 
as compared to isolated galaxies (+0.63\,dex,  +0.45\,dex, respectively).
The density distributions in Case 4 
are more similar with the isolated galaxies, with $\Delta$log($\rho/\rho_{0}$) of +0.07 dex only. 
We suspect that our observed higher fractions of  passive galaxies
in Case 1 \& Case 2 are a result of
their overall denser environment.
The lack of a significantly higher fractions of  passive galaxies in Case 3
is a result of the morphology-based case definition, which excludes ETGs 
from Case 3 classification, as discussed in Sec.~\ref{subsec:agnfrac} .

\subsection{Selection bias of the AGN sample}
\label{subsec:agnbias}
In this paper, we use optical emission-line ratio and line width to select AGNs.
This method is based on the different emission line properties in AGNs' NLRs and \ion{H}{2} regions.
The emission from NLR could be contaminated by the broad emission lines or strong nuclear starbursts.
Therefore, our BPT-selection 
is biased against AGNs with strong broad-line component
or with strong central star formation \citep[e.g.][]{2015ApJ...811...26T}. 
The AGNs missed due to dust extinction or dilution from star formation are known to
lie along or above the SFMS \citep[e.g.][]{2017ApJS..233...19C}, 
which possibly contributes to the lack of AGNs above the SFMS in our Figure~\ref{fig:sfms}. 
Another selection bias of the BPT method 
is against AGNs with quenched host galaxy that has no or weak emission-lines due to lack of recent star formation
\citep[e.g.][]{2014ARA&A..52..589H}.
For instance, 
radio selected AGNs are doomed to be left out in our sample (e.g. the majority of radio AGNs have no emission lines 
from \citet{2020ApJ...901..159C}, see Sec.~\ref{subsec:bpt}). 
In addition,
the MaNGA survey aims to study the resolved properties of nearby galaxies. 
The most luminous quasars are therefore
not a preferred target as they easily outshine the host galaxies, 
making the data analysis difficult \citep{2017AJ....154...86W}. 
As a result, 
our sample is biased towards AGNs with median to low luminosity and low SFR,
landing them in the transition region in Figure~\ref{fig:sfms}.

\subsection{Caveats in the analysis of mergers}
\label{subsec:caveats}
When it comes to the late stage of merging, 
there are several caveats in both sample selection and the analysis.
First caveat is the merger classification.
As mentioned in Sec.~\ref{subsec:pairselect},
we visually classify the
late-stage merger systems missed by the pair selection based on physical separation and velocity offset.
Mergers with high inclination would be missed, 
while secularly evolved irregular galaxies could also contaminate the merger sample.
In addition, 
some galaxy pairs may not follow the Toomre Sequence \citep{1972ApJ...178..623T}
and should not be included in merger-sequence related analysis.
For example, \citet{2012ApJ...751...17S} have simulated and found that there are at least 
20\%-30\% flybys in galaxy pair samples from large surveys, 
which may smear out the actual evolutionary trends of the true merging pairs.

Moreover, the distortions in galaxy pairs may affect our analysis of their host galaxies.
The radii generated from typical ellipsoid model may not be appropriate for galaxies with bridges or tails,
resulting in inaccurate $R_{e}$ estimate.
The overlapping region between galaxies may contaminate the measurements of their properties.
Lastly, even though we require our isolated control sample
to show no distortion in SDSS images and have no spectroscopic companion,
it is possible that their SFR are affected by hidden minor mergers or flybys.

\section{Summary} \label{sec:conclusion}
In this work, we select 1156 local galaxies
in pair or merger systems from the MaNGA MPL-6
and classify them into 4 categories (cases), presumably representing various stages along the merger sequence.
Then we identify 61 AGNs in these pair systems using the BPT and WHAN diagrams
and compare them with isolated AGNs and SFGs
via both global and resolved properties.
We calculate the AGN fractions along the merger sequence,
analyze their global SFR-$M_{*}$ relation,
$\rm \Sigma$[\ion{O}{3}], and their resolved radial profiles of the sSFR, Balmer decrement, and Dn4000.
Our main conclusions are as follows:

(1) The AGN fraction of galaxies in pair or merger systems is consistent with that in isolated galaxies ($\sim$5\%).
This in agreement with several previous SDSS works
that found no significant AGN fraction change in galaxy pairs
\citep[e.g.][]{2007MNRAS.375.1017A,2008AJ....135.1877E,2010MNRAS.401.1552D}.
Besides, we do not find any evolution in AGN fractions for the different merger cases.
More passive galaxies and fewer SFGs are found in galaxy pairs,
especially in early merger stages, possibly due to their denser environment.

(2) As for the global properties, AGNs tend to locate in the transition region between main sequence galaxies and passive galaxies, partly due to selection bias. 
Compared to isolated AGNs, 
AGNs in pairs have similar stellar mass, global SFR, and $\rm \Sigma$[\ion{O}{3}]. 

(3) The resolved sSFR of AGN host galaxies, regardless in pairs or isolated,  
show an increase from the center to outskirts.
This supports the `inside-out' quenching scenario in AGN host galaxies.
Unlike the higher mass AGNs, 
AGNs with lower stellar mass (log($M_{*}/$$\rm M_{\sun}$)\,$<$\,10.5) 
show a different sSFR radial profile that is flat across all radius.
We find no sSFR difference between AGNs in pairs and isolated AGNs.

(4) The Balmer decrements of AGN host galaxies show an inside-out decrease,
indicating more dust attenuation in the central regions.

(5) The Dn4000 radial profile for AGNs decreases from the center to the outskirts,
and suggests older stellar populations in the galaxies' central regions with no recent star formation,
which is consistent with the sSFR results,
and similar to quenched galaxies reported earlier.
AGNs with lower stellar mass (log($M_{*}/$$\rm M_{\sun}$)\,$<$\,10.5) 
show a different Dn4000 radial profile that is flat across all radius.

(6) At all radii, AGNs have significantly lower sSFR and higher Dn4000 than SFGs, 
regardless of whether they are in pairs or isolated galaxies. 
They also show steeper Balmer decrement radial profiles. 
The enhanced SF in SFG pairs are not found in AGN pairs. 
Galaxy interactions enhance the sSFR of SFGs at all radii, 
especially in the central region, resulting in higher sSFR and lower Dn4000. 
While in AGNs and passive galaxies, no significant change in sSFR or Dn4000 is found between pairs and isolated galaxies.

\acknowledgments
We thank the anonymous referee for their helpful comments
that help improve the presentation of the paper.
We also thank Kevin Xu, Nicholas Fraser Boardman, Chuan He and Zijian Li for helpful discussions. 
Support for this work is provided by the Chinese National Nature Science Foundation grant No. 10878003.
This work was supported in part by the National Key R\&D Program of China via grant No.2017YFA0402703
and by NSFC grants 11433003, 11822303, 11773020, 11733002, 11933003, 11373034, 11803044, and 11673028.
Additional support came from the Chinese Academy of Sciences (CAS) through a grant to the South America Center for Astronomy (CASSACA)
in Santiago, Chile.
This project makes use of the MaNGA-PIPE3D data products, and we thank the IA-UNAM MaNGA team for creating this catalog,
and the CONACyT-180125 project for supporting them.

Funding for the Sloan Digital Sky Survey IV has been provided by the Alfred P. Sloan Foundation, the U.S. Department of Energy Office of Science, and the Participating Institutions. SDSS-IV acknowledges
support and resources from the Center for High-Performance Computing at
the University of Utah. The SDSS website is \url{www.sdss.org}.
SDSS-IV is managed by the Astrophysical Research Consortium for the 
Participating Institutions of the SDSS Collaboration including the 
Brazilian Participation Group, the Carnegie Institution for Science, 
Carnegie Mellon University, the Chilean Participation Group, the French Participation Group, Harvard-Smithsonian Center for Astrophysics, 
Instituto de Astrof\'isica de Canarias, The Johns Hopkins University, Kavli Institute for the Physics and Mathematics of the Universe (IPMU) / 
University of Tokyo, the Korean Participation Group, Lawrence Berkeley National Laboratory, 
Leibniz Institut f\"ur Astrophysik Potsdam (AIP),  
Max-Planck-Institut f\"ur Astronomie (MPIA Heidelberg), 
Max-Planck-Institut f\"ur Astrophysik (MPA Garching), 
Max-Planck-Institut f\"ur Extraterrestrische Physik (MPE), 
National Astronomical Observatories of China, New Mexico State University, 
New York University, University of Notre Dame, 
Observat\'ario Nacional / MCTI, The Ohio State University, 
Pennsylvania State University, Shanghai Astronomical Observatory, 
United Kingdom Participation Group,
Universidad Nacional Aut\'onoma de M\'exico, University of Arizona, 
University of Colorado Boulder, University of Oxford, University of Portsmouth, 
University of Utah, University of Virginia, University of Washington, University of Wisconsin, 
Vanderbilt University, and Yale University.

\software{astropy \citep{2013A&A...558A..33A},  
          Marvin \citep{2019AJ....158...74C}
          }

\appendix
\setcounter{figure}{0}
\renewcommand{\thefigure}{A\arabic{figure}}

\section{SFR in AGNs and Passive Galaxies}
\subsection{Decomposition of the $\rm H\alpha$ emission}\label{ap:decompose}
\begin{figure*}
  \plotone{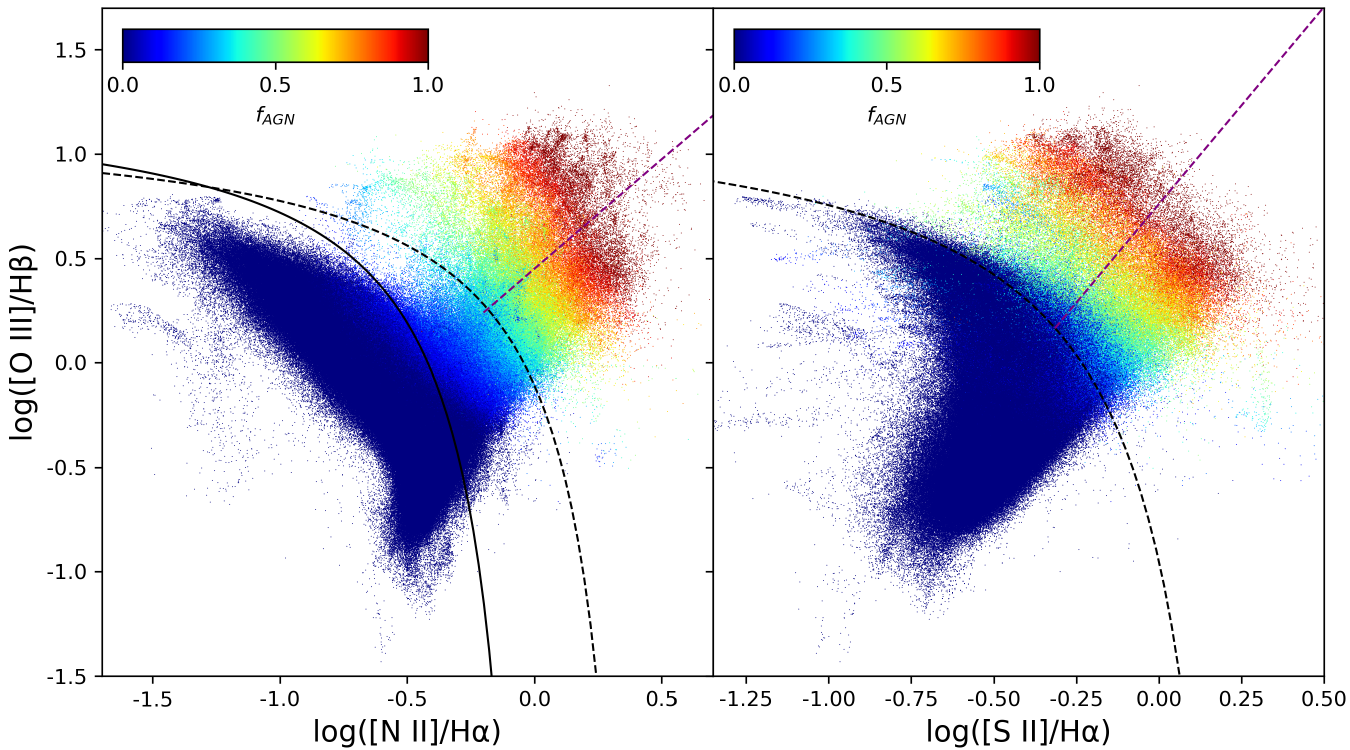}
  \caption{Spaxels' location in the BPT diagrams and their AGN fraction to the $\rm H\alpha$ emission ($f_{AGN}$). All spaxels have robust S/N and are coded by the value of $f_{AGN}$ derived from Equation~\ref{eq:fagn}. $f_{AGN}$ increase from 0 in the star-forming region to 1 towards the edge of AGN sequence.}
  \label{fig:decompose}
\end{figure*}

In this Appendix we describe how we do the AGN-SF decomposition of the dust-corrected $\rm H\alpha$,
as mentioned in Sec.~\ref{subsec:measure}.
We require the S/N of [\ion{O}{3}]$_{\lambda 5007}$, $\rm H\beta$,
[\ion{N}{2}]$_{\lambda 6584}$, ${\rm H\alpha}$, [\ion{S}{2}]$_{\lambda\lambda 6716,6731}$ should all be greater than 5.
We use these emission-lines and Equation~\ref{eq:fagn} to calculate the contribution of AGN to $\rm H\alpha$ emission ($\rm H\alpha _{AGN}/H\alpha _{total}$, $f_{AGN}$).
Then we plot all the qualifying spaxels in the BPT and modified BPT diagrams, as shown in Figure~\ref{fig:decompose}, coded by their $f_{AGN}$ values.
In the star-forming regions of the two diagrams, the AGN's contribution to $\rm H\alpha$ emission is negligible.
Along the Ke01 maximum starburst line (black dashed) in the [\ion{N}{2}]-BPT diagram, the $f_{AGN}$ is about 40\%, 
consistent with the approximation ($\sim$50\%) in \citet{2009MNRAS.397..135K}, derived from the SDSS single fiber spectra.
The 100\% AGN boundary is better defined in the [\ion{S}{2}]-BPT diagram than that in the [\ion{N}{2}]-BPT diagram.

\subsection{The sSFR-Dn4000 relation}\label{ap:dn4kssfr}
\begin{figure}
  \epsscale{1.1}  
  \plotone{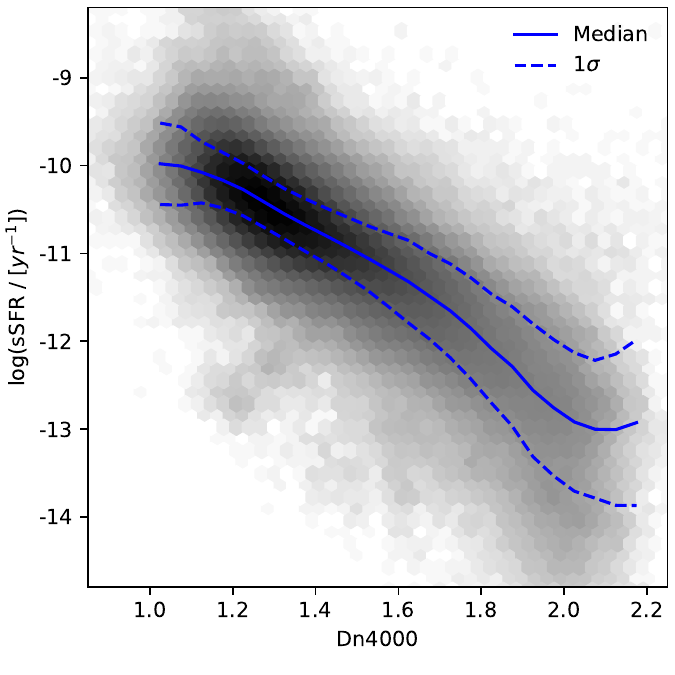}
  \caption{The two dimensional histogram of resolved sSFR versus Dn4000 for all MPL-6 spaxels with enough emission line S/N.
  The sSFR is calculated from the AGN-removed $\rm H\alpha$ luminosity,
  as described in Sec.~\ref{subsec:measure}.
  The gray colors represent the number density in a logarithm scale.
  The median relation is shown as the blue solid line.
  Blue dahsed lines represent the 1$\sigma$ dispersion.
  The average standard deviation of sSFR is about 0.6 dex.
  We note that the standard deviation is higher (about 0.8 dex) for high Dn4000 spaxels (Dn4000$>$1.7).
}\label{fig:dnssfrcal}
\end{figure}

For MaNGA spaxels with low S/N or without ${\rm H\alpha}$ emission,
SFR cannot be derived directly from the ${\rm H\alpha}$ emission. 
Inspired by \citet{2018MNRAS.476..580S,2019ApJ...877..132W};
and \citet{2020MNRAS.492...96B},
here we adopt Dn4000 as a proxy of sSFR. 
We derive the sSFR-Dn4000 correlation
based on 
all MPL-6 spaxels with robust stellar mass, dust-corrected ${\rm H\alpha}$ luminosity, and Dn4000 values,
regardless of their galaxy type. 
Figure~\ref{fig:dnssfrcal} shows the actual distribution
of the reliable Dn4000 and sSFR distributions for all spaxels,
and the derived median and 1\,$\sigma$ dispersion, with a Dn4000 bin size of 0.05. 
The 5th order polynomial fit can be expressed as:
$y = 19.0x^5 -145.0x^4+473.5x^3+651.9x^2+478.1x-147.6$,
where y = sSFR, and x = Dn4000. 
The average 1$\sigma$ dispersion for sSFR is $\sim$\,0.6 dex.
We note that this function can only be used in the Dn4000 range between 1.0 and 2.1.
Our derived correlation shows a similar negative trend as found in previous works,
though the gradient factor and uncertainties varies from study to study. 
The difference may rise from the different SFR estimators used. 
For instance,  single fiber ${\rm H\alpha}$ luminosity was used in \citet{2004MNRAS.351.1151B},
while UV photometry in \citet{2007ApJS..173..267S}, 
and IFS ${\rm H\alpha}$ luminosity in \citet{2018MNRAS.476..580S,2018ApJ...856..137W,2020MNRAS.492...96B}.
Here we use the dust-corrected, AGN-removed IFU ${\rm H\alpha}$ for each spaxel.

\bibliography{article}{}

\begin{thebibliography}{}
\expandafter\ifx\csname natexlab\endcsname\relax\def\natexlab#1{#1}\fi
\providecommand{\url}[1]{\href{#1}{#1}}
\providecommand{\dodoi}[1]{doi:~\href{http://doi.org/#1}{\nolinkurl{#1}}}
\providecommand{\doeprint}[1]{\href{http://ascl.net/#1}{\nolinkurl{http://ascl.net/#1}}}
\providecommand{\doarXiv}[1]{\href{https://arxiv.org/abs/#1}{\nolinkurl{https://arxiv.org/abs/#1}}}

\bibitem[{{Abazajian} {et~al.}(2009){Abazajian}, {Adelman-McCarthy},
  {Ag{\"u}eros}, {Allam}, {Allende Prieto}, {An}, {Anderson}, {Anderson},
  {Annis}, {Bahcall}, {Bailer-Jones}, {Barentine}, {Bassett}, {Becker},
  {Beers}, {Bell}, {Belokurov}, {Berlind}, {Berman}, {Bernardi}, {Bickerton},
  {Bizyaev}, {Blakeslee}, {Blanton}, {Bochanski}, {Boroski}, {Brewington},
  {Brinchmann}, {Brinkmann}, {Brunner}, {Budav{\'a}ri}, {Carey}, {Carliles},
  {Carr}, {Castander}, {Cinabro}, {Connolly}, {Csabai}, {Cunha}, {Czarapata},
  {Davenport}, {de Haas}, {Dilday}, {Doi}, {Eisenstein}, {Evans}, {Evans},
  {Fan}, {Friedman}, {Frieman}, {Fukugita}, {G{\"a}nsicke}, {Gates},
  {Gillespie}, {Gilmore}, {Gonzalez}, {Gonzalez}, {Grebel}, {Gunn},
  {Gy{\"o}ry}, {Hall}, {Harding}, {Harris}, {Harvanek}, {Hawley}, {Hayes},
  {Heckman}, {Hendry}, {Hennessy}, {Hindsley}, {Hoblitt}, {Hogan}, {Hogg},
  {Holtzman}, {Hyde}, {Ichikawa}, {Ichikawa}, {Im}, {Ivezi{\'c}}, {Jester},
  {Jiang}, {Johnson}, {Jorgensen}, {Juri{\'c}}, {Kent}, {Kessler}, {Kleinman},
  {Knapp}, {Konishi}, {Kron}, {Krzesinski}, {Kuropatkin}, {Lampeitl},
  {Lebedeva}, {Lee}, {Lee}, {French Leger}, {L{\'e}pine}, {Li}, {Lima}, {Lin},
  {Long}, {Loomis}, {Loveday}, {Lupton}, {Magnier}, {Malanushenko},
  {Malanushenko}, {Mand elbaum}, {Margon}, {Marriner}, {Mart{\'\i}nez-Delgado},
  {Matsubara}, {McGehee}, {McKay}, {Meiksin}, {Morrison}, {Mullally}, {Munn},
  {Murphy}, {Nash}, {Nebot}, {Neilsen}, {Newberg}, {Newman}, {Nichol},
  {Nicinski}, {Nieto-Santisteban}, {Nitta}, {Okamura}, {Oravetz}, {Ostriker},
  {Owen}, {Padmanabhan}, {Pan}, {Park}, {Pauls}, {Peoples}, {Percival}, {Pier},
  {Pope}, {Pourbaix}, {Price}, {Purger}, {Quinn}, {Raddick}, {Re Fiorentin},
  {Richards}, {Richmond}, {Riess}, {Rix}, {Rockosi}, {Sako}, {Schlegel},
  {Schneider}, {Scholz}, {Schreiber}, {Schwope}, {Seljak}, {Sesar}, {Sheldon},
  {Shimasaku}, {Sibley}, {Simmons}, {Sivarani}, {Allyn Smith}, {Smith},
  {Smol{\v{c}}i{\'c}}, {Snedden}, {Stebbins}, {Steinmetz}, {Stoughton},
  {Strauss}, {SubbaRao}, {Suto}, {Szalay}, {Szapudi}, {Szkody}, {Tanaka},
  {Tegmark}, {Teodoro}, {Thakar}, {Tremonti}, {Tucker}, {Uomoto}, {Vanden
  Berk}, {Vandenberg}, {Vidrih}, {Vogeley}, {Voges}, {Vogt}, {Wadadekar},
  {Watters}, {Weinberg}, {West}, {White}, {Wilhite}, {Wonders}, {Yanny},
  {Yocum}, {York}, {Zehavi}, {Zibetti}, \& {Zucker}}]{2009ApJS..182..543A}
{Abazajian}, K.~N., {Adelman-McCarthy}, J.~K., {Ag{\"u}eros}, M.~A., {et~al.}
  2009, \apjs, 182, 543, \dodoi{10.1088/0067-0049/182/2/543}

\bibitem[{{Ackermann} {et~al.}(2018){Ackermann}, {Schawinski}, {Zhang},
  {Weigel}, \& {Turp}}]{2018MNRAS.479..415A}
{Ackermann}, S., {Schawinski}, K., {Zhang}, C., {Weigel}, A.~K., \& {Turp},
  M.~D. 2018, \mnras, 479, 415, \dodoi{10.1093/mnras/sty1398}

\bibitem[{{Alonso} {et~al.}(2007){Alonso}, {Lambas}, {Tissera}, \&
  {Coldwell}}]{2007MNRAS.375.1017A}
{Alonso}, M.~S., {Lambas}, D.~G., {Tissera}, P., \& {Coldwell}, G. 2007,
  \mnras, 375, 1017, \dodoi{10.1111/j.1365-2966.2007.11367.x}

\bibitem[{{Alonso} {et~al.}(2018){Alonso}, {Coldwell}, {Duplancic}, {Mesa}, \&
  {Lambas}}]{2018A&A...618A.149A}
{Alonso}, S., {Coldwell}, G., {Duplancic}, F., {Mesa}, V., \& {Lambas}, D.~G.
  2018, \aap, 618, A149, \dodoi{10.1051/0004-6361/201832796}

\bibitem[{{Alpaslan} {et~al.}(2015){Alpaslan}, {Driver}, {Robotham},
  {Obreschkow}, {Andrae}, {Cluver}, {Kelvin}, {Lange}, {Owers}, {Taylor},
  {Andrews}, {Bamford}, {Bland-Hawthorn}, {Brough}, {Brown}, {Colless},
  {Davies}, {Eardley}, {Grootes}, {Hopkins}, {Kennedy}, {Liske},
  {Lara-L{\'o}pez}, {L{\'o}pez-S{\'a}nchez}, {Loveday}, {Madore}, {Mahajan},
  {Meyer}, {Moffett}, {Norberg}, {Penny}, {Pimbblet}, {Popescu}, {Seibert}, \&
  {Tuffs}}]{2015MNRAS.451.3249A}
{Alpaslan}, M., {Driver}, S., {Robotham}, A. S.~G., {et~al.} 2015, \mnras, 451,
  3249, \dodoi{10.1093/mnras/stv1176}

\bibitem[{{Argudo-Fern{\'a}ndez} {et~al.}(2016){Argudo-Fern{\'a}ndez}, {Shen},
  {Sabater}, {Duarte Puertas}, {Verley}, \& {Yang}}]{2016A&A...592A..30A}
{Argudo-Fern{\'a}ndez}, M., {Shen}, S., {Sabater}, J., {et~al.} 2016, \aap,
  592, A30, \dodoi{10.1051/0004-6361/201628232}

\bibitem[{{Astropy Collaboration} {et~al.}(2013){Astropy Collaboration},
  {Robitaille}, {Tollerud}, {Greenfield}, {Droettboom}, {Bray}, {Aldcroft},
  {Davis}, {Ginsburg}, {Price-Whelan}, {Kerzendorf}, {Conley}, {Crighton},
  {Barbary}, {Muna}, {Ferguson}, {Grollier}, {Parikh}, {Nair}, {Unther},
  {Deil}, {Woillez}, {Conseil}, {Kramer}, {Turner}, {Singer}, {Fox}, {Weaver},
  {Zabalza}, {Edwards}, {Azalee Bostroem}, {Burke}, {Casey}, {Crawford},
  {Dencheva}, {Ely}, {Jenness}, {Labrie}, {Lim}, {Pierfederici}, {Pontzen},
  {Ptak}, {Refsdal}, {Servillat}, \& {Streicher}}]{2013A&A...558A..33A}
{Astropy Collaboration}, {Robitaille}, T.~P., {Tollerud}, E.~J., {et~al.} 2013,
  \aap, 558, A33, \dodoi{10.1051/0004-6361/201322068}

\bibitem[{{Baldwin} {et~al.}(1981){Baldwin}, {Phillips}, \&
  {Terlevich}}]{1981PASP...93....5B}
{Baldwin}, J.~A., {Phillips}, M.~M., \& {Terlevich}, R. 1981, \pasp, 93, 5,
  \dodoi{10.1086/130766}

\bibitem[{{Barnes}(1988)}]{1988ApJ...331..699B}
{Barnes}, J.~E. 1988, \apj, 331, 699, \dodoi{10.1086/166593}

\bibitem[{{Barnes} \& {Hernquist}(1992)}]{1992ARA&A..30..705B}
{Barnes}, J.~E., \& {Hernquist}, L. 1992, \araa, 30, 705,
  \dodoi{10.1146/annurev.aa.30.090192.003421}

\bibitem[{{Barnes} \& {Hernquist}(1991)}]{1991ApJ...370L..65B}
{Barnes}, J.~E., \& {Hernquist}, L.~E. 1991, \apjl, 370, L65,
  \dodoi{10.1086/185978}

\bibitem[{{Barrera-Ballesteros}
  {et~al.}(2015{\natexlab{a}}){Barrera-Ballesteros}, {S{\'a}nchez},
  {Garc{\'\i}a-Lorenzo}, {Falc{\'o}n-Barroso}, {Mast}, {Garc{\'\i}a-Benito},
  {Husemann}, {van de Ven}, {Iglesias-P{\'a}ramo}, {Rosales-Ortega},
  {P{\'e}rez-Torres}, {M{\'a}rquez}, {Kehrig}, {Marino}, {Vilchez}, {Galbany},
  {L{\'o}pez-S{\'a}nchez}, {Walcher}, \& {Califa
  Collaboration}}]{2015A&A...579A..45B}
{Barrera-Ballesteros}, J.~K., {S{\'a}nchez}, S.~F., {Garc{\'\i}a-Lorenzo}, B.,
  {et~al.} 2015{\natexlab{a}}, \aap, 579, A45,
  \dodoi{10.1051/0004-6361/201425397}

\bibitem[{{Barrera-Ballesteros}
  {et~al.}(2015{\natexlab{b}}){Barrera-Ballesteros}, {Garc{\'\i}a-Lorenzo},
  {Falc{\'o}n-Barroso}, {van de Ven}, {Lyubenova}, {Wild}, {M{\'e}ndez-Abreu},
  {S{\'a}nchez}, {Marquez}, {Masegosa}, {Monreal-Ibero}, {Ziegler}, {del Olmo},
  {Verdes-Montenegro}, {Garc{\'\i}a-Benito}, {Husemann}, {Mast}, {Kehrig},
  {Iglesias-Paramo}, {Marino}, {Aguerri}, {Walcher}, {V{\'\i}lchez}, {Bomans},
  {Cortijo-Ferrero}, {Gonz{\'a}lez Delgado}, {Bland-Hawthorn}, {McIntosh}, \&
  {Bekerait{\.{e}}}}]{2015A&A...582A..21B}
{Barrera-Ballesteros}, J.~K., {Garc{\'\i}a-Lorenzo}, B., {Falc{\'o}n-Barroso},
  J., {et~al.} 2015{\natexlab{b}}, \aap, 582, A21,
  \dodoi{10.1051/0004-6361/201424935}

\bibitem[{{Barton} {et~al.}(2000){Barton}, {Geller}, \&
  {Kenyon}}]{2000ApJ...530..660B}
{Barton}, E.~J., {Geller}, M.~J., \& {Kenyon}, S.~J. 2000, \apj, 530, 660,
  \dodoi{10.1086/308392}

\bibitem[{{Belfiore} {et~al.}(2018){Belfiore}, {Maiolino}, {Bundy}, {Masters},
  {Bershady}, {Oyarz{\'u}n}, {Lin}, {Cano-Diaz}, {Wake}, {Spindler}, {Thomas},
  {Brownstein}, {Drory}, \& {Yan}}]{2018MNRAS.477.3014B}
{Belfiore}, F., {Maiolino}, R., {Bundy}, K., {et~al.} 2018, \mnras, 477, 3014,
  \dodoi{10.1093/mnras/sty768}

\bibitem[{{Belfiore} {et~al.}(2019){Belfiore}, {Westfall}, {Schaefer},
  {Cappellari}, {Ji}, {Bershady}, {Tremonti}, {Law}, {Yan}, {Bundy}, {Shetty},
  {Drory}, {Thomas}, {Emsellem}, \& {S{\'a}nchez}}]{2019AJ....158..160B}
{Belfiore}, F., {Westfall}, K.~B., {Schaefer}, A., {et~al.} 2019, \aj, 158,
  160, \dodoi{10.3847/1538-3881/ab3e4e}

\bibitem[{{Bennert} {et~al.}(2008){Bennert}, {Canalizo}, {Jungwiert},
  {Stockton}, {Schweizer}, {Peng}, \& {Lacy}}]{2008ApJ...677..846B}
{Bennert}, N., {Canalizo}, G., {Jungwiert}, B., {et~al.} 2008, \apj, 677, 846,
  \dodoi{10.1086/529068}

\bibitem[{{Best} {et~al.}(2005){Best}, {Kauffmann}, {Heckman}, \&
  {Ivezi{\'c}}}]{2005MNRAS.362....9B}
{Best}, P.~N., {Kauffmann}, G., {Heckman}, T.~M., \& {Ivezi{\'c}}, {\v{Z}}.
  2005, \mnras, 362, 9, \dodoi{10.1111/j.1365-2966.2005.09283.x}

\bibitem[{{Binette} {et~al.}(1994){Binette}, {Magris}, {Stasi{\'n}ska}, \&
  {Bruzual}}]{1994A&A...292...13B}
{Binette}, L., {Magris}, C.~G., {Stasi{\'n}ska}, G., \& {Bruzual}, A.~G. 1994,
  \aap, 292, 13

\bibitem[{{Bing} {et~al.}(2019){Bing}, {Shi}, {Chen}, {S{\'a}nchez},
  {Maiolino}, {Riffel}, {Riffel}, {Wylezalek}, {Bizyaev}, {Pan}, \&
  {Drory}}]{2019MNRAS.482..194B}
{Bing}, L., {Shi}, Y., {Chen}, Y., {et~al.} 2019, \mnras, 482, 194,
  \dodoi{10.1093/mnras/sty2662}

\bibitem[{{Blanton} {et~al.}(2017){Blanton}, {Bershady}, {Abolfathi},
  {Albareti}, {Allende Prieto}, {Almeida}, {Alonso-Garc{\'\i}a}, {Anders},
  {Anderson}, {Andrews}, {Aquino-Ort{\'\i}z}, {Arag{\'o}n-Salamanca},
  {Argudo-Fern{\'a}ndez}, {Armengaud}, {Aubourg}, {Avila-Reese}, {Badenes},
  {Bailey}, {Barger}, {Barrera-Ballesteros}, {Bartosz}, {Bates}, {Baumgarten},
  {Bautista}, {Beaton}, {Beers}, {Belfiore}, {Bender}, {Berlind}, {Bernardi},
  {Beutler}, {Bird}, {Bizyaev}, {Blanc}, {Blomqvist}, {Bolton}, {Boquien},
  {Borissova}, {van den Bosch}, {Bovy}, {Brandt}, {Brinkmann}, {Brownstein},
  {Bundy}, {Burgasser}, {Burtin}, {Busca}, {Cappellari}, {Delgado Carigi},
  {Carlberg}, {Carnero Rosell}, {Carrera}, {Chanover}, {Cherinka}, {Cheung},
  {G{\'o}mez Maqueo Chew}, {Chiappini}, {Choi}, {Chojnowski}, {Chuang},
  {Chung}, {Cirolini}, {Clerc}, {Cohen}, {Comparat}, {da Costa}, {Cousinou},
  {Covey}, {Crane}, {Croft}, {Cruz-Gonzalez}, {Garrido Cuadra}, {Cunha},
  {Damke}, {Darling}, {Davies}, {Dawson}, {de la Macorra}, {Dell'Agli}, {De
  Lee}, {Delubac}, {Di Mille}, {Diamond-Stanic}, {Cano-D{\'\i}az}, {Donor},
  {Downes}, {Drory}, {du Mas des Bourboux}, {Duckworth}, {Dwelly}, {Dyer},
  {Ebelke}, {Eigenbrot}, {Eisenstein}, {Emsellem}, {Eracleous}, {Escoffier},
  {Evans}, {Fan}, {Fern{\'a}ndez-Alvar}, {Fernandez-Trincado}, {Feuillet},
  {Finoguenov}, {Fleming}, {Font-Ribera}, {Fredrickson}, {Freischlad},
  {Frinchaboy}, {Fuentes}, {Galbany}, {Garcia-Dias},
  {Garc{\'\i}a-Hern{\'a}ndez}, {Gaulme}, {Geisler}, {Gelfand},
  {Gil-Mar{\'\i}n}, {Gillespie}, {Goddard}, {Gonzalez-Perez}, {Grabowski},
  {Green}, {Grier}, {Gunn}, {Guo}, {Guy}, {Hagen}, {Hahn}, {Hall}, {Harding},
  {Hasselquist}, {Hawley}, {Hearty}, {Gonzalez Hern{\'a}ndez}, {Ho}, {Hogg},
  {Holley-Bockelmann}, {Holtzman}, {Holzer}, {Huehnerhoff}, {Hutchinson},
  {Hwang}, {Ibarra-Medel}, {da Silva Ilha}, {Ivans}, {Ivory}, {Jackson},
  {Jensen}, {Johnson}, {Jones}, {J{\"o}nsson}, {Jullo}, {Kamble}, {Kinemuchi},
  {Kirkby}, {Kitaura}, {Klaene}, {Knapp}, {Kneib}, {Kollmeier}, {Lacerna},
  {Lane}, {Lang}, {Law}, {Lazarz}, {Lee}, {Le Goff}, {Liang}, {Li}, {Li},
  {Lian}, {Lima}, {Lin}, {Lin}, {Bertran de Lis}, {Liu}, {de Icaza Lizaola},
  {Long}, {Lucatello}, {Lundgren}, {MacDonald}, {Deconto Machado}, {MacLeod},
  {Mahadevan}, {Geimba Maia}, {Maiolino}, {Majewski}, {Malanushenko},
  {Malanushenko}, {Manchado}, {Mao}, {Maraston}, {Marques-Chaves}, {Masseron},
  {Masters}, {McBride}, {McDermid}, {McGrath}, {McGreer}, {Medina Pe{\~n}a},
  {Melendez}, {Merloni}, {Merrifield}, {Meszaros}, {Meza}, {Minchev},
  {Minniti}, {Miyaji}, {More}, {Mulchaey}, {M{\"u}ller-S{\'a}nchez}, {Muna},
  {Munoz}, {Myers}, {Nair}, {Nandra}, {Correa do Nascimento}, {Negrete},
  {Ness}, {Newman}, {Nichol}, {Nidever}, {Nitschelm}, {Ntelis}, {O'Connell},
  {Oelkers}, {Oravetz}, {Oravetz}, {Pace}, {Padilla}, {Palanque-Delabrouille},
  {Alonso Palicio}, {Pan}, {Parejko}, {Parikh}, {P{\^a}ris}, {Park}, {Patten},
  {Peirani}, {Pellejero-Ibanez}, {Penny}, {Percival}, {Perez-Fournon},
  {Petitjean}, {Pieri}, {Pinsonneault}, {Pisani}, {Poleski}, {Prada},
  {Prakash}, {Queiroz}, {Raddick}, {Raichoor}, {Barboza Rembold}, {Richstein},
  {Riffel}, {Riffel}, {Rix}, {Robin}, {Rockosi}, {Rodr{\'\i}guez-Torres},
  {Roman-Lopes}, {Rom{\'a}n-Z{\'u}{\~n}iga}, {Rosado}, {Ross}, {Rossi}, {Ruan},
  {Ruggeri}, {Rykoff}, {Salazar-Albornoz}, {Salvato}, {S{\'a}nchez}, {Aguado},
  {S{\'a}nchez-Gallego}, {Santana}, {Santiago}, {Sayres}, {Schiavon}, {da Silva
  Schimoia}, {Schlafly}, {Schlegel}, {Schneider}, {Schultheis}, {Schuster},
  {Schwope}, {Seo}, {Shao}, {Shen}, {Shetrone}, {Shull}, {Simon}, {Skinner},
  {Skrutskie}, {Slosar}, {Smith}, {Sobeck}, {Sobreira}, {Somers}, {Souto},
  {Stark}, {Stassun}, {Stauffer}, {Steinmetz}, {Storchi-Bergmann},
  {Streblyanska}, {Stringfellow}, {Su{\'a}rez}, {Sun}, {Suzuki}, {Szigeti},
  {Taghizadeh-Popp}, {Tang}, {Tao}, {Tayar}, {Tembe}, {Teske}, {Thakar},
  {Thomas}, {Thompson}, {Tinker}, {Tissera}, {Tojeiro}, {Hernandez Toledo}, {de
  la Torre}, {Tremonti}, {Troup}, {Valenzuela}, {Martinez Valpuesta},
  {Vargas-Gonz{\'a}lez}, {Vargas-Maga{\~n}a}, {Vazquez}, {Villanova}, {Vivek},
  {Vogt}, {Wake}, {Walterbos}, {Wang}, {Weaver}, {Weijmans}, {Weinberg},
  {Westfall}, {Whelan}, {Wild}, {Wilson}, {Wood-Vasey}, {Wylezalek}, {Xiao},
  {Yan}, {Yang}, {Ybarra}, {Y{\`e}che}, {Zakamska}, {Zamora}, {Zarrouk},
  {Zasowski}, {Zhang}, {Zhao}, {Zheng}, {Zheng}, {Zhou}, {Zhou}, {Zhu},
  {Zoccali}, \& {Zou}}]{2017AJ....154...28B}
{Blanton}, M.~R., {Bershady}, M.~A., {Abolfathi}, B., {et~al.} 2017, \aj, 154,
  28, \dodoi{10.3847/1538-3881/aa7567}

\bibitem[{{Bluck} {et~al.}(2020){Bluck}, {Maiolino}, {S{\'a}nchez}, {Ellison},
  {Thorp}, {Piotrowska}, {Teimoorinia}, \& {Bundy}}]{2020MNRAS.492...96B}
{Bluck}, A. F.~L., {Maiolino}, R., {S{\'a}nchez}, S.~F., {et~al.} 2020, \mnras,
  492, 96, \dodoi{10.1093/mnras/stz3264}

\bibitem[{{Blumenthal} \& {Barnes}(2018)}]{2018MNRAS.479.3952B}
{Blumenthal}, K.~A., \& {Barnes}, J.~E. 2018, \mnras, 479, 3952,
  \dodoi{10.1093/mnras/sty1605}

\bibitem[{{Bottrell} {et~al.}(2019){Bottrell}, {Hani}, {Teimoorinia},
  {Ellison}, {Moreno}, {Torrey}, {Hayward}, {Thorp}, {Simard}, \&
  {Hernquist}}]{2019MNRAS.490.5390B}
{Bottrell}, C., {Hani}, M.~H., {Teimoorinia}, H., {et~al.} 2019, \mnras, 490,
  5390, \dodoi{10.1093/mnras/stz2934}

\bibitem[{{Brinchmann} {et~al.}(2004){Brinchmann}, {Charlot}, {White},
  {Tremonti}, {Kauffmann}, {Heckman}, \& {Brinkmann}}]{2004MNRAS.351.1151B}
{Brinchmann}, J., {Charlot}, S., {White}, S.~D.~M., {et~al.} 2004, \mnras, 351,
  1151, \dodoi{10.1111/j.1365-2966.2004.07881.x}

\bibitem[{{Bundy} {et~al.}(2015){Bundy}, {Bershady}, {Law}, {Yan}, {Drory},
  {MacDonald}, {Wake}, {Cherinka}, {S{\'a}nchez-Gallego}, {Weijmans}, {Thomas},
  {Tremonti}, {Masters}, {Coccato}, {Diamond-Stanic}, {Arag{\'o}n-Salamanca},
  {Avila-Reese}, {Badenes}, {Falc{\'o}n-Barroso}, {Belfiore}, {Bizyaev},
  {Blanc}, {Bland-Hawthorn}, {Blanton}, {Brownstein}, {Byler}, {Cappellari},
  {Conroy}, {Dutton}, {Emsellem}, {Etherington}, {Frinchaboy}, {Fu}, {Gunn},
  {Harding}, {Johnston}, {Kauffmann}, {Kinemuchi}, {Klaene}, {Knapen},
  {Leauthaud}, {Li}, {Lin}, {Maiolino}, {Malanushenko}, {Malanushenko}, {Mao},
  {Maraston}, {McDermid}, {Merrifield}, {Nichol}, {Oravetz}, {Pan}, {Parejko},
  {Sanchez}, {Schlegel}, {Simmons}, {Steele}, {Steinmetz}, {Thanjavur},
  {Thompson}, {Tinker}, {van den Bosch}, {Westfall}, {Wilkinson}, {Wright},
  {Xiao}, \& {Zhang}}]{2015ApJ...798....7B}
{Bundy}, K., {Bershady}, M.~A., {Law}, D.~R., {et~al.} 2015, \apj, 798, 7,
  \dodoi{10.1088/0004-637X/798/1/7}

\bibitem[{{Calzetti} {et~al.}(2000){Calzetti}, {Armus}, {Bohlin}, {Kinney},
  {Koornneef}, \& {Storchi-Bergmann}}]{2000ApJ...533..682C}
{Calzetti}, D., {Armus}, L., {Bohlin}, R.~C., {et~al.} 2000, \apj, 533, 682,
  \dodoi{10.1086/308692}

\bibitem[{{Cano-D{\'\i}az} {et~al.}(2019){Cano-D{\'\i}az}, {{\'A}vila-Reese},
  {S{\'a}nchez}, {Hern{\'a}ndez-Toledo}, {Rodr{\'\i}guez-Puebla}, {Boquien}, \&
  {Ibarra-Medel}}]{2019MNRAS.488.3929C}
{Cano-D{\'\i}az}, M., {{\'A}vila-Reese}, V., {S{\'a}nchez}, S.~F., {et~al.}
  2019, \mnras, 488, 3929, \dodoi{10.1093/mnras/stz1894}

\bibitem[{{Cao} {et~al.}(2016){Cao}, {Xu}, {Domingue}, {Buat}, {Cheng}, {Gao},
  {Huang}, {Jarrett}, {Lisenfeld}, {Lu}, {Mazzarella}, {Sun}, {Wu}, {Yun},
  {Ronca}, \& {Jacques}}]{2016ApJS..222...16C}
{Cao}, C., {Xu}, C.~K., {Domingue}, D., {et~al.} 2016, \apjs, 222, 16,
  \dodoi{10.3847/0067-0049/222/2/16}

\bibitem[{{Capelo} {et~al.}(2017){Capelo}, {Dotti}, {Volonteri}, {Mayer},
  {Bellovary}, \& {Shen}}]{2017MNRAS.469.4437C}
{Capelo}, P.~R., {Dotti}, M., {Volonteri}, M., {et~al.} 2017, \mnras, 469,
  4437, \dodoi{10.1093/mnras/stx1067}

\bibitem[{{Cappellari}(2017)}]{2017MNRAS.466..798C}
{Cappellari}, M. 2017, \mnras, 466, 798, \dodoi{10.1093/mnras/stw3020}

\bibitem[{{Cappellari} \& {Emsellem}(2004)}]{2004PASP..116..138C}
{Cappellari}, M., \& {Emsellem}, E. 2004, \pasp, 116, 138,
  \dodoi{10.1086/381875}

\bibitem[{{Chang} {et~al.}(2017){Chang}, {Le Floc'h}, {Juneau}, {da Cunha},
  {Salvato}, {Civano}, {Marchesi}, {Ilbert}, {Toba}, {Lim}, {Tang}, {Wang},
  {Ferraro}, {Urry}, {Griffiths}, \& {Kartaltepe}}]{2017ApJS..233...19C}
{Chang}, Y.-Y., {Le Floc'h}, E., {Juneau}, S., {et~al.} 2017, \apjs, 233, 19,
  \dodoi{10.3847/1538-4365/aa97da}

\bibitem[{{Cherinka} {et~al.}(2019){Cherinka}, {Andrews},
  {S{\'a}nchez-Gallego}, {Brownstein}, {Argudo-Fern{\'a}ndez}, {Blanton},
  {Bundy}, {Jones}, {Masters}, {Law}, {Rowlands}, {Weijmans}, {Westfall}, \&
  {Yan}}]{2019AJ....158...74C}
{Cherinka}, B., {Andrews}, B.~H., {S{\'a}nchez-Gallego}, J., {et~al.} 2019,
  \aj, 158, 74, \dodoi{10.3847/1538-3881/ab2634}

\bibitem[{{Cid Fernandes} {et~al.}(2011){Cid Fernandes}, {Stasi{\'n}ska},
  {Mateus}, \& {Vale Asari}}]{2011MNRAS.413.1687C}
{Cid Fernandes}, R., {Stasi{\'n}ska}, G., {Mateus}, A., \& {Vale Asari}, N.
  2011, \mnras, 413, 1687, \dodoi{10.1111/j.1365-2966.2011.18244.x}

\bibitem[{{Cid Fernandes} {et~al.}(2010){Cid Fernandes}, {Stasi{\'n}ska},
  {Schlickmann}, {Mateus}, {Vale Asari}, {Schoenell}, \&
  {Sodr{\'e}}}]{2010MNRAS.403.1036C}
{Cid Fernandes}, R., {Stasi{\'n}ska}, G., {Schlickmann}, M.~S., {et~al.} 2010,
  \mnras, 403, 1036, \dodoi{10.1111/j.1365-2966.2009.16185.x}

\bibitem[{{Coil} {et~al.}(2017){Coil}, {Mendez}, {Eisenstein}, \&
  {Moustakas}}]{2017ApJ...838...87C}
{Coil}, A.~L., {Mendez}, A.~J., {Eisenstein}, D.~J., \& {Moustakas}, J. 2017,
  \apj, 838, 87, \dodoi{10.3847/1538-4357/aa63ec}

\bibitem[{{Coldwell} \& {Lambas}(2006)}]{2006MNRAS.371..786C}
{Coldwell}, G.~V., \& {Lambas}, D.~G. 2006, \mnras, 371, 786,
  \dodoi{10.1111/j.1365-2966.2006.10712.x}

\bibitem[{{Comerford} {et~al.}(2020){Comerford}, {Negus},
  {M{\"u}ller-S{\'a}nchez}, {Eracleous}, {Wylezalek}, {Storchi-Bergmann},
  {Greene}, {Barrows}, {Nevin}, {Roy}, \& {Stemo}}]{2020ApJ...901..159C}
{Comerford}, J.~M., {Negus}, J., {M{\"u}ller-S{\'a}nchez}, F., {et~al.} 2020,
  \apj, 901, 159, \dodoi{10.3847/1538-4357/abb2ae}

\bibitem[{{Cox} {et~al.}(2008){Cox}, {Jonsson}, {Somerville}, {Primack}, \&
  {Dekel}}]{2008MNRAS.384..386C}
{Cox}, T.~J., {Jonsson}, P., {Somerville}, R.~S., {Primack}, J.~R., \& {Dekel},
  A. 2008, \mnras, 384, 386, \dodoi{10.1111/j.1365-2966.2007.12730.x}

\bibitem[{{Darg} {et~al.}(2010){Darg}, {Kaviraj}, {Lintott}, {Schawinski},
  {Sarzi}, {Bamford}, {Silk}, {Andreescu}, {Murray}, {Nichol}, {Raddick},
  {Slosar}, {Szalay}, {Thomas}, \& {Vandenberg}}]{2010MNRAS.401.1552D}
{Darg}, D.~W., {Kaviraj}, S., {Lintott}, C.~J., {et~al.} 2010, \mnras, 401,
  1552, \dodoi{10.1111/j.1365-2966.2009.15786.x}

\bibitem[{{de la Torre} {et~al.}(2011){de la Torre}, {Le F{\`e}vre},
  {Porciani}, {Guzzo}, {Meneux}, {Abbas}, {Tasca}, {Carollo}, {Contini},
  {Kneib}, {Lilly}, {Mainieri}, {Renzini}, {Scodeggio}, {Zamorani}, {Bardelli},
  {Bolzonella}, {Bongiorno}, {Caputi}, {Coppa}, {Cucciati}, {de Ravel},
  {Franzetti}, {Garilli}, {Halliday}, {Iovino}, {Kampczyk}, {Knobel},
  {Koekemoer}, {Kova{\v{c}}}, {Lamareille}, {Le Borgne}, {Le Brun}, {Maier},
  {Mignoli}, {Pell{\'o}}, {Peng}, {Perez-Montero}, {Ricciardelli}, {Silverman},
  {Tanaka}, {Tresse}, {Vergani}, {Zucca}, {Bottini}, {Cappi}, {Cassata},
  {Cimatti}, {Leauthaud}, {Maccagni}, {Marinoni}, {McCracken}, {Memeo},
  {Oesch}, {Pozzetti}, \& {Scaramella}}]{2011MNRAS.412..825D}
{de la Torre}, S., {Le F{\`e}vre}, O., {Porciani}, C., {et~al.} 2011, \mnras,
  412, 825, \dodoi{10.1111/j.1365-2966.2010.17939.x}

\bibitem[{{Di Matteo} {et~al.}(2007){Di Matteo}, {Combes}, {Melchior}, \&
  {Semelin}}]{2007A&A...468...61D}
{Di Matteo}, P., {Combes}, F., {Melchior}, A.~L., \& {Semelin}, B. 2007, \aap,
  468, 61, \dodoi{10.1051/0004-6361:20066959}

\bibitem[{{Di Matteo} {et~al.}(2005){Di Matteo}, {Springel}, \&
  {Hernquist}}]{2005Natur.433..604D}
{Di Matteo}, T., {Springel}, V., \& {Hernquist}, L. 2005, \nat, 433, 604,
  \dodoi{10.1038/nature03335}

\bibitem[{{Dom{\'\i}nguez S{\'a}nchez} {et~al.}(2018){Dom{\'\i}nguez
  S{\'a}nchez}, {Huertas-Company}, {Bernardi}, {Tuccillo}, \&
  {Fischer}}]{2018MNRAS.476.3661D}
{Dom{\'\i}nguez S{\'a}nchez}, H., {Huertas-Company}, M., {Bernardi}, M.,
  {Tuccillo}, D., \& {Fischer}, J.~L. 2018, \mnras, 476, 3661,
  \dodoi{10.1093/mnras/sty338}

\bibitem[{{Donley} {et~al.}(2018){Donley}, {Kartaltepe}, {Kocevski}, {Salvato},
  {Santini}, {Suh}, {Civano}, {Koekemoer}, {Trump}, {Brusa}, {Cardamone},
  {Castro}, {Cisternas}, {Conselice}, {Croton}, {Hathi}, {Liu}, {Lucas},
  {Nair}, {Rosario}, {Sanders}, {Simmons}, {Villforth}, {Alexander}, {Bell},
  {Faber}, {Grogin}, {Lotz}, {McIntosh}, \& {Nagao}}]{2018ApJ...853...63D}
{Donley}, J.~L., {Kartaltepe}, J., {Kocevski}, D., {et~al.} 2018, \apj, 853,
  63, \dodoi{10.3847/1538-4357/aa9ffa}

\bibitem[{{Dopita} {et~al.}(2013){Dopita}, {Sutherland}, {Nicholls}, {Kewley},
  \& {Vogt}}]{2013ApJS..208...10D}
{Dopita}, M.~A., {Sutherland}, R.~S., {Nicholls}, D.~C., {Kewley}, L.~J., \&
  {Vogt}, F. P.~A. 2013, \apjs, 208, 10, \dodoi{10.1088/0067-0049/208/1/10}

\bibitem[{{Drory} {et~al.}(2015){Drory}, {MacDonald}, {Bershady}, {Bundy},
  {Gunn}, {Law}, {Smith}, {Stoll}, {Tremonti}, {Wake}, {Yan}, {Weijmans},
  {Byler}, {Cherinka}, {Cope}, {Eigenbrot}, {Harding}, {Holder}, {Huehnerhoff},
  {Jaehnig}, {Jansen}, {Klaene}, {Paat}, {Percival}, \&
  {Sayres}}]{2015AJ....149...77D}
{Drory}, N., {MacDonald}, N., {Bershady}, M.~A., {et~al.} 2015, \aj, 149, 77,
  \dodoi{10.1088/0004-6256/149/2/77}

\bibitem[{{Ellison} {et~al.}(2013){Ellison}, {Mendel}, {Patton}, \&
  {Scudder}}]{2013MNRAS.435.3627E}
{Ellison}, S.~L., {Mendel}, J.~T., {Patton}, D.~R., \& {Scudder}, J.~M. 2013,
  \mnras, 435, 3627, \dodoi{10.1093/mnras/stt1562}

\bibitem[{{Ellison} {et~al.}(2011){Ellison}, {Patton}, {Mendel}, \&
  {Scudder}}]{2011MNRAS.418.2043E}
{Ellison}, S.~L., {Patton}, D.~R., {Mendel}, J.~T., \& {Scudder}, J.~M. 2011,
  \mnras, 418, 2043, \dodoi{10.1111/j.1365-2966.2011.19624.x}

\bibitem[{{Ellison} {et~al.}(2008){Ellison}, {Patton}, {Simard}, \&
  {McConnachie}}]{2008AJ....135.1877E}
{Ellison}, S.~L., {Patton}, D.~R., {Simard}, L., \& {McConnachie}, A.~W. 2008,
  \aj, 135, 1877, \dodoi{10.1088/0004-6256/135/5/1877}

\bibitem[{{Ellison} {et~al.}(2018){Ellison}, {S{\'a}nchez}, {Ibarra-Medel},
  {Antonio}, {Mendel}, \& {Barrera-Ballesteros}}]{2018MNRAS.474.2039E}
{Ellison}, S.~L., {S{\'a}nchez}, S.~F., {Ibarra-Medel}, H., {et~al.} 2018,
  \mnras, 474, 2039, \dodoi{10.1093/mnras/stx2882}

\bibitem[{{Ellison} {et~al.}(2019){Ellison}, {Viswanathan}, {Patton},
  {Bottrell}, {McConnachie}, {Gwyn}, \& {Cuillandre}}]{2019MNRAS.487.2491E}
{Ellison}, S.~L., {Viswanathan}, A., {Patton}, D.~R., {et~al.} 2019, \mnras,
  487, 2491, \dodoi{10.1093/mnras/stz1431}

\bibitem[{{Falc{\'o}n-Barroso} {et~al.}(2011){Falc{\'o}n-Barroso},
  {S{\'a}nchez-Bl{\'a}zquez}, {Vazdekis}, {Ricciardelli}, {Cardiel}, {Cenarro},
  {Gorgas}, \& {Peletier}}]{2011A&A...532A..95F}
{Falc{\'o}n-Barroso}, J., {S{\'a}nchez-Bl{\'a}zquez}, P., {Vazdekis}, A.,
  {et~al.} 2011, \aap, 532, A95, \dodoi{10.1051/0004-6361/201116842}

\bibitem[{{Fan} {et~al.}(2016){Fan}, {Han}, {Fang}, {Gao}, {Zhang}, {Jiang},
  {Wu}, {Yang}, \& {Li}}]{2016ApJ...822L..32F}
{Fan}, L., {Han}, Y., {Fang}, G., {et~al.} 2016, \apjl, 822, L32,
  \dodoi{10.3847/2041-8205/822/2/L32}

\bibitem[{{Feng} {et~al.}(2020){Feng}, {Shen}, {Yuan}, {Riffel}, \&
  {Pan}}]{2020ApJ...892L..20F}
{Feng}, S., {Shen}, S.-Y., {Yuan}, F.-T., {Riffel}, R.~A., \& {Pan}, K. 2020,
  \apjl, 892, L20, \dodoi{10.3847/2041-8213/ab7dba}

\bibitem[{{Feng} {et~al.}(2019){Feng}, {Shen}, {Yuan}, {Luo}, {Zhang}, {Wang},
  {Wang}, {Li}, {Hou}, {Kong}, {Guo}, \& {Zuo}}]{2019ApJ...880..114F}
{Feng}, S., {Shen}, S.-Y., {Yuan}, F.-T., {et~al.} 2019, \apj, 880, 114,
  \dodoi{10.3847/1538-4357/ab24da}

\bibitem[{{Fu} {et~al.}(2018){Fu}, {Steffen}, {Gross}, {Dai}, {Isbell}, {Lin},
  {Wake}, {Xue}, {Bizyaev}, \& {Pan}}]{2018ApJ...856...93F}
{Fu}, H., {Steffen}, J.~L., {Gross}, A.~C., {et~al.} 2018, \apj, 856, 93,
  \dodoi{10.3847/1538-4357/aab364}

\bibitem[{{Gabor} {et~al.}(2016){Gabor}, {Capelo}, {Volonteri}, {Bournaud},
  {Bellovary}, {Governato}, \& {Quinn}}]{2016A&A...592A..62G}
{Gabor}, J.~M., {Capelo}, P.~R., {Volonteri}, M., {et~al.} 2016, \aap, 592,
  A62, \dodoi{10.1051/0004-6361/201527143}

\bibitem[{{Gao} {et~al.}(2020){Gao}, {Wang}, {Pearson}, {Gordon}, {Holwerda},
  {Hopkins}, {Brown}, {Bland -Hawthorn}, \& {Owers}}]{2020A&A...637A..94G}
{Gao}, F., {Wang}, L., {Pearson}, W.~J., {et~al.} 2020, \aap, 637, A94,
  \dodoi{10.1051/0004-6361/201937178}

\bibitem[{{Geller} {et~al.}(2006){Geller}, {Kenyon}, {Barton}, {Jarrett}, \&
  {Kewley}}]{2006AJ....132.2243G}
{Geller}, M.~J., {Kenyon}, S.~J., {Barton}, E.~J., {Jarrett}, T.~H., \&
  {Kewley}, L.~J. 2006, \aj, 132, 2243, \dodoi{10.1086/508258}

\bibitem[{{Glikman} {et~al.}(2015){Glikman}, {Simmons}, {Mailly}, {Schawinski},
  {Urry}, \& {Lacy}}]{2015ApJ...806..218G}
{Glikman}, E., {Simmons}, B., {Mailly}, M., {et~al.} 2015, \apj, 806, 218,
  \dodoi{10.1088/0004-637X/806/2/218}

\bibitem[{{Goulding} {et~al.}(2018){Goulding}, {Greene}, {Bezanson}, {Greco},
  {Johnson}, {Leauthaud}, {Matsuoka}, {Medezinski}, \&
  {Price-Whelan}}]{2018PASJ...70S..37G}
{Goulding}, A.~D., {Greene}, J.~E., {Bezanson}, R., {et~al.} 2018, \pasj, 70,
  S37, \dodoi{10.1093/pasj/psx135}

\bibitem[{{Groves} {et~al.}(2004){Groves}, {Dopita}, \&
  {Sutherland}}]{2004ApJS..153....9G}
{Groves}, B.~A., {Dopita}, M.~A., \& {Sutherland}, R.~S. 2004, \apjs, 153, 9,
  \dodoi{10.1086/421113}

\bibitem[{{Guo} {et~al.}(2019){Guo}, {Peng}, {Shao}, {Fu}, {Catinella},
  {Cortese}, {Yuan}, {Yan}, {Zhang}, \& {Dou}}]{2019ApJ...870...19G}
{Guo}, K., {Peng}, Y., {Shao}, L., {et~al.} 2019, \apj, 870, 19,
  \dodoi{10.3847/1538-4357/aaee88}

\bibitem[{{Heckman} \& {Best}(2014)}]{2014ARA&A..52..589H}
{Heckman}, T.~M., \& {Best}, P.~N. 2014, \araa, 52, 589,
  \dodoi{10.1146/annurev-astro-081913-035722}

\bibitem[{{Hopkins} {et~al.}(2006{\natexlab{a}}){Hopkins}, {Hernquist}, {Cox},
  {Di Matteo}, {Robertson}, \& {Springel}}]{2006ApJS..163....1H}
{Hopkins}, P.~F., {Hernquist}, L., {Cox}, T.~J., {et~al.} 2006{\natexlab{a}},
  \apjs, 163, 1, \dodoi{10.1086/499298}

\bibitem[{{Hopkins} {et~al.}(2006{\natexlab{b}}){Hopkins}, {Somerville},
  {Hernquist}, {Cox}, {Robertson}, \& {Li}}]{2006ApJ...652..864H}
{Hopkins}, P.~F., {Somerville}, R.~S., {Hernquist}, L., {et~al.}
  2006{\natexlab{b}}, \apj, 652, 864, \dodoi{10.1086/508503}

\bibitem[{{Hou} {et~al.}(2020){Hou}, {Li}, \& {Liu}}]{2020ApJ...900...79H}
{Hou}, M., {Li}, Z., \& {Liu}, X. 2020, \apj, 900, 79,
  \dodoi{10.3847/1538-4357/aba4a7}

\bibitem[{{Hwang} {et~al.}(2011){Hwang}, {Elbaz}, {Dickinson}, {Charmandaris},
  {Daddi}, {Le Borgne}, {Buat}, {Magdis}, {Altieri}, {Aussel}, {Coia},
  {Dannerbauer}, {Dasyra}, {Kartaltepe}, {Leiton}, {Magnelli}, {Popesso}, \&
  {Valtchanov}}]{2011A&A...535A..60H}
{Hwang}, H.~S., {Elbaz}, D., {Dickinson}, M., {et~al.} 2011, \aap, 535, A60,
  \dodoi{10.1051/0004-6361/201117476}

\bibitem[{{Ji} \& {Yan}(2020)}]{2020MNRAS.499.5749J}
{Ji}, X., \& {Yan}, R. 2020, \mnras, 499, 5749, \dodoi{10.1093/mnras/staa3259}

\bibitem[{{Kartaltepe} {et~al.}(2015){Kartaltepe}, {Mozena}, {Kocevski},
  {McIntosh}, {Lotz}, {Bell}, {Faber}, {Ferguson}, {Koo}, {Bassett}, {Bernyk},
  {Blancato}, {Bournaud}, {Cassata}, {Castellano}, {Cheung}, {Conselice},
  {Croton}, {Dahlen}, {de Mello}, {DeGroot}, {Donley}, {Guedes}, {Grogin},
  {Hathi}, {Hilton}, {Hollon}, {Koekemoer}, {Liu}, {Lucas}, {Martig},
  {McGrath}, {McPartland}, {Mobasher}, {Morlock}, {O'Leary}, {Peth}, {Pforr},
  {Pillepich}, {Rosario}, {Soto}, {Straughn}, {Telford}, {Sunnquist}, {Trump},
  {Weiner}, {Wuyts}, {Inami}, {Kassin}, {Lani}, {Poole}, \&
  {Rizer}}]{2015ApJS..221...11K}
{Kartaltepe}, J.~S., {Mozena}, M., {Kocevski}, D., {et~al.} 2015, \apjs, 221,
  11, \dodoi{10.1088/0067-0049/221/1/11}

\bibitem[{{Kauffmann} \& {Haehnelt}(2000)}]{2000MNRAS.311..576K}
{Kauffmann}, G., \& {Haehnelt}, M. 2000, \mnras, 311, 576,
  \dodoi{10.1046/j.1365-8711.2000.03077.x}

\bibitem[{{Kauffmann} \& {Heckman}(2009)}]{2009MNRAS.397..135K}
{Kauffmann}, G., \& {Heckman}, T.~M. 2009, \mnras, 397, 135,
  \dodoi{10.1111/j.1365-2966.2009.14960.x}

\bibitem[{{Kauffmann} {et~al.}(2004){Kauffmann}, {White}, {Heckman},
  {M{\'e}nard}, {Brinchmann}, {Charlot}, {Tremonti}, \&
  {Brinkmann}}]{2004MNRAS.353..713K}
{Kauffmann}, G., {White}, S. D.~M., {Heckman}, T.~M., {et~al.} 2004, \mnras,
  353, 713, \dodoi{10.1111/j.1365-2966.2004.08117.x}

\bibitem[{{Kauffmann} {et~al.}(2003){Kauffmann}, {Heckman}, {Tremonti},
  {Brinchmann}, {Charlot}, {White}, {Ridgway}, {Brinkmann}, {Fukugita}, {Hall},
  {Ivezi{\'c}}, {Richards}, \& {Schneider}}]{2003MNRAS.346.1055K}
{Kauffmann}, G., {Heckman}, T.~M., {Tremonti}, C., {et~al.} 2003, \mnras, 346,
  1055, \dodoi{10.1111/j.1365-2966.2003.07154.x}

\bibitem[{{Keel} {et~al.}(1985){Keel}, {Kennicutt}, {Hummel}, \& {van der
  Hulst}}]{1985AJ.....90..708K}
{Keel}, W.~C., {Kennicutt}, R.~C., J., {Hummel}, E., \& {van der Hulst}, J.~M.
  1985, \aj, 90, 708, \dodoi{10.1086/113779}

\bibitem[{{Kennicutt} {et~al.}(1987){Kennicutt}, {Keel}, {van der Hulst},
  {Hummel}, \& {Roettiger}}]{1987AJ.....93.1011K}
{Kennicutt}, Robert~C., J., {Keel}, W.~C., {van der Hulst}, J.~M., {Hummel},
  E., \& {Roettiger}, K.~A. 1987, \aj, 93, 1011, \dodoi{10.1086/114384}

\bibitem[{{Kennicutt} \& {Evans}(2012)}]{2012ARA&A..50..531K}
{Kennicutt}, R.~C., \& {Evans}, N.~J. 2012, \araa, 50, 531,
  \dodoi{10.1146/annurev-astro-081811-125610}

\bibitem[{{Kewley} \& {Dopita}(2002)}]{2002ApJS..142...35K}
{Kewley}, L.~J., \& {Dopita}, M.~A. 2002, \apjs, 142, 35,
  \dodoi{10.1086/341326}

\bibitem[{{Kewley} {et~al.}(2001){Kewley}, {Dopita}, {Sutherland}, {Heisler},
  \& {Trevena}}]{2001ApJ...556..121K}
{Kewley}, L.~J., {Dopita}, M.~A., {Sutherland}, R.~S., {Heisler}, C.~A., \&
  {Trevena}, J. 2001, \apj, 556, 121, \dodoi{10.1086/321545}

\bibitem[{{Knapen} \& {James}(2009)}]{2009ApJ...698.1437K}
{Knapen}, J.~H., \& {James}, P.~A. 2009, \apj, 698, 1437,
  \dodoi{10.1088/0004-637X/698/2/1437}

\bibitem[{{Lackner} {et~al.}(2014){Lackner}, {Silverman}, {Salvato},
  {Kampczyk}, {Kartaltepe}, {Sanders}, {Capak}, {Civano}, {Halliday}, {Ilbert},
  {Jahnke}, {Koekemoer}, {Lee}, {Le F{\`e}vre}, {Liu}, {Scoville}, {Sheth}, \&
  {Toft}}]{2014AJ....148..137L}
{Lackner}, C.~N., {Silverman}, J.~D., {Salvato}, M., {et~al.} 2014, \aj, 148,
  137, \dodoi{10.1088/0004-6256/148/6/137}

\bibitem[{{Lambas} {et~al.}(2003){Lambas}, {Tissera}, {Alonso}, \&
  {Coldwell}}]{2003MNRAS.346.1189L}
{Lambas}, D.~G., {Tissera}, P.~B., {Alonso}, M.~S., \& {Coldwell}, G. 2003,
  \mnras, 346, 1189, \dodoi{10.1111/j.1365-2966.2003.07179.x}

\bibitem[{{Larson} \& {Tinsley}(1978)}]{1978ApJ...219...46L}
{Larson}, R.~B., \& {Tinsley}, B.~M. 1978, \apj, 219, 46,
  \dodoi{10.1086/155753}

\bibitem[{{Law} {et~al.}(2015){Law}, {Yan}, {Bershady}, {Bundy}, {Cherinka},
  {Drory}, {MacDonald}, {S{\'a}nchez-Gallego}, {Wake}, {Weijmans}, {Blanton},
  {Klaene}, {Moran}, {Sanchez}, \& {Zhang}}]{2015AJ....150...19L}
{Law}, D.~R., {Yan}, R., {Bershady}, M.~A., {et~al.} 2015, \aj, 150, 19,
  \dodoi{10.1088/0004-6256/150/1/19}

\bibitem[{{Li} {et~al.}(2008{\natexlab{a}}){Li}, {Kauffmann}, {Heckman},
  {Jing}, \& {White}}]{2008MNRAS.385.1903L}
{Li}, C., {Kauffmann}, G., {Heckman}, T.~M., {Jing}, Y.~P., \& {White}, S.
  D.~M. 2008{\natexlab{a}}, \mnras, 385, 1903,
  \dodoi{10.1111/j.1365-2966.2008.13000.x}

\bibitem[{{Li} {et~al.}(2008{\natexlab{b}}){Li}, {Kauffmann}, {Heckman},
  {White}, \& {Jing}}]{2008MNRAS.385.1915L}
{Li}, C., {Kauffmann}, G., {Heckman}, T.~M., {White}, S. D.~M., \& {Jing},
  Y.~P. 2008{\natexlab{b}}, \mnras, 385, 1915,
  \dodoi{10.1111/j.1365-2966.2008.13023.x}

\bibitem[{{Li} {et~al.}(2006){Li}, {Kauffmann}, {Jing}, {White}, {B{\"o}rner},
  \& {Cheng}}]{2006MNRAS.368...21L}
{Li}, C., {Kauffmann}, G., {Jing}, Y.~P., {et~al.} 2006, \mnras, 368, 21,
  \dodoi{10.1111/j.1365-2966.2006.10066.x}

\bibitem[{{Li} {et~al.}(2021){Li}, {Shi}, {Bizyaev}, {Duckworth}, {Yan},
  {Chen}, {Bing}, {Chen}, {Yu}, \& {Riffel}}]{2021MNRAS.501...14L}
{Li}, S.-l., {Shi}, Y., {Bizyaev}, D., {et~al.} 2021, \mnras, 501, 14,
  \dodoi{10.1093/mnras/staa3618}

\bibitem[{{Lin} {et~al.}(2004){Lin}, {Koo}, {Willmer}, {Patton}, {Conselice},
  {Yan}, {Coil}, {Cooper}, {Davis}, {Faber}, {Gerke}, {Guhathakurta}, \&
  {Newman}}]{2004ApJ...617L...9L}
{Lin}, L., {Koo}, D.~C., {Willmer}, C. N.~A., {et~al.} 2004, \apjl, 617, L9,
  \dodoi{10.1086/427183}

\bibitem[{{Lin} {et~al.}(2007){Lin}, {Koo}, {Weiner}, {Chiueh}, {Coil}, {Lotz},
  {Conselice}, {Willner}, {Smith}, {Guhathakurta}, {Huang}, {Le Floc'h},
  {Noeske}, {Willmer}, {Cooper}, \& {Phillips}}]{2007ApJ...660L..51L}
{Lin}, L., {Koo}, D.~C., {Weiner}, B.~J., {et~al.} 2007, \apjl, 660, L51,
  \dodoi{10.1086/517919}

\bibitem[{{Lin} {et~al.}(2017){Lin}, {Belfiore}, {Pan}, {Bothwell}, {Hsieh},
  {Huang}, {Xiao}, {S{\'a}nchez}, {Hsieh}, {Masters}, {Ramya}, {Lin}, {Hsu},
  {Li}, {Maiolino}, {Bundy}, {Bizyaev}, {Drory}, {Ibarra-Medel}, {Lacerna},
  {Haines}, {Smethurst}, {Stark}, \& {Thomas}}]{2017ApJ...851...18L}
{Lin}, L., {Belfiore}, F., {Pan}, H.-A., {et~al.} 2017, \apj, 851, 18,
  \dodoi{10.3847/1538-4357/aa96ae}

\bibitem[{{Liu} {et~al.}(2012){Liu}, {Shen}, \&
  {Strauss}}]{2012ApJ...745...94L}
{Liu}, X., {Shen}, Y., \& {Strauss}, M.~A. 2012, \apj, 745, 94,
  \dodoi{10.1088/0004-637X/745/1/94}

\bibitem[{{Moreno} {et~al.}(2015){Moreno}, {Torrey}, {Ellison}, {Patton},
  {Bluck}, {Bansal}, \& {Hernquist}}]{2015MNRAS.448.1107M}
{Moreno}, J., {Torrey}, P., {Ellison}, S.~L., {et~al.} 2015, \mnras, 448, 1107,
  \dodoi{10.1093/mnras/stv094}

\bibitem[{{Muldrew} {et~al.}(2012){Muldrew}, {Croton}, {Skibba}, {Pearce},
  {Ann}, {Baldry}, {Brough}, {Choi}, {Conselice}, {Cowan}, {Gallazzi}, {Gray},
  {Gr{\"u}tzbauch}, {Li}, {Park}, {Pilipenko}, {Podgorzec}, {Robotham},
  {Wilman}, {Yang}, {Zhang}, \& {Zibetti}}]{2012MNRAS.419.2670M}
{Muldrew}, S.~I., {Croton}, D.~J., {Skibba}, R.~A., {et~al.} 2012, \mnras, 419,
  2670, \dodoi{10.1111/j.1365-2966.2011.19922.x}

\bibitem[{{Noeske} {et~al.}(2007){Noeske}, {Weiner}, {Faber}, {Papovich},
  {Koo}, {Somerville}, {Bundy}, {Conselice}, {Newman}, {Schiminovich}, {Le
  Floc'h}, {Coil}, {Rieke}, {Lotz}, {Primack}, {Barmby}, {Cooper}, {Davis},
  {Ellis}, {Fazio}, {Guhathakurta}, {Huang}, {Kassin}, {Martin}, {Phillips},
  {Rich}, {Small}, {Willmer}, \& {Wilson}}]{2007ApJ...660L..43N}
{Noeske}, K.~G., {Weiner}, B.~J., {Faber}, S.~M., {et~al.} 2007, \apjl, 660,
  L43, \dodoi{10.1086/517926}

\bibitem[{{Norberg} {et~al.}(2002){Norberg}, {Baugh}, {Hawkins}, {Maddox},
  {Madgwick}, {Lahav}, {Cole}, {Frenk}, {Baldry}, {Bland-Hawthorn}, {Bridges},
  {Cannon}, {Colless}, {Collins}, {Couch}, {Dalton}, {De Propris}, {Driver},
  {Efstathiou}, {Ellis}, {Glazebrook}, {Jackson}, {Lewis}, {Lumsden},
  {Peacock}, {Peterson}, {Sutherland}, \& {Taylor}}]{2002MNRAS.332..827N}
{Norberg}, P., {Baugh}, C.~M., {Hawkins}, E., {et~al.} 2002, \mnras, 332, 827,
  \dodoi{10.1046/j.1365-8711.2002.05348.x}

\bibitem[{{Oke} \& {Gunn}(1983)}]{1983ApJ...266..713O}
{Oke}, J.~B., \& {Gunn}, J.~E. 1983, \apj, 266, 713, \dodoi{10.1086/160817}

\bibitem[{{Osterbrock} \& {Ferland}(2006)}]{2006agna.book.....O}
{Osterbrock}, D.~E., \& {Ferland}, G.~J. 2006, {Astrophysics of gaseous nebulae
  and active galactic nuclei} (University Science Books)

\bibitem[{{Padovani} {et~al.}(2017){Padovani}, {Alexander}, {Assef}, {De
  Marco}, {Giommi}, {Hickox}, {Richards}, {Smol{\v{c}}i{\'c}},
  {Hatziminaoglou}, {Mainieri}, \& {Salvato}}]{2017A&ARv..25....2P}
{Padovani}, P., {Alexander}, D.~M., {Assef}, R.~J., {et~al.} 2017, \aapr, 25,
  2, \dodoi{10.1007/s00159-017-0102-9}

\bibitem[{{Pan} {et~al.}(2018){Pan}, {Lin}, {Hsieh}, {S{\'a}nchez},
  {Ibarra-Medel}, {Boquien}, {Lacerna}, {Argudo-Fern{\'a}ndez}, {Bizyaev},
  {Cano-D{\'\i}az}, {Drory}, {Gao}, {Masters}, {Pan}, {Tabor}, {Tissera}, \&
  {Xiao}}]{2018ApJ...854..159P}
{Pan}, H.-A., {Lin}, L., {Hsieh}, B.-C., {et~al.} 2018, \apj, 854, 159,
  \dodoi{10.3847/1538-4357/aaa9bc}

\bibitem[{{Pan} {et~al.}(2019){Pan}, {Lin}, {Hsieh}, {Barrera-Ballesteros},
  {S{\'a}nchez}, {Hsu}, {Keenan}, {Tissera}, {Boquien}, {Dai}, {Knapen},
  {Riffel}, {Argudo-Fern{\'a}ndez}, {Xiao}, \& {Yuan}}]{2019ApJ...881..119P}
---. 2019, \apj, 881, 119, \dodoi{10.3847/1538-4357/ab311c}

\bibitem[{{Patton} \& {Atfield}(2008)}]{2008ApJ...685..235P}
{Patton}, D.~R., \& {Atfield}, J.~E. 2008, \apj, 685, 235,
  \dodoi{10.1086/590542}

\bibitem[{{Patton} {et~al.}(2011){Patton}, {Ellison}, {Simard}, {McConnachie},
  \& {Mendel}}]{2011MNRAS.412..591P}
{Patton}, D.~R., {Ellison}, S.~L., {Simard}, L., {McConnachie}, A.~W., \&
  {Mendel}, J.~T. 2011, \mnras, 412, 591,
  \dodoi{10.1111/j.1365-2966.2010.17932.x}

\bibitem[{{Patton} {et~al.}(2005){Patton}, {Grant}, {Simard}, {Pritchet},
  {Carlberg}, \& {Borne}}]{2005AJ....130.2043P}
{Patton}, D.~R., {Grant}, J.~K., {Simard}, L., {et~al.} 2005, \aj, 130, 2043,
  \dodoi{10.1086/491672}

\bibitem[{{Patton} {et~al.}(2002){Patton}, {Pritchet}, {Carlberg}, {Marzke},
  {Yee}, {Hall}, {Lin}, {Morris}, {Sawicki}, {Shepherd}, \&
  {Wirth}}]{2002ApJ...565..208P}
{Patton}, D.~R., {Pritchet}, C.~J., {Carlberg}, R.~G., {et~al.} 2002, \apj,
  565, 208, \dodoi{10.1086/324543}

\bibitem[{{Pearson} {et~al.}(2019){Pearson}, {Wang}, {Trayford}, {Petrillo}, \&
  {van der Tak}}]{2019A&A...626A..49P}
{Pearson}, W.~J., {Wang}, L., {Trayford}, J.~W., {Petrillo}, C.~E., \& {van der
  Tak}, F.~F.~S. 2019, \aap, 626, A49, \dodoi{10.1051/0004-6361/201935355}

\bibitem[{{Rembold} {et~al.}(2017){Rembold}, {Shimoia}, {Storchi-Bergmann},
  {Riffel}, {Riffel}, {Mallmann}, {do Nascimento}, {Moreira}, {Ilha},
  {Machado}, {Cirolini}, {da Costa}, {Maia}, {Santiago}, {Schneider},
  {Wylezalek}, {Bizyaev}, {Pan}, \&
  {M{\"u}ller-S{\'a}nchez}}]{2017MNRAS.472.4382R}
{Rembold}, S.~B., {Shimoia}, J.~S., {Storchi-Bergmann}, T., {et~al.} 2017,
  \mnras, 472, 4382, \dodoi{10.1093/mnras/stx2264}

\bibitem[{{Rosario} {et~al.}(2016){Rosario}, {Mendel}, {Ellison}, {Lutz}, \&
  {Trump}}]{2016MNRAS.457.2703R}
{Rosario}, D.~J., {Mendel}, J.~T., {Ellison}, S.~L., {Lutz}, D., \& {Trump},
  J.~R. 2016, \mnras, 457, 2703, \dodoi{10.1093/mnras/stw096}

\bibitem[{{Saintonge} {et~al.}(2017){Saintonge}, {Catinella}, {Tacconi},
  {Kauffmann}, {Genzel}, {Cortese}, {Dav{\'e}}, {Fletcher},
  {Graci{\'a}-Carpio}, {Kramer}, {Heckman}, {Janowiecki}, {Lutz}, {Rosario},
  {Schiminovich}, {Schuster}, {Wang}, {Wuyts}, {Borthakur}, {Lamperti}, \&
  {Roberts-Borsani}}]{2017ApJS..233...22S}
{Saintonge}, A., {Catinella}, B., {Tacconi}, L.~J., {et~al.} 2017, \apjs, 233,
  22, \dodoi{10.3847/1538-4365/aa97e0}

\bibitem[{{Salim} {et~al.}(2007){Salim}, {Rich}, {Charlot}, {Brinchmann},
  {Johnson}, {Schiminovich}, {Seibert}, {Mallery}, {Heckman}, {Forster},
  {Friedman}, {Martin}, {Morrissey}, {Neff}, {Small}, {Wyder}, {Bianchi},
  {Donas}, {Lee}, {Madore}, {Milliard}, {Szalay}, {Welsh}, \&
  {Yi}}]{2007ApJS..173..267S}
{Salim}, S., {Rich}, R.~M., {Charlot}, S., {et~al.} 2007, \apjs, 173, 267,
  \dodoi{10.1086/519218}

\bibitem[{{Salpeter}(1955)}]{1955ApJ...121..161S}
{Salpeter}, E.~E. 1955, \apj, 121, 161, \dodoi{10.1086/145971}

\bibitem[{{S{\'a}nchez}(2020)}]{2020ARA&A..58...99S}
{S{\'a}nchez}, S.~F. 2020, \araa, 58, 99,
  \dodoi{10.1146/annurev-astro-012120-013326}

\bibitem[{{S{\'a}nchez} {et~al.}(2016{\natexlab{a}}){S{\'a}nchez}, {P{\'e}rez},
  {S{\'a}nchez-Bl{\'a}zquez}, {Gonz{\'a}lez}, {Ros{\'a}les-Ortega},
  {Cano-D{\'\i}az}, {L{\'o}pez-Cob{\'a}}, {Marino}, {Gil de Paz}, {Moll{\'a}},
  {L{\'o}pez-S{\'a}nchez}, {Ascasibar}, \&
  {Barrera-Ballesteros}}]{2016RMxAA..52...21S}
{S{\'a}nchez}, S.~F., {P{\'e}rez}, E., {S{\'a}nchez-Bl{\'a}zquez}, P., {et~al.}
  2016{\natexlab{a}}, \rmxaa, 52, 21.
\newblock \doarXiv{1509.08552}

\bibitem[{{S{\'a}nchez} {et~al.}(2016{\natexlab{b}}){S{\'a}nchez}, {P{\'e}rez},
  {S{\'a}nchez-Bl{\'a}zquez}, {Garc{\'\i}a-Benito}, {Ibarra-Mede},
  {Gonz{\'a}lez}, {Rosales-Ortega}, {S{\'a}nchez-Menguiano}, {Ascasibar},
  {Bitsakis}, {Law}, {Cano-D{\'\i}az}, {L{\'o}pez-Cob{\'a}}, {Marino}, {Gil de
  Paz}, {L{\'o}pez-S{\'a}nchez}, {Barrera-Ballesteros}, {Galbany}, {Mast},
  {Abril-Melgarejo}, \& {Roman-Lopes}}]{2016RMxAA..52..171S}
---. 2016{\natexlab{b}}, \rmxaa, 52, 171.
\newblock \doarXiv{1602.01830}

\bibitem[{{S{\'a}nchez} {et~al.}(2018){S{\'a}nchez}, {Avila-Reese},
  {Hernandez-Toledo}, {Cortes-Su{\'a}rez}, {Rodr{\'\i}guez-Puebla},
  {Ibarra-Medel}, {Cano-D{\'\i}az}, {Barrera-Ballesteros}, {Negrete},
  {Calette}, {de Lorenzo-C{\'a}ceres}, {Ortega-Minakata}, {Aquino},
  {Valenzuela}, {Clemente}, {Storchi-Bergmann}, {Riffel}, {Schimoia}, {Riffel},
  {Rembold}, {Brownstein}, {Pan}, {Yates}, {Mallmann}, \&
  {Bitsakis}}]{2018RMxAA..54..217S}
{S{\'a}nchez}, S.~F., {Avila-Reese}, V., {Hernandez-Toledo}, H., {et~al.} 2018,
  \rmxaa, 54, 217.
\newblock \doarXiv{1709.05438}

\bibitem[{{S{\'a}nchez-Bl{\'a}zquez} {et~al.}(2006){S{\'a}nchez-Bl{\'a}zquez},
  {Peletier}, {Jim{\'e}nez-Vicente}, {Cardiel}, {Cenarro},
  {Falc{\'o}n-Barroso}, {Gorgas}, {Selam}, \& {Vazdekis}}]{2006MNRAS.371..703S}
{S{\'a}nchez-Bl{\'a}zquez}, P., {Peletier}, R.~F., {Jim{\'e}nez-Vicente}, J.,
  {et~al.} 2006, \mnras, 371, 703, \dodoi{10.1111/j.1365-2966.2006.10699.x}

\bibitem[{{Sanders} \& {Mirabel}(1996)}]{1996ARA&A..34..749S}
{Sanders}, D.~B., \& {Mirabel}, I.~F. 1996, \araa, 34, 749,
  \dodoi{10.1146/annurev.astro.34.1.749}

\bibitem[{{Satyapal} {et~al.}(2014){Satyapal}, {Ellison}, {McAlpine}, {Hickox},
  {Patton}, \& {Mendel}}]{2014MNRAS.441.1297S}
{Satyapal}, S., {Ellison}, S.~L., {McAlpine}, W., {et~al.} 2014, \mnras, 441,
  1297, \dodoi{10.1093/mnras/stu650}

\bibitem[{{Schmidt} {et~al.}(2013){Schmidt}, {Rix}, {da Cunha}, {Brammer},
  {Cox}, {van Dokkum}, {F{\"o}rster Schreiber}, {Franx}, {Fumagalli},
  {Jonsson}, {Lundgren}, {Maseda}, {Momcheva}, {Nelson}, {Skelton}, {van der
  Wel}, \& {Whitaker}}]{2013MNRAS.432..285S}
{Schmidt}, K.~B., {Rix}, H.-W., {da Cunha}, E., {et~al.} 2013, \mnras, 432,
  285, \dodoi{10.1093/mnras/stt459}

\bibitem[{{Schmitt}(2001)}]{2001AJ....122.2243S}
{Schmitt}, H.~R. 2001, \aj, 122, 2243, \dodoi{10.1086/323547}

\bibitem[{{Scudder} {et~al.}(2015){Scudder}, {Ellison}, {Momjian}, {Rosenberg},
  {Torrey}, {Patton}, {Fertig}, \& {Mendel}}]{2015MNRAS.449.3719S}
{Scudder}, J.~M., {Ellison}, S.~L., {Momjian}, E., {et~al.} 2015, \mnras, 449,
  3719, \dodoi{10.1093/mnras/stv588}

\bibitem[{{Secrest} {et~al.}(2020){Secrest}, {Ellison}, {Satyapal}, \&
  {Blecha}}]{2020MNRAS.499.2380S}
{Secrest}, N.~J., {Ellison}, S.~L., {Satyapal}, S., \& {Blecha}, L. 2020,
  \mnras, 499, 2380, \dodoi{10.1093/mnras/staa1692}

\bibitem[{{Shah} {et~al.}(2020){Shah}, {Kartaltepe}, {Magagnoli}, {Cox},
  {Wetherell}, {Vanderhoof}, {Calabro}, {Chartab}, {Conselice}, {Croton},
  {Donley}, {de Groot}, {de la Vega}, {Hathi}, {Ilbert}, {Inami}, {Kocevski},
  {Koekemoer}, {Lemaux}, {Mantha}, {Marchesi}, {Martig}, {Masters}, {McGrath},
  {McIntosh}, {Moreno}, {Nayyeri}, {Pampliega}, {Salvato}, {Snyder},
  {Straughn}, {Treister}, \& {Weston}}]{2020ApJ...904..107S}
{Shah}, E.~A., {Kartaltepe}, J.~S., {Magagnoli}, C.~T., {et~al.} 2020, \apj,
  904, 107, \dodoi{10.3847/1538-4357/abbf59}

\bibitem[{{Shao} {et~al.}(2015){Shao}, {Li}, {Kauffmann}, \&
  {Wang}}]{2015MNRAS.448L..72S}
{Shao}, L., {Li}, C., {Kauffmann}, G., \& {Wang}, J. 2015, \mnras, 448, L72,
  \dodoi{10.1093/mnrasl/slu197}

\bibitem[{{Shen} {et~al.}(2016){Shen}, {Argudo-Fern{\'a}ndez}, {Chen}, {Chen},
  {Feng}, {Hou}, {Hou}, {Jiang}, {Jing}, {Kong}, {Luo}, {Luo}, {Shao}, {Wang},
  {Wang}, {Wang}, {Wu}, {Wu}, {Yang}, {Yang}, {Yuan}, {Yuan}, {Zhang}, {Zhang},
  \& {Zhang}}]{2016RAA....16...43S}
{Shen}, S.-Y., {Argudo-Fern{\'a}ndez}, M., {Chen}, L., {et~al.} 2016, Research
  in Astronomy and Astrophysics, 16, 43, \dodoi{10.1088/1674-4527/16/3/043}

\bibitem[{{Silva} {et~al.}(2021){Silva}, {Marchesini}, {Silverman}, {Martis},
  {Iono}, {Espada}, \& {Skelton}}]{2021ApJ...909..124S}
{Silva}, A., {Marchesini}, D., {Silverman}, J.~D., {et~al.} 2021, \apj, 909,
  124, \dodoi{10.3847/1538-4357/abdbb1}

\bibitem[{{Silverman} {et~al.}(2011){Silverman}, {Kampczyk}, {Jahnke},
  {Andrae}, {Lilly}, {Elvis}, {Civano}, {Mainieri}, {Vignali}, {Zamorani},
  {Nair}, {Le F{\`e}vre}, {de Ravel}, {Bardelli}, {Bongiorno}, {Bolzonella},
  {Cappi}, {Caputi}, {Carollo}, {Contini}, {Coppa}, {Cucciati}, {de la Torre},
  {Franzetti}, {Garilli}, {Halliday}, {Hasinger}, {Iovino}, {Knobel},
  {Koekemoer}, {Kova{\v{c}}}, {Lamareille}, {Le Borgne}, {Le Brun}, {Maier},
  {Mignoli}, {Pello}, {P{\'e}rez-Montero}, {Ricciardelli}, {Peng}, {Scodeggio},
  {Tanaka}, {Tasca}, {Tresse}, {Vergani}, {Zucca}, {Brusa}, {Cappelluti},
  {Comastri}, {Finoguenov}, {Fu}, {Gilli}, {Hao}, {Ho}, \&
  {Salvato}}]{2011ApJ...743....2S}
{Silverman}, J.~D., {Kampczyk}, P., {Jahnke}, K., {et~al.} 2011, \apj, 743, 2,
  \dodoi{10.1088/0004-637X/743/1/2}

\bibitem[{{Sinha} \& {Holley-Bockelmann}(2012)}]{2012ApJ...751...17S}
{Sinha}, M., \& {Holley-Bockelmann}, K. 2012, \apj, 751, 17,
  \dodoi{10.1088/0004-637X/751/1/17}

\bibitem[{{Skibba} {et~al.}(2009){Skibba}, {Bamford}, {Nichol}, {Lintott},
  {Andreescu}, {Edmondson}, {Murray}, {Raddick}, {Schawinski}, {Slosar},
  {Szalay}, {Thomas}, \& {Vandenberg}}]{2009MNRAS.399..966S}
{Skibba}, R.~A., {Bamford}, S.~P., {Nichol}, R.~C., {et~al.} 2009, \mnras, 399,
  966, \dodoi{10.1111/j.1365-2966.2009.15334.x}

\bibitem[{{Smith} \& {Struck}(2010)}]{2010AJ....140.1975S}
{Smith}, B.~J., \& {Struck}, C. 2010, \aj, 140, 1975,
  \dodoi{10.1088/0004-6256/140/6/1975}

\bibitem[{{Spindler} {et~al.}(2018){Spindler}, {Wake}, {Belfiore}, {Bershady},
  {Bundy}, {Drory}, {Masters}, {Thomas}, {Westfall}, \&
  {Wild}}]{2018MNRAS.476..580S}
{Spindler}, A., {Wake}, D., {Belfiore}, F., {et~al.} 2018, \mnras, 476, 580,
  \dodoi{10.1093/mnras/sty247}

\bibitem[{{Steffen} {et~al.}(2021){Steffen}, {Fu}, {Comerford}, {Dai}, {Feng},
  {Gross}, \& {Xue}}]{2021ApJ...909..120S}
{Steffen}, J.~L., {Fu}, H., {Comerford}, J.~M., {et~al.} 2021, \apj, 909, 120,
  \dodoi{10.3847/1538-4357/abe2a5}

\bibitem[{{Thorp} {et~al.}(2019){Thorp}, {Ellison}, {Simard}, {S{\'a}nchez}, \&
  {Antonio}}]{2019MNRAS.482L..55T}
{Thorp}, M.~D., {Ellison}, S.~L., {Simard}, L., {S{\'a}nchez}, S.~F., \&
  {Antonio}, B. 2019, \mnras, 482, L55, \dodoi{10.1093/mnrasl/sly185}

\bibitem[{{Toomre}(1977)}]{1977egsp.conf..401T}
{Toomre}, A. 1977, in Evolution of Galaxies and Stellar Populations, ed. B.~M.
  {Tinsley} \& D.~C. {Larson}, Richard B.~Gehret, 401

\bibitem[{{Toomre} \& {Toomre}(1972)}]{1972ApJ...178..623T}
{Toomre}, A., \& {Toomre}, J. 1972, \apj, 178, 623, \dodoi{10.1086/151823}

\bibitem[{{Torrey} {et~al.}(2012){Torrey}, {Cox}, {Kewley}, \&
  {Hernquist}}]{2012ApJ...746..108T}
{Torrey}, P., {Cox}, T.~J., {Kewley}, L., \& {Hernquist}, L. 2012, \apj, 746,
  108, \dodoi{10.1088/0004-637X/746/1/108}

\bibitem[{{Trump} {et~al.}(2015){Trump}, {Sun}, {Zeimann}, {Luck}, {Bridge},
  {Grier}, {Hagen}, {Juneau}, {Montero-Dorta}, {Rosario}, {Brandt},
  {Ciardullo}, \& {Schneider}}]{2015ApJ...811...26T}
{Trump}, J.~R., {Sun}, M., {Zeimann}, G.~R., {et~al.} 2015, \apj, 811, 26,
  \dodoi{10.1088/0004-637X/811/1/26}

\bibitem[{{Urrutia} {et~al.}(2008){Urrutia}, {Lacy}, \&
  {Becker}}]{2008ApJ...674...80U}
{Urrutia}, T., {Lacy}, M., \& {Becker}, R.~H. 2008, \apj, 674, 80,
  \dodoi{10.1086/523959}

\bibitem[{{Veilleux} {et~al.}(2002){Veilleux}, {Kim}, \&
  {Sanders}}]{2002ApJS..143..315V}
{Veilleux}, S., {Kim}, D.~C., \& {Sanders}, D.~B. 2002, \apjs, 143, 315,
  \dodoi{10.1086/343844}

\bibitem[{{Veilleux} \& {Osterbrock}(1987)}]{1987ApJS...63..295V}
{Veilleux}, S., \& {Osterbrock}, D.~E. 1987, \apjs, 63, 295,
  \dodoi{10.1086/191166}

\bibitem[{{Veilleux} {et~al.}(2009){Veilleux}, {Kim}, {Rupke}, {Peng},
  {Tacconi}, {Genzel}, {Lutz}, {Sturm}, {Contursi}, {Schweitzer}, {Dasyra},
  {Ho}, {Sanders}, \& {Burkert}}]{2009ApJ...701..587V}
{Veilleux}, S., {Kim}, D.~C., {Rupke}, D.~S.~N., {et~al.} 2009, \apj, 701, 587,
  \dodoi{10.1088/0004-637X/701/1/587}

\bibitem[{{Violino} {et~al.}(2018){Violino}, {Ellison}, {Sargent}, {Coppin},
  {Scudder}, {Mendel}, \& {Saintonge}}]{2018MNRAS.476.2591V}
{Violino}, G., {Ellison}, S.~L., {Sargent}, M., {et~al.} 2018, \mnras, 476,
  2591, \dodoi{10.1093/mnras/sty345}

\bibitem[{{Wake} {et~al.}(2017){Wake}, {Bundy}, {Diamond-Stanic}, {Yan},
  {Blanton}, {Bershady}, {S{\'a}nchez-Gallego}, {Drory}, {Jones}, {Kauffmann},
  {Law}, {Li}, {MacDonald}, {Masters}, {Thomas}, {Tinker}, {Weijmans}, \&
  {Brownstein}}]{2017AJ....154...86W}
{Wake}, D.~A., {Bundy}, K., {Diamond-Stanic}, A.~M., {et~al.} 2017, \aj, 154,
  86, \dodoi{10.3847/1538-3881/aa7ecc}

\bibitem[{{Walmsley} {et~al.}(2019){Walmsley}, {Ferguson}, {Mann}, \&
  {Lintott}}]{2019MNRAS.483.2968W}
{Walmsley}, M., {Ferguson}, A. M.~N., {Mann}, R.~G., \& {Lintott}, C.~J. 2019,
  \mnras, 483, 2968, \dodoi{10.1093/mnras/sty3232}

\bibitem[{{Wang} {et~al.}(2019){Wang}, {Lilly}, {Pezzulli}, \&
  {Matthee}}]{2019ApJ...877..132W}
{Wang}, E., {Lilly}, S.~J., {Pezzulli}, G., \& {Matthee}, J. 2019, \apj, 877,
  132, \dodoi{10.3847/1538-4357/ab1c5b}

\bibitem[{{Wang} {et~al.}(2018){Wang}, {Li}, {Xiao}, {Lin}, {Bershady}, {Law},
  {Merrifield}, {Sanchez}, {Riffel}, {Riffel}, \& {Yan}}]{2018ApJ...856..137W}
{Wang}, E., {Li}, C., {Xiao}, T., {et~al.} 2018, \apj, 856, 137,
  \dodoi{10.3847/1538-4357/aab263}

\bibitem[{{Wang} {et~al.}(2009){Wang}, {Mo}, {Jing}, {Guo}, {van den Bosch}, \&
  {Yang}}]{2009MNRAS.394..398W}
{Wang}, H., {Mo}, H.~J., {Jing}, Y.~P., {et~al.} 2009, \mnras, 394, 398,
  \dodoi{10.1111/j.1365-2966.2008.14301.x}

\bibitem[{{Westfall} {et~al.}(2019){Westfall}, {Cappellari}, {Bershady},
  {Bundy}, {Belfiore}, {Ji}, {Law}, {Schaefer}, {Shetty}, {Tremonti}, {Yan},
  {Andrews}, {Brownstein}, {Cherinka}, {Coccato}, {Drory}, {Maraston},
  {Parikh}, {S{\'a}nchez-Gallego}, {Thomas}, {Weijmans}, {Barrera-Ballesteros},
  {Du}, {Goddard}, {Li}, {Masters}, {Ibarra Medel}, {S{\'a}nchez}, {Yang},
  {Zheng}, \& {Zhou}}]{2019AJ....158..231W}
{Westfall}, K.~B., {Cappellari}, M., {Bershady}, M.~A., {et~al.} 2019, \aj,
  158, 231, \dodoi{10.3847/1538-3881/ab44a2}

\bibitem[{{Weston} {et~al.}(2017){Weston}, {McIntosh}, {Brodwin}, {Mann},
  {Cooper}, {McConnell}, \& {Nielsen}}]{2017MNRAS.464.3882W}
{Weston}, M.~E., {McIntosh}, D.~H., {Brodwin}, M., {et~al.} 2017, \mnras, 464,
  3882, \dodoi{10.1093/mnras/stw2620}

\bibitem[{{Wild} {et~al.}(2014){Wild}, {Rosales-Ortega}, {Falc{\'o}n-Barroso},
  {Garc{\'\i}a-Benito}, {Gallazzi}, {Gonz{\'a}lez Delgado}, {Bekerait{\'e}},
  {Pasquali}, {Johansson}, {Garc{\'\i}a Lorenzo}, {van de Ven}, {Pawlik},
  {Per{\'e}z}, {Monreal-Ibero}, {Lyubenova}, {Cid Fernandes},
  {M{\'e}ndez-Abreu}, {Barrera-Ballesteros}, {Kehrig}, {Iglesias-P{\'a}ramo},
  {Bomans}, {M{\'a}rquez}, {Johnson}, {Kennicutt}, {Husemann}, {Mast},
  {S{\'a}nchez}, {Walcher}, {Alves}, {Aguerri}, {Alonso Herrero},
  {Bland-Hawthorn}, {Catal{\'a}n-Torrecilla}, {Florido}, {Gomes}, {Jahnke},
  {L{\'o}pez-S{\'a}nchez}, {de Lorenzo-C{\'a}ceres}, {Marino},
  {M{\'a}rmol-Queralt{\'o}}, {Olden}, {del Olmo}, {Papaderos}, {Quirrenbach},
  {V{\'\i}lchez}, \& {Ziegler}}]{2014A&A...567A.132W}
{Wild}, V., {Rosales-Ortega}, F., {Falc{\'o}n-Barroso}, J., {et~al.} 2014,
  \aap, 567, A132, \dodoi{10.1051/0004-6361/201321624}

\bibitem[{{Willett} {et~al.}(2013){Willett}, {Lintott}, {Bamford}, {Masters},
  {Simmons}, {Casteels}, {Edmondson}, {Fortson}, {Kaviraj}, {Keel}, {Melvin},
  {Nichol}, {Raddick}, {Schawinski}, {Simpson}, {Skibba}, {Smith}, \&
  {Thomas}}]{2013MNRAS.435.2835W}
{Willett}, K.~W., {Lintott}, C.~J., {Bamford}, S.~P., {et~al.} 2013, \mnras,
  435, 2835, \dodoi{10.1093/mnras/stt1458}

\bibitem[{{Woods} \& {Geller}(2007)}]{2007AJ....134..527W}
{Woods}, D.~F., \& {Geller}, M.~J. 2007, \aj, 134, 527, \dodoi{10.1086/519381}

\bibitem[{{Woods} {et~al.}(2006){Woods}, {Geller}, \&
  {Barton}}]{2006AJ....132..197W}
{Woods}, D.~F., {Geller}, M.~J., \& {Barton}, E.~J. 2006, \aj, 132, 197,
  \dodoi{10.1086/504834}

\bibitem[{{Woods} {et~al.}(2010){Woods}, {Geller}, {Kurtz}, {Westra},
  {Fabricant}, \& {Dell'Antonio}}]{2010AJ....139.1857W}
{Woods}, D.~F., {Geller}, M.~J., {Kurtz}, M.~J., {et~al.} 2010, \aj, 139, 1857,
  \dodoi{10.1088/0004-6256/139/5/1857}

\bibitem[{{Wyder} {et~al.}(2007){Wyder}, {Martin}, {Schiminovich}, {Seibert},
  {Budav{\'a}ri}, {Treyer}, {Barlow}, {Forster}, {Friedman}, {Morrissey},
  {Neff}, {Small}, {Bianchi}, {Donas}, {Heckman}, {Lee}, {Madore}, {Milliard},
  {Rich}, {Szalay}, {Welsh}, \& {Yi}}]{2007ApJS..173..293W}
{Wyder}, T.~K., {Martin}, D.~C., {Schiminovich}, D., {et~al.} 2007, \apjs, 173,
  293, \dodoi{10.1086/521402}

\bibitem[{{Wylezalek} {et~al.}(2018){Wylezalek}, {Zakamska}, {Greene},
  {Riffel}, {Drory}, {Andrews}, {Merloni}, \& {Thomas}}]{2018MNRAS.474.1499W}
{Wylezalek}, D., {Zakamska}, N.~L., {Greene}, J.~E., {et~al.} 2018, \mnras,
  474, 1499, \dodoi{10.1093/mnras/stx2784}

\bibitem[{{Xu} \& {Sulentic}(1991)}]{1991ApJ...374..407X}
{Xu}, C., \& {Sulentic}, J.~W. 1991, \apj, 374, 407, \dodoi{10.1086/170132}

\bibitem[{{Xu} {et~al.}(2010){Xu}, {Domingue}, {Cheng}, {Lu}, {Huang}, {Gao},
  {Mazzarella}, {Cutri}, {Sun}, \& {Surace}}]{2010ApJ...713..330X}
{Xu}, C.~K., {Domingue}, D., {Cheng}, Y.-W., {et~al.} 2010, \apj, 713, 330,
  \dodoi{10.1088/0004-637X/713/1/330}

\bibitem[{{Yan} \& {Blanton}(2012)}]{2012ApJ...747...61Y}
{Yan}, R., \& {Blanton}, M.~R. 2012, \apj, 747, 61,
  \dodoi{10.1088/0004-637X/747/1/61}

\bibitem[{{Yan} {et~al.}(2016{\natexlab{a}}){Yan}, {Tremonti}, {Bershady},
  {Law}, {Schlegel}, {Bundy}, {Drory}, {MacDonald}, {Bizyaev}, {Blanc},
  {Blanton}, {Cherinka}, {Eigenbrot}, {Gunn}, {Harding}, {Hogg},
  {S{\'a}nchez-Gallego}, {S{\'a}nchez}, {Wake}, {Weijmans}, {Xiao}, \&
  {Zhang}}]{2016AJ....151....8Y}
{Yan}, R., {Tremonti}, C., {Bershady}, M.~A., {et~al.} 2016{\natexlab{a}}, \aj,
  151, 8, \dodoi{10.3847/0004-6256/151/1/8}

\bibitem[{{Yan} {et~al.}(2016{\natexlab{b}}){Yan}, {Bundy}, {Law}, {Bershady},
  {Andrews}, {Cherinka}, {Diamond-Stanic}, {Drory}, {MacDonald},
  {S{\'a}nchez-Gallego}, {Thomas}, {Wake}, {Weijmans}, {Westfall}, {Zhang},
  {Arag{\'o}n-Salamanca}, {Belfiore}, {Bizyaev}, {Blanc}, {Blanton},
  {Brownstein}, {Cappellari}, {D'Souza}, {Emsellem}, {Fu}, {Gaulme}, {Graham},
  {Goddard}, {Gunn}, {Harding}, {Jones}, {Kinemuchi}, {Li}, {Li}, {Maiolino},
  {Mao}, {Maraston}, {Masters}, {Merrifield}, {Oravetz}, {Pan}, {Parejko},
  {Sanchez}, {Schlegel}, {Simmons}, {Thanjavur}, {Tinker}, {Tremonti}, {van den
  Bosch}, \& {Zheng}}]{2016AJ....152..197Y}
{Yan}, R., {Bundy}, K., {Law}, D.~R., {et~al.} 2016{\natexlab{b}}, \aj, 152,
  197, \dodoi{10.3847/0004-6256/152/6/197}

\bibitem[{{Yang} {et~al.}(2007){Yang}, {Mo}, {van den Bosch}, {Pasquali}, {Li},
  \& {Barden}}]{2007ApJ...671..153Y}
{Yang}, X., {Mo}, H.~J., {van den Bosch}, F.~C., {et~al.} 2007, \apj, 671, 153,
  \dodoi{10.1086/522027}

\bibitem[{{Yuan} {et~al.}(2018){Yuan}, {Argudo-Fern{\'a}ndez}, {Shen}, {Hao},
  {Jiang}, {Yin}, {Boquien}, \& {Lin}}]{2018A&A...613A..13Y}
{Yuan}, F.-T., {Argudo-Fern{\'a}ndez}, M., {Shen}, S., {et~al.} 2018, \aap,
  613, A13, \dodoi{10.1051/0004-6361/201731865}

\bibitem[{{Yuan} {et~al.}(2012){Yuan}, {Takeuchi}, {Matsuoka}, {Buat},
  {Burgarella}, \& {Iglesias-P{\'a}ramo}}]{2012A&A...548A.117Y}
{Yuan}, F.~T., {Takeuchi}, T.~T., {Matsuoka}, Y., {et~al.} 2012, \aap, 548,
  A117, \dodoi{10.1051/0004-6361/201220451}

\bibitem[{{Zehavi} {et~al.}(2005){Zehavi}, {Zheng}, {Weinberg}, {Frieman},
  {Berlind}, {Blanton}, {Scoccimarro}, {Sheth}, {Strauss}, {Kayo}, {Suto},
  {Fukugita}, {Nakamura}, {Bahcall}, {Brinkmann}, {Gunn}, {Hennessy},
  {Ivezi{\'c}}, {Knapp}, {Loveday}, {Meiksin}, {Schlegel}, {Schneider},
  {Szapudi}, {Tegmark}, {Vogeley}, {York}, \& {SDSS
  Collaboration}}]{2005ApJ...630....1Z}
{Zehavi}, I., {Zheng}, Z., {Weinberg}, D.~H., {et~al.} 2005, \apj, 630, 1,
  \dodoi{10.1086/431891}

\end{thebibliography}
\bibliographystyle{aasjournal}

\end{document}